\documentclass[12pt,preprint]{aastex}

\usepackage{float}
\usepackage{times}
\usepackage{graphicx}
\usepackage{amssymb}
\usepackage{color}

\newcommand{\msun}{$M_\odot$}

\newcommand{\mbh}{$M_{\rm BH}$}
\newcommand{\sig}{$\sigma$}
\newcommand{\msig}{$M_{\rm BH}$-$\sigma$}
\newcommand{\mlum}{${M_{\rm BH}}$-$L$}
\newcommand{\degree}{$^{\circ}$}
\newcommand{\ml}{$\Upsilon$}
\newcommand{\mlssp}{$\Upsilon_{\rm SSP}$}

\newcommand{\kms}{km~s$^{-1}$}
\newcommand{\kmseq}{{\rm km~s}^{-1}}

\def\aj{AJ}%
\def\apj{ApJ}%
\def\apjl{ApJ}%
\def\apjs{ApJS}%
\def\apss{Ap\&SS}%
\def\aap{A\&A}%
\def\aaps{A\&AS}%
\def\mnras{MNRAS}%
\def\pasp{PASP}%
\def\nat{Nature}%
\def\procspie{Proc.~SPIE}%

\long\def\symbolfootnote[#1]#2{\begingroup%
\def\thefootnote{\fnsymbol{footnote}}\footnote[#1]{#2}\endgroup}

\shorttitle{The Influence of DM on \mbh}
\shortauthors{Rusli et al.}

\begin{document}

\title{The Influence of Dark Matter Halos on Dynamical Estimates of
  Black Hole Mass: Ten New Measurements for High-$\sigma$
  Early-Type Galaxies \thanks{Based on observations at the European
    Southern Observatory Very Large Telescope [082.B-0037(A),
    083.B-0126(A), 082.B-0037(B), and 086.B-0085(A)]. This
    paper includes data taken at The McDonald Observatory of The
    University of Texas at Austin.}}

\author{S.P. Rusli\altaffilmark{1,2}, J. Thomas\altaffilmark{1,2}, R.P. Saglia\altaffilmark{1,2}, M. Fabricius\altaffilmark{1,2}, P. Erwin\altaffilmark{1,2}, R. Bender\altaffilmark{1,2},  N. Nowak\altaffilmark{3}, C.H. Lee\altaffilmark{2}, A. Riffeser\altaffilmark{2}, R. Sharp\altaffilmark{4,5}}

\altaffiltext{1}{Max-Planck-Institut f\"{u}r extraterrestrische Physik, Giessenbachstrasse, D-85748 Garching, Germany}
\altaffiltext{2}{Universit\"{a}ts-Sternwarte M\"{u}nchen, Scheinerstrasse 1, D-81679 M\"{u}nchen, Germany}
\altaffiltext{3}{Max-Planck-Institut f\"{u}r Physik, F\"{o}hringer Ring 6, D-80805 M\"{u}nchen, Germany}
\altaffiltext{4}{Anglo-Australian Observatory, PO Box 296, Epping, NSW 1710, Australia}
\altaffiltext{5}{Research School of Astronomy \& Astrophysics, The Australian National University, Cotter Road, Weston Creek, ACT 2611, Australia}

\begin{abstract}
  Adaptive-Optics assisted SINFONI observations of the central regions
  of ten early-type galaxies are presented. Based primarily on the
  SINFONI kinematics, ten black hole masses occupying the high-mass
  regime of the \msig\ relation are derived using three-integral
  Schwarzschild models. The effect of dark matter inclusion on the
  black hole mass is explored. The omission of a dark matter halo in
  the model results in a higher stellar mass-to-light ratio,
  especially when extensive kinematic data are used in the
  model. However, when the diameter of the sphere of influence --
  computed using the black hole mass derived without a dark halo -- is
  at least 10 times the PSF FWHM during the observations, it is safe
  to exclude a dark matter component in the dynamical modeling,
  i.e. the change in black hole mass is negligible. When the spatial
  resolution is marginal, restricting the mass-to-light ratio to the
  right value returns the correct \mbh\ although dark halo is not
  present in the model. Compared to the \msig\ and \mlum\ relations of
  McConnell et al. (2011), the ten black holes are all more massive
  than expected from the luminosities and seven black hole masses are
  higher than expected from the stellar velocity dispersions of the
  host bulges. Using new fitted relations which include the ten
  galaxies, we find that the space density of the most massive black
  holes ($M_{\rm BH} \gtrsim 10^9$ \msun) estimated from the \mlum\
  relation is higher than the estimate based on the \msig\ relation
  and the latter is higher than model predictions based on quasar
  counts, each by about an order of magnitude.
\end{abstract}


\maketitle

\section{Introduction}

Accurate black hole mass measurements are important because they are
the building blocks of the empirical relations between the central
black hole and the properties of the host bulge, e.g. the stellar
velocity dispersion $\sigma$ (\citealt{Ferrarese-00};
\citealt{Gebhardtletter-00}) and the luminosity $L$
(\citealt{Kormendy-95}; \citealt{Magorrian-98}). The interpretations
of the relations are crucial for understanding the growth of black
holes, their density distribution and ultimately their roles in the
formation and evolution of galaxies. It is therefore necessary to
establish unbiased and well-defined correlations, to determine if
there is truly a physical connection between the black hole and the
host galaxy (see the recent review by \citealt{KormendyHo-13}).

The upper end of these relations is the most critical part. The galaxy
sample that shapes the relations in this regime is still relatively
sparse. Based on the black hole sample compiled in
\citet{McConnell-11b}, there are only 15 galaxies with \sig\ above 250
\kms. The \msig\ and \mlum\ relations contradict each other in
predicting the mass function of the most massive black holes
\citep{Lauer-07}, with \mlum\ giving a higher density of black
holes. In addition to that, the \msig\ relation implies that the
  largest black holes powering distant quasars are rare or absent in
  the local universe. \citet{Salviander-08} find the highest $\sigma$
  in SDSS DR2 galaxies at $z<0.3$ to be 444 \kms. According to the
  latest version of the \msig\ relation \citep{McConnell-11b}, \mbh\
  of $> 10^{10}$\msun\ requires $\sigma \gtrsim430$ \kms. These
issues can be resolved if the \msig\ relation is curved upwards or has
large intrinsic scatter at the high mass end. The very recent
discovery of two black hole masses of $10^{10}$ \msun\ in BCGs
supports this view \citep{McConnell-11b}. The finding also implies
that BCGs might be the host of quasar remnants. With these new
indications, enlarging the size of the black hole sample at the
high-\sig\ end becomes even more important. On top of that, the
accuracy of the measurements also needs to be improved.

Recent publications on dynamical masses of black holes in the center
of galaxies focus on the need for including a dark matter halo (DM) in
the modeling (\citealt{Gebhardt-09}; \citealt{Shen-10};
\citealt{McConnell-11a}; \citealt{Schulze-11};
\citealt{Gebhardt-11}). These investigations were initiated by
\citet{Gebhardt-09}. They find that the black hole mass (\mbh)
estimate in M87 increases by more than a factor of two when the dark
halo is taken into account. They argue that this is due to the
degeneracy of the mass components, i.e. black hole, stars and DM. In
dynamical modeling, only the total enclosed mass is
constrained. Omitting the dark halo component in the model forces the
actual stellar mass budget to account for the galaxy's dark matter as
well, i.e. increasing the stellar mass-to-light ratio (\ml). Since
\ml\ is assumed to be constant at all radii, \mbh\ has to decrease to
compensate for the higher stellar mass. The effect of including DM is
thought to be important for massive galaxies with shallow luminosity
profiles in the center: the line-of-sight kinematics close to the
black hole is more affected by the contribution from the stars at
large radii where DM is dominant.

Subsequent papers have presented several new \mbh\ measurements with and
without DM in the models and have shown how much \mbh\ changes with the
inclusion of DM. The results vary. \citet{Shen-10} revisit NGC\,4649
and see only a negligible increase in \mbh. \citet{McConnell-11a}
report a new measurement for NGC\,6086 and observe that \mbh\ becomes
larger by a factor of six when DM is included. \citet{Schulze-11} reanalyze \mbh\ of 12
galaxies, spanning a wide range of stellar velocity dispersions, and
find an average increase of 20 percent; all the changes in \mbh\ are
well within the measurement errors. M87 is remodeled by
\citet{Gebhardt-11} using high-spatial-resolution data and they find
similar \mbh\ with and without DM. The authors attribute the
consistent \mbh\ to data which resolves the sphere of
influence (SoI) and thereby breaks the degeneracy between \ml\ and
\mbh. The spatial resolution required to guarantee an unbiased \mbh\
without the need to include DM is, however, not clear. To avoid
systematic biases in the black-hole bulge relations, understanding the
effect of DM on \mbh, and thereby providing unbiased \mbh\
measurements, is essential.

We present ten new black hole measurements, primarily based on
integral-field unit (IFU) data obtained using SINFONI with adaptive
optics. For each galaxy, we address the question of how \mbh\ would
change when DM is present in the modeling. The ten galaxies
increase the number of \mbh\ measurements at the very high mass end of
the SMBH distribution. Nine of the galaxies have velocity dispersions
greater than 250 \kms, while only one has a dispersion less than 200
\kms. They belong to a class of galaxies whose \mbh\ values are
thought to be most affected by the inclusion or exclusion of DM halos
in the models. This work provides the first direct measurements of the
black hole masses for each galaxy in the sample. The galaxies and
their properties are listed in Table \ref{sample}.

The structure of the paper is as follows. The SINFONI observations and the
data reduction are described in Section \ref{sinfoniobsanddatared}. We
present SINFONI kinematics and describe the additional kinematic data
that we use in the modeling in Section \ref{kinematicdata}. The
surface photometry and the light distribution are discussed in Section
\ref{lightdistrib}. The details of the dynamical modeling and how we
include the DM halo can be found in Section \ref{dynmodels}. Section
\ref{bhmasses} presents the resulting black hole masses, together with
the corresponding mass-to-light ratios. We discuss the influence of
including the DM on \mbh\ and \ml\ in Section \ref{bhduetodm}. The
black hole-bulge scaling relations are addressed in Section
\ref{bhbulgerel}. We conclude the paper by summarizing the results in
Section \ref{summ}.

\begin{table*}[h!]
{\small
\caption{The Sample \label{sample}}
\begin{tabular}{llllllll}
Galaxy & Type & Distance (Mpc) & ${\rm M_B}$ (mag) &  ${\rm M_V}$ (mag) & $R_e$ (\arcsec) & $\sigma_0$ (\kms) & $\sigma_e$ (\kms)\\
\hline
NGC\,1374 & E   & 19.23\,$^*$ & -19.00  & -20.37 & 24.4 & $186.4\pm3.9$  & $166.8\pm3.4$  \\
NGC\,1407 & E   & 28.05\,$^*$ & -21.49  & -22.73 & 70.3 & $270.6\pm6.1$  & $276.1\pm1.8$ \\
NGC\,1550 & E   & 51.57       & -21.13  & -22.30 & 25.5 & $308.0\pm6.2$  & $270.1\pm10.4$ \\
NGC\,3091 & E   & 51.25       & -21.76  & -22.66 & 32.9 & $321.4\pm9.3$  & $297.2\pm11.7$ \\ 
NGC\,4472 & E   & 17.14       & -21.79  & -22.86 & 225.5 & $293.8\pm2.9$  & $300.2\pm7.4$ \\
NGC\,4751 & E/S0& 26.92       & -19.71  & -20.75 & 22.8 & $349.2\pm10.3$ & $355.4\pm13.6$\\
NGC\,5328 & E   & 64.10       & -21.76  & -22.80 & 22.2 & $313.4\pm11.6$ & $332.9 \pm 1.9$\\
NGC\,5516 & E  & 58.44       & -21.87  & -22.87 & 22.1 & $307.3\pm11.9$ & $328.2\pm11.4$\\
NGC\,6861 & E/S0& 27.30\,$^*$ & -21.14  & -21.39 & 17.7 & $414.0\pm20.0$ & $388.8\pm2.6$\\
NGC\,7619 & E   & 51.52\,$^*$ & -22.01  & -22.86 & 36.9 & $322.4\pm5.6$  & $292.2\pm5.0$ \\
\hline
\end{tabular}
\tablecomments{The properties of the sample galaxies. Types are from HyperLeda (http://leda.univ-lyon1.fr), 
    except NGC\,5516 which we classify as E based on our kinematics and photometry. Asterisks in col. 3 mark distances 
  which are adopted from the SBF survey \citep{Tonry-01} after applying the Cepheid zero-point 
  correction of -0.06 mag \citep{Mei-05}; the others are calculated from the radial velocity corrected for the infall 
  velocity of Local Group to the Virgo cluster (HyperLeda) assuming $H_0 = 72$ \kms, except for NGC4472 where we take 
  the SBF distance from \citet{Mei-07}. 
  ${\rm M_B}$ is the absolute magnitude in the B-band taken from HyperLeda, used only to calculate the appropriate 
    dark halo parameters based on the scaling relation in \citet{Thomas-09} -- see equations \ref{eq1} and \ref{eq2}.
  ${\rm M_V}$ is the absolute magnitude measured in the V-band. For all galaxies but NGC\,4751 and NGC\,5516, ${\rm M_V}$ 
  is derived from the ${\rm B^0_T}$ magnitude and B-V color of RC3. For NGC\,4751, ${\rm M_V}$ is calculated from a V 
  magnitude of 11.80 from NED and for NGC\,5516, ${\rm M_V}$ is obtained from total B magnitude and B-V color from HyperLeda. 
  All ${\rm M_V}$ values are corrected for the foreground Galactic extinction of \citet{Schlegel-98}. 
  $R_e$ is the effective radius from RC3 (except for NGC\,4472, where we use the value of \citealt{Caon-94}), 
  $\sigma_0$ is the central velocity dispersion from HyperLeda and $\sigma_e$ is the velocity dispersion within $R_e$ 
  (see Section \ref{bhbulgerel}).}
}
\end{table*}

\section{SINFONI Observations and Data Reduction}
\label{sinfoniobsanddatared}
The primary data used in this work were obtained using SINFONI,
mounted at the UT4 of the VLT. SINFONI (Spectrograph for INtegral
Field Observations in the Near Infrared) consists of the near IR
Integral Field Spectrograph SPIFFI \citep{Eisenhauer-03b} and the
curvature-sensing adaptive optics module MACAO \citep{Bonnet-03}. The
image slicer in SPIFFI divides the field of view (FoV) into 32
slitlets. Preslit-optics enable the observer to set the width of the
slitlet (pixel scale) to 250 milli-arcsec (mas), 100 mas or 25 mas,
corresponding to $8\times8$~arcsec$^2$, $3\times3$~arcsec$^2$, or
$0.8\times0.8$~arcsec$^2$ FoV, respectively. The velocity
  resolutions of the three pixel scales are 60 \kms\ (250-mas), 53 \kms
  (100-mas) and 45 \kms\ (25-mas). Each slitlet is imaged onto 64
detector pixels resulting in rectangular spatial pixels
(spaxels). There are four gratings on the filter wheel covering J, H,
K and H+K band.
 
Each galaxy was observed in multiple ten-minute exposures, following
the Object-Sky-Object (O-S-O) pattern. Immediately before or after two
pattern sets were completed, we observed a telluric standard star and
a PSF standard star; the latter is used to estimate the point spread
function (PSF) during the galaxy observation. The telluric stars were
selected to be early-type stars (B stars) at approximately the same
airmass as the galaxy. For the PSF star, a nearby (single) star with a
similar brightness (R magnitude) and colour (B-R) was chosen. This
  star was observed following the O-S-O sequence as well with a
  typical exposure time of 1 minute. The resulting datacube was then
  collapsed along the spectral dimension to generate the PSF image. In
  the case of multiple observations we averaged the corresponding PSF
  star observations, weighted by the exposure time, and normalized the
  combined image to be used for the seeing correction in dynamical
  modeling and in the photometry (in the case of NGC\,1550). The
Object exposures of each galaxy and each PSF star were dithered by a
few spaxels to achieve a full sampling of the spatial axis
perpendicular to the slitlets, that is, sampling the rectangular
spaxels into two square pixels.

We used only the K-band grating (1.95-2.45 microns) for all
galaxies. For observations with adaptive optics, the seeing correction
can be performed by using the nucleus of the galaxy as the natural
guide star (NGS mode), or using the artificial/laser guide star (LGS
mode) created by the laser system PARSEC (\citealt{Rabien-04};
\citealt{Bonaccini-02}). For the latter, the galaxy nucleus acts as the
tip-tilt reference ``star''. For galaxy nuclei that were bright enough
($R < 14$ mag), we used the NGS mode. The PSF star corresponding to
each galaxy was observed using the same mode.

The observations were carried out in four separate runs in 2008 and
2009. The majority of the observations used the 100-mas pixel scale
which allowed us to resolve the expected SoI of the black hole and to
achieve a sufficiently high signal-to-noise ratio (S/N) with a
reasonable exposure time. When the weather condition worsened, we
switched to the 250-mas pixel scale. These data are useful in
providing kinematic constraints on a larger spatial scale for the
dynamical models. The details of the observations are summarized in
Table \ref{obsruns}.

Most of the data reduction steps were performed using a custom
pipeline which incorporated elements from the ESO SINFONI Pipeline
\citep{Modigliani-07}, which was written based on the SPIFFI Data
Reduction Software SPRED (\citealt{Schreiber-04};
\citealt{Abuter-06}). The bias subtraction during the data processing
left artifacts that appeared as dark stripes on SINFONI raw frames,
both science and calibration frames. Using an IDL code provided by
ESO, we removed these dark lines prior to the data reduction. We then
used the recipes in the ESO pipeline for the standard reduction
cascade, involving dark subtraction, correction for optical distortion
and detector non-linearity, flat-fielding, wavelength calibration, sky
subtraction (which includes a sky-subtraction method of
\citealp{Davies-07}) on the science, PSF and the telluric standard
frames and construction of datacubes. For the sky subtraction, we
selected a sky frame that was observed closest in time for each
standard and science frame. The science datacubes were further
telluric-corrected using the reduced standard telluric star
spectrum. This was done by first dividing out the stellar continuum
using a blackbody curve, specified by the stellar temperature. After
normalization, the resulting spectrum was used to divide the galaxy
spectra. Finally, the individual science datacubes were combined into
a single datacube, by taking into account the dither pattern during
the observations. The telluric corrections and the cube-combining step
were done using SPRED.

\begin{table*}[h!]
\caption{Details of the Observation Runs \label{obsruns}}
{\tiny
\begin{tabular}{cclccccc}
Galaxy  & Date             & Program ID       & Instrument  & Pixel       & $T_{\rm exp}$  & AO mode & PSF \\
        &                  &                  & PA (\degree)& scale (mas) & (min)         &         & FWHM (\arcsec)\\
\hline
NGC\,1374 & 2008 Nov 27           & 082.B-0037(A) & 120.0         & 100 & 80     & NGS     & 0.15  \\
NGC\,1407 & 2008 Nov 23,25        & 082.B-0037(A) &  40.0         & 100 & 200    & LGS     & 0.19  \\
NGC\,1550 & 2008 Nov 26,27        & 082.B-0037(A) &  27.8         & 100 & 120    & LGS     & 0.17  \\
NGC\,3091 & 2008 Nov 24,25        & 082.B-0037(A) & 144.3         & 100 & 80     & NGS     & 0.17  \\
          & 2009 Apr 19,20,22     & 083.B-0126(A) & 144.3         & 100 & 40     & NGS     & 0.17  \\
          & 2008 Nov 25,26,27     & 082.B-0037(A) & 144.3         & 250 & 60     & No AO   & 0.67  \\
NGC\,4472 & 2009 Apr 24           & 083.B-0126(A) & 160.0         & 250 & 60     & NGS     & 0.47  \\
NGC\,4751 & 2009 Mar 20,22        & 082.B-0037(B) & 174.6         & 100 & 80     & LGS     & 0.22  \\
NGC\,5328 & 2008 Mar 11           & 080.B-0336(A) &   0.0         & 100 & 10     & NGS     & 0.14  \\
          & 2009 Mar 10           & 082.B-0037(B) &   0.0         & 100 & 10     & NGS     & 0.14  \\
          & 2009 Apr 24,25        & 083.B-0126(A) &   0.0         & 100 & 130    & NGS     & 0.14  \\
NGC\,5516 & 2009 Mar 21,22,23     & 082.B-0037(B) &   0.0         & 100 & 140    & LGS     & 0.14  \\
NGC\,6861 & 2011 Jun 24,25, Aug 8 & 086.B-0085(A) & 142.0         & 250 & 125    & NGS     & 0.38  \\
NGC\,7619 & 2008 Nov 23,25,27     & 082.B-0037(A) &  30.0         & 100 & 120    & LGS     & 0.18  \\
\hline
\end{tabular}
}
\tablecomments{The pixel scale is stated in milli-arcsec (mas). $T_{\rm
exp}$ is the combined exposure time on-source (excluding sky exposures). AO
mode is the mode of the adaptive optics; No AO means that the observation
is seeing-limited. PSF FWHM is the full-width at half maximum derived from
the PSF star observed next in time to the galaxy. For the 100-mas data of
NGC\,3091 and NGC\,5328, the given FWHM is based on the combined PSF from
all runs.\\}
\end{table*}

\section{Kinematic Data}
\label{kinematicdata}
\subsection{SINFONI kinematics}
\label{sinfonikinematics}
\begin{figure*}
\centering
  \includegraphics[scale=0.77]{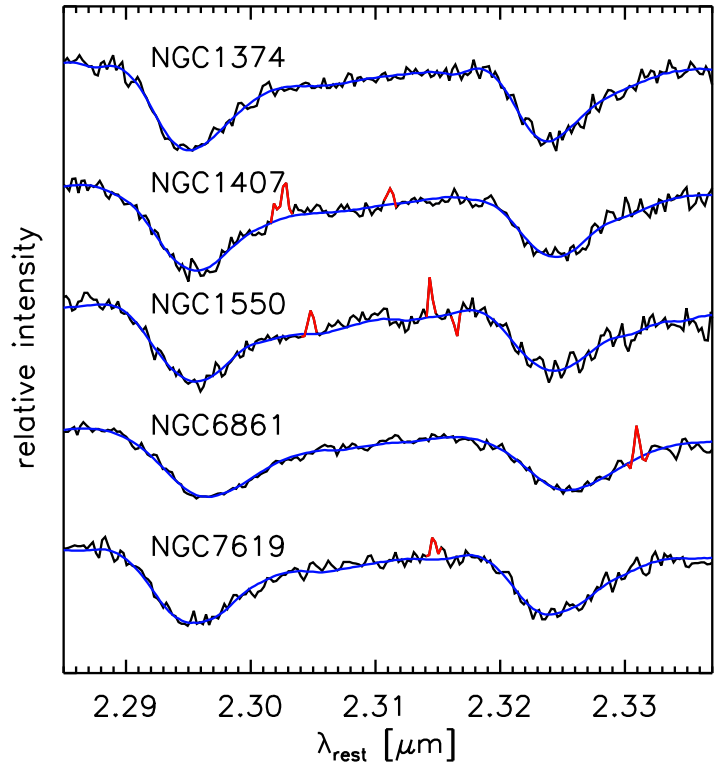}
  \caption[]{Examples of SINFONI spectra after spatial binning and
    continuum removal. For each galaxy above we display the spectrum
    derived from one of the central bins, with $\lambda_\mathrm{rest}$ 
    denoting the rest-frame wavelength. The blue lines show the best fits,
    that are performed masking the red parts.}
\label{sinfospectra}
\end{figure*}

\begin{figure*}
\centering
  \includegraphics[scale=0.77]{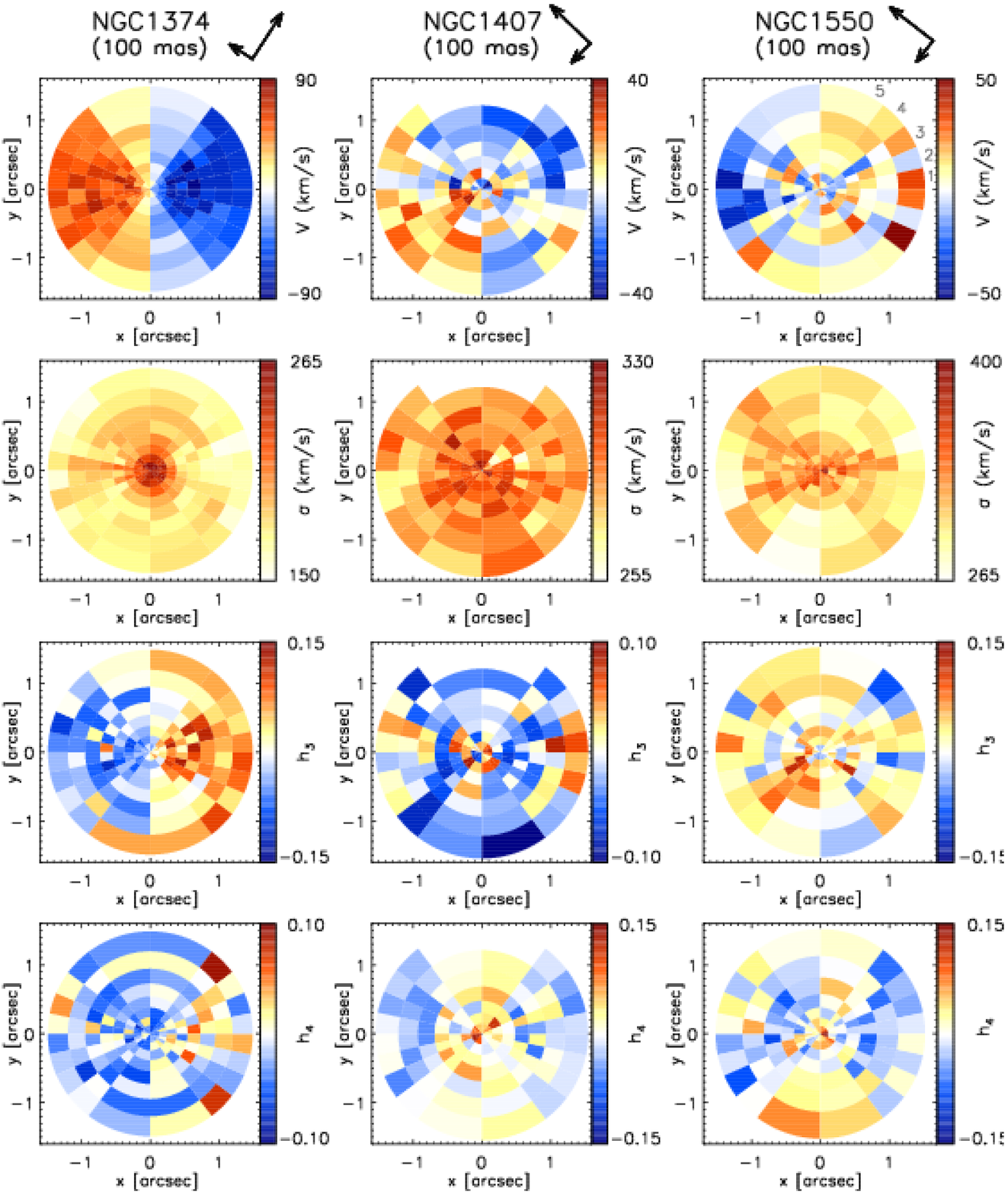}
  \caption[]{The two-dimensional kinematic maps of NGC\,1374,
    NGC\,1407 and NGC\,1550, derived from SINFONI data and presented
    in terms of $V$, \sig, $h_3$ and $h_4$. The major and minor axes
    of each galaxy are aligned with the abscissa and ordinate of the
    coordinate system. To identify the angular bins, we show the iv
    number in the velocity map of NGC\,1550 with a grey color. The
    midpoint angles for iv=1 to iv=5 are 5.8\degree, 17.6\degree,
    30.2\degree, 45.0\degree and 71.6\degree. The arrows next to the
    galaxy name indicate the North (long arrow) and East (short arrow)
    directions.\\}
\label{kinmap1}
\end{figure*}

\begin{figure*}
\centering
  \includegraphics[scale=0.77]{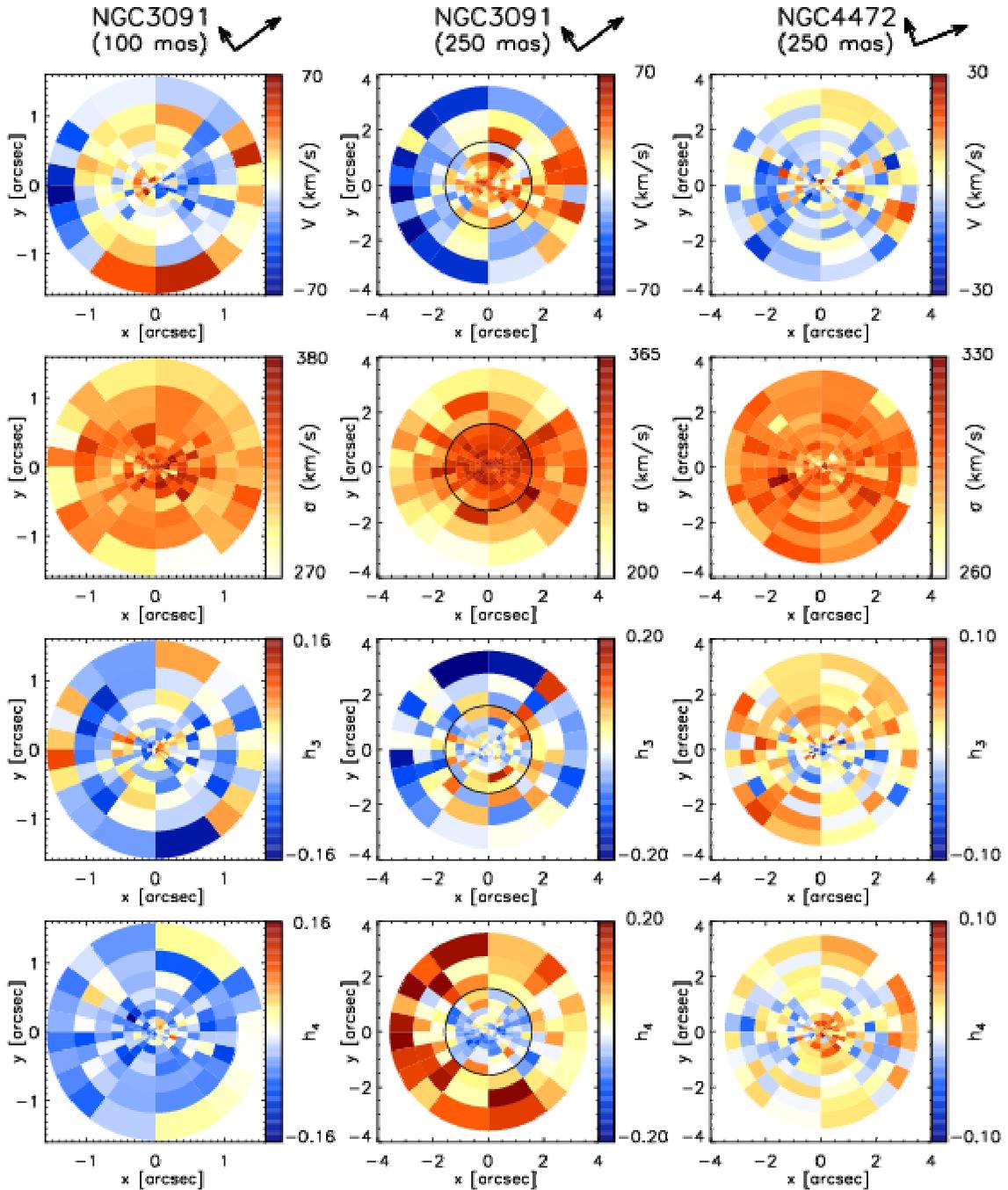}
  \caption[]{The same as Fig. \ref{kinmap1} for NGC\,3091 and
    NGC\,4472. The scope of the 100-mas data of NGC\,3091 is outlined
    in the 250-mas map (the black circle). \\}
\label{kinmap2}
\end{figure*}

\begin{figure*}
\centering
  \includegraphics[scale=0.77]{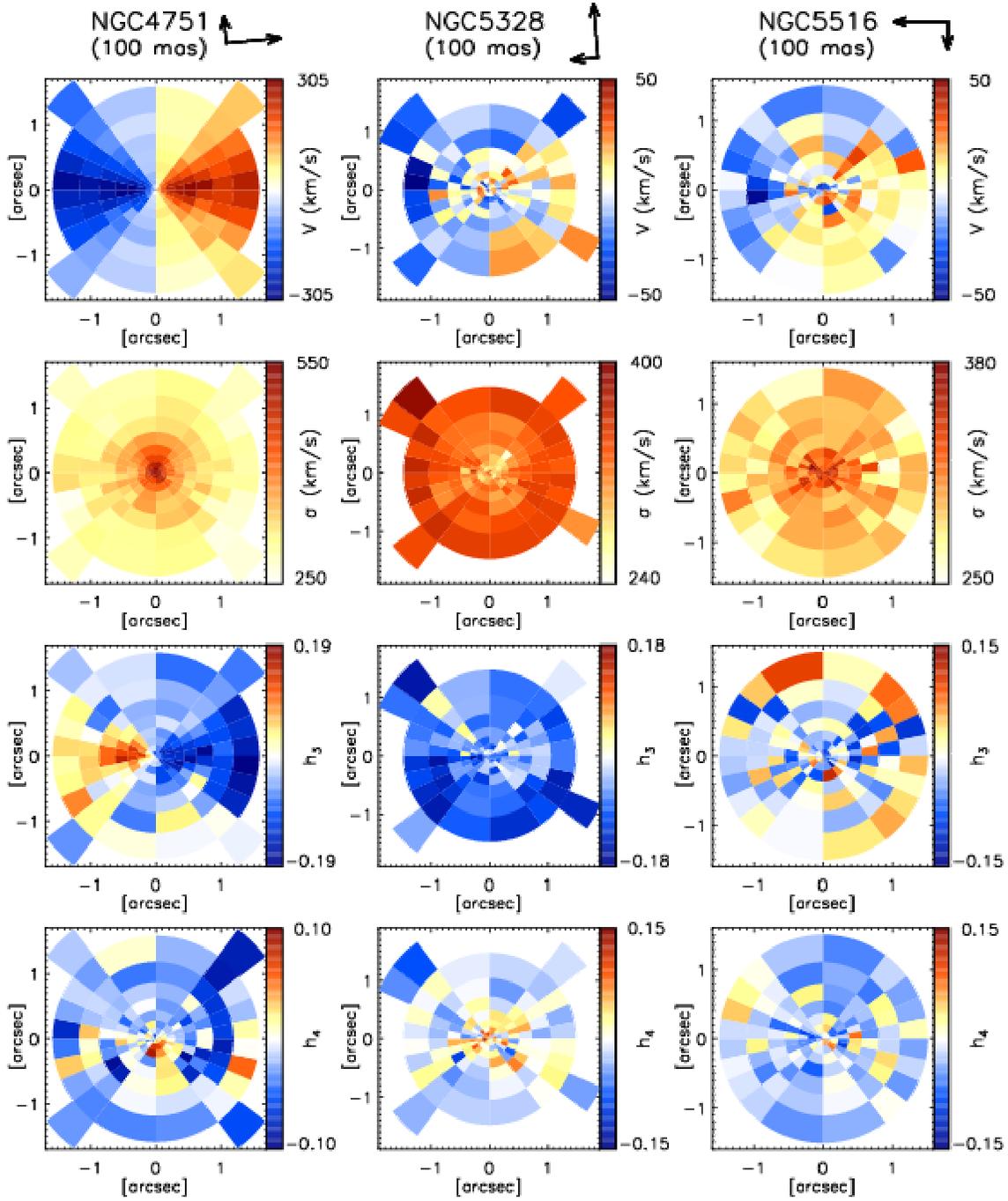}
  \caption[]{The same as Fig. \ref{kinmap1} for NGC\,4751, NGC\,5328
    and NGC\,5516. \\}
\label{kinmap3}
\end{figure*}

\begin{figure*}
\centering
  \includegraphics[scale=0.77]{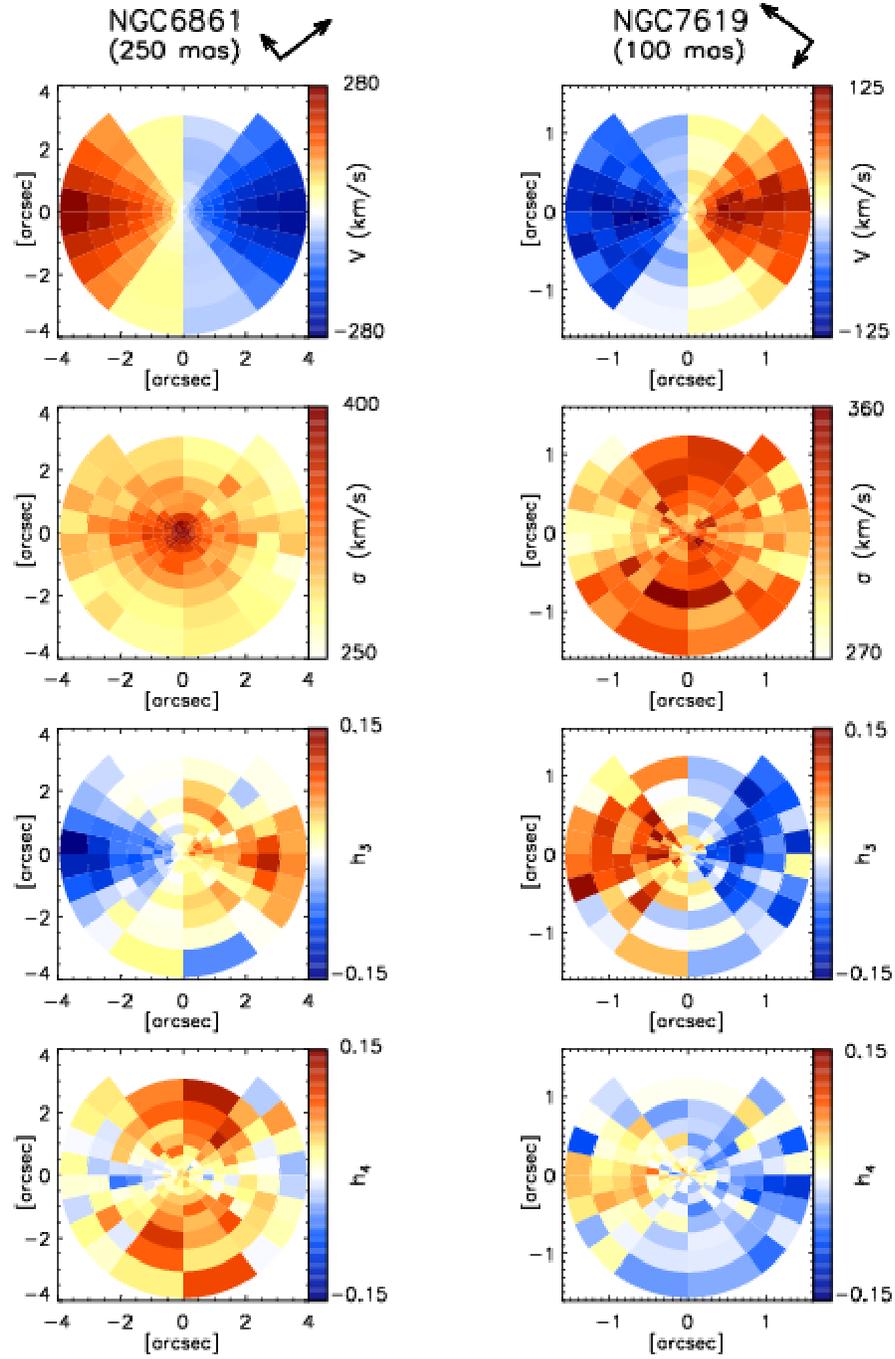}
  \caption[]{The same as Fig. \ref{kinmap1} for NGC\,6861and NGC\,7619.\\}
\label{kinmap4}
\end{figure*}

We spatially combined the spectra of individual pixels into radial and
angular bins following \citet{Gebhardt-03} to improve S/N. The binning
scheme divides each galaxy into four quadrants separated by the major
and minor axes. Every quadrant consists of five angular bins whose
centers are at the angles 5.8\degree(iv=1), 17.6\degree(iv=2),
30.2\degree(iv=3), 45.0\degree(iv=4) and 71.6\degree(iv=5). Each
angular bin is divided into 7-12 radial bins, depending on the field
of view (FoV), resolution and S/N (see
Figs. \ref{kinmap1}-\ref{kinmap4}). Using the combined spectra, we
derived the kinematics non parametrically using a Maximum Penalized
Likelihood (MPL) method \citep{Gebhardt-00}, resulting in a
line-of-sight velocity distribution (LOSVD) for each spatial bin. We
fitted the first two CO bandheads with the convolved stellar template
spectrum, which was a weighted combination of several stellar
spectra. These were stars of K and M spectral type, observed with the
same instrumental setup as the galaxy. Fig. \ref{sinfospectra}
  shows some examples of the binned SINFONI spectra.

To avoid template mismatch, we chose only stars with equivalent width or line strength within the range of those derived from the galaxy spectra, i.e. we selected only stars that were representative of the stellar population of the galaxy. For this purpose, we used the line-strength index for the first CO band-head at 2.29 microns proposed by \citet{Marmol-08}. This index definition has very little sensitivity to many aspects, most importantly S/N and velocity dispersion broadening (tested up to 400 \kms), which makes it well-suited for our work.

We derived the kinematics from the normalized spectra as follows. With
the MPL method, an initial LOSVD was generated and then convolved with
a linearly combined spectrum of template stars. The LOSVD and the
weights of the template stars were iteratively adjusted until the
convolved spectrum matched the galaxy spectrum, achieved by minimizing
$\chi_P^2=\chi^2+\alpha P$, where $\alpha$ is the smoothing parameter
and $P$, the penalty function, is the integral of squared second
derivative of the LOSVD. To estimate the smoothing parameter, we
created a large set of model galaxy spectra with an appropriate \sig\
and added varying amounts of noise to them. The smoothing required to
recover the input LOSVD for each S/N was then used to infer the
smoothing for the real galaxy spectra (see Appendix B of
\citet{Nowak-08} for details).

The S/N was calculated after continuum normalisation as the
  inverse of the rms value obtained from the spectral fitting. For
  reliable kinematic results, we avoided having S/N smaller than 30
  for each spatial bin. For an initial estimate of S/N we ran MPL
  using a fixed smoothing parameter for all spectra. The resulting fit
  (rms and thus the initial S/N) was then used to determine the
  appropriate smoothing parameter for each spectrum. When necessary,
  the binning was (re)adjusted to optimize the S/N.

The errors in the LOSVD were determined through a Monte Carlo
simulation \citep{Gebhardt-00}. The best-fitting combination of template spectra was convolved
with the measured LOSVD and gaussian noise was added to create 100
different realizations of a galaxy spectrum. The LOSVDs of the
realisations were measured and used to compute the 68 percent
confidence interval. In Figs. \ref{kinmap1}-\ref{kinmap4}, we
  present the two-dimensional kinematics of each galaxy, parametrised
  in the four Gauss-Hermite moments.  The errors of the velocity and
  $\sigma$ measurements are about 10 \kms, while the uncertainties of
  $h_3$ and $h_4$ are around 0.02, on average.


Four galaxies exhibit regular rotation: NGC\,1374, NGC\,4751,
  NGC\,6861 and NGC\,7619. In these galaxies, $h_3$ is anti-correlated
  with the rotational velocity, matching a widely observed trend
  (e.g. \citealt{Bender-94,Krajnovic-08,Krajnovic-11,Fabricius-12}). Of
  the four galaxies, NGC1374 has the lowest rotation; it also has the
  lowest velocity dispersion among the galaxies in the sample. The
  $\sigma$ profile of NGC\,1374 shows a rather strong and clear
  gradient, peaking at around 260 \kms\ in the center. NGC\,4751 and
  NGC\,6861 show the fastest rotation and the highest dispersion. A
  very steep gradient of $\sigma$ is seen in NGC\,4751: from just
  below 500 \kms\ in the center down to 250 \kms within the 1.5 arcsec
  radius. In NGC\,7619, the velocity dispersion profile decreases more
  steeply along the major axis than along the minor axis and the
  majority of the LOSVDs are flat-topped as indicated by negative
  values in the $h_4$ map.

Five galaxies are dispersion-dominated with rotational velocity
  not more than 50 \kms\ and no obvious rotation axis: NGC\,1407,
  NGC\,1550, NGC\,4472, NGC\,5328 and NGC\,5516. In this group,
  NGC\,1550, NGC\,5328 and NGC\,5516 have the highest $\sigma$,
  although none are higher than NGC\,4751. In NGC\,1550 and NGC\,5516
  $\sigma$ increases towards the centre, while in NGC\,5328 it decreases.
  The $\sigma$ gradients in NGC\,1407 and NGC\,4472 are shallow. The
lack of rotation in NGC\,4472 is in agreement with
  \citet{vanderMarel-90} who found no rotation within a radius of 5
  arcsec.

For NGC\,5328, the MPL fit yields a negative $h_3$ throughout the SINFONI FoV. This is
  atypical for early-type galaxies, regardless of whether they are rotating
  systems (where $h_3$ is usually anti-correlated with the velocity)
  or dispersion dominated systems ($h_3 \sim 0$). With a systemic
  velocity of 4740\,\kms, NGC\,5328 has the largest redshift of all
  the galaxies in our sample. This means that the second CO bandhead is shifted into
  the red end of the SINFONI spectral range, which is significantly
  contaminated by residual OH emission. We therefore used only the
  first bandhead in the kinematic fitting. Given the single bandhead
  fit, we can not exclude the possibility that the negative $h_3$ may
  be the result of a faint sky emission line that is still strong
  enough to affect the extracted Gauss-Hermite moments. In the models,
  we therefore exclude the measured $h_3$ and $h_4$.

In some galaxies, like NGC\,1407 and NGC\,4472, one observes a
  slight trend for $h_4$ to increase towards the centre. In the
  analytical toy models of \citet{Baess-05} (which, however, assume
  spherical symmetry and isotropy) $h_4$ tends to increase with the
  mass of the central black hole (relative to the rest of the system).
  However, the absolute values also
  depend on the steepness of the light profile and the increase in
  $h_4$ with the mass of the central black-hole is, even at central
  radii, not everywhere monotonic. It is therefore not clear if the
  observed central $h_4$ values are more indicative of the orbital or
  of the mass structure in the centres of the galaxies.

NGC\,3091 displays a signature of a kinematically decoupled
  core. The stars at $r<1$ arcsec appear to be rotating in the
  opposite direction from the stars at larger radii. The $h_4$ within
  $\sim$1.5 arcsec is predominantly negative, but becomes positive
  outside this radius, which provides another evidence for a decoupled
  core. The 100-mas and 250-mas datasets for NGC\,3091 overlap in the
  inner 1.5 arcsec radius. For the purpose of the modelling, we used
  only the data with higher spatial resolution up to the 6th radial
  bin ($r < 0.65$\arcsec) and used the 250-mas data to cover the
  region with 0.65\arcsec $< r <$ 4\arcsec. 

\begin{figure}
 \centering
 \includegraphics[scale=0.5]{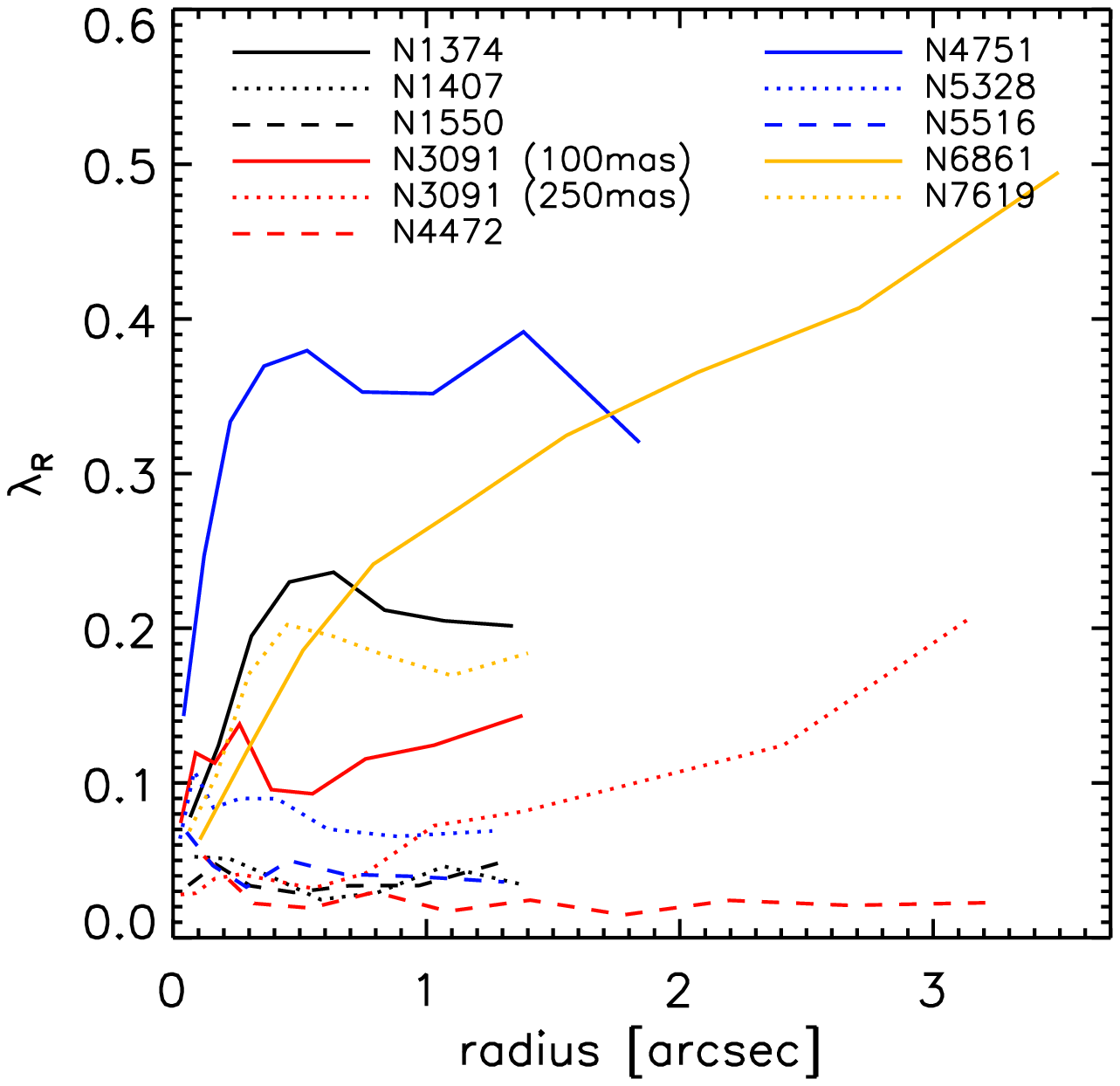} 
 \caption{The $\lambda_R$ parameter of \citet{Emsellem-07} for the sample galaxies.}
\label{lambdar}
\end{figure}

In Fig.~\ref{lambdar} we show the $\lambda_R$ parameter of
  \citet{Emsellem-07} for our galaxies. A global classification into
  slow and fast rotators based on this plot is difficult due to the
  small field of view (e.g. \citealt{Emsellem-11}). Note also, that the
  field of view is not much larger than the PSF, which affects the derived 
  $\lambda_R$ as seen clearly for NGC\,3091.

\subsection{Additional kinematics}
\label{additionalkinematics}
In addition to the SINFONI data, we used a set of ground-based
kinematic data at larger radii for the dynamical modeling of each
galaxy. These data were necessary to get a good handle on \ml\ and
DM. Most of the datasets came in the form of long slit data and were
taken from the literature. All these auxiliary data were available in
the form of Gauss-Hermite moments with the respective errors. From
these moments, we constructed the LOSVDs with the uncertainties
measured through 100 Monte Carlo realizations. We used these LOSVDs in
the modeling.

{\bf NGC\,1374}.  The auxiliary dataset for this galaxy is long slit
data along the major axis (up to $\sim$20 arcsec) taken with the EMMI
spectrograph at the ESO NTT. For the details on the observation, reduction and the
kinematic analysis, see Appendix \ref{appendix13746861}. There are
other long slit data available in the literature,
i.e. \citet{Longo-94} and \citet{Donofrio-95}. We did not attempt to
include these data because they did not provide $h_3$ and $h_4$
information and the data were less extended than the long-slit data
that we use in this work.

{\bf NGC\,1407}. 
We used the slit data from \citet{Spolaor-08a}, who provided kinematics along the major axis up to 40 arcsec, a little more than 0.5$R_e$. 

{\bf NGC\,1550}. We used long-slit kinematic data from
\citet{Simien-00}, who provided only $V$ and $\sigma$
measurements. Using this dataset, we generated LOSVDs by setting $h_3$
and $h_4$ to zero with uncertainties of 0.1, allowing the models some
extra freedom in fitting the LOSVD without biasing them into negative
or positive $h_3$ and $h_4$ values. These data extend out to 26 arcsec
along the major axis. The slit was placed at a position angle of
38\degree. The SINFONI data was adjusted to this PA to align the slit
with the photometric major axis.

{\bf NGC\,3091}. For this galaxy, we obtained a set of new IFU data
using VIRUS-W at the McDonald 2.7m telescope (see
Appendix \ref{appendix3091}). Each fiber has a size of 3.14 arcsec in
diameter, much larger than the model bin size in the central 10
arcsec. We therefore ignored the data inside 10 arcsec. Since SINFONI
data are available only out to 4 arcsec, the region between 4 and 10
arcsec is not covered by any data. Between 10 and 50 arcsec, where the
S/N is sufficiently good, we averaged the spectra of the fibers that
fell on the same model bin and derived the kinematics from these
spectra. The VIRUS-W data was binned into radial and angular bins,
like the SINFONI data.

{\bf NGC\,4472}. We used long slit data from \citet{Bender-94}, who presented kinematics of NGC\,4472 out to $\sim75$ arcsec along the major axis and out to 60 arcsec along the minor axis. 

\begin{figure}
 \centering
 \includegraphics[scale=0.5]{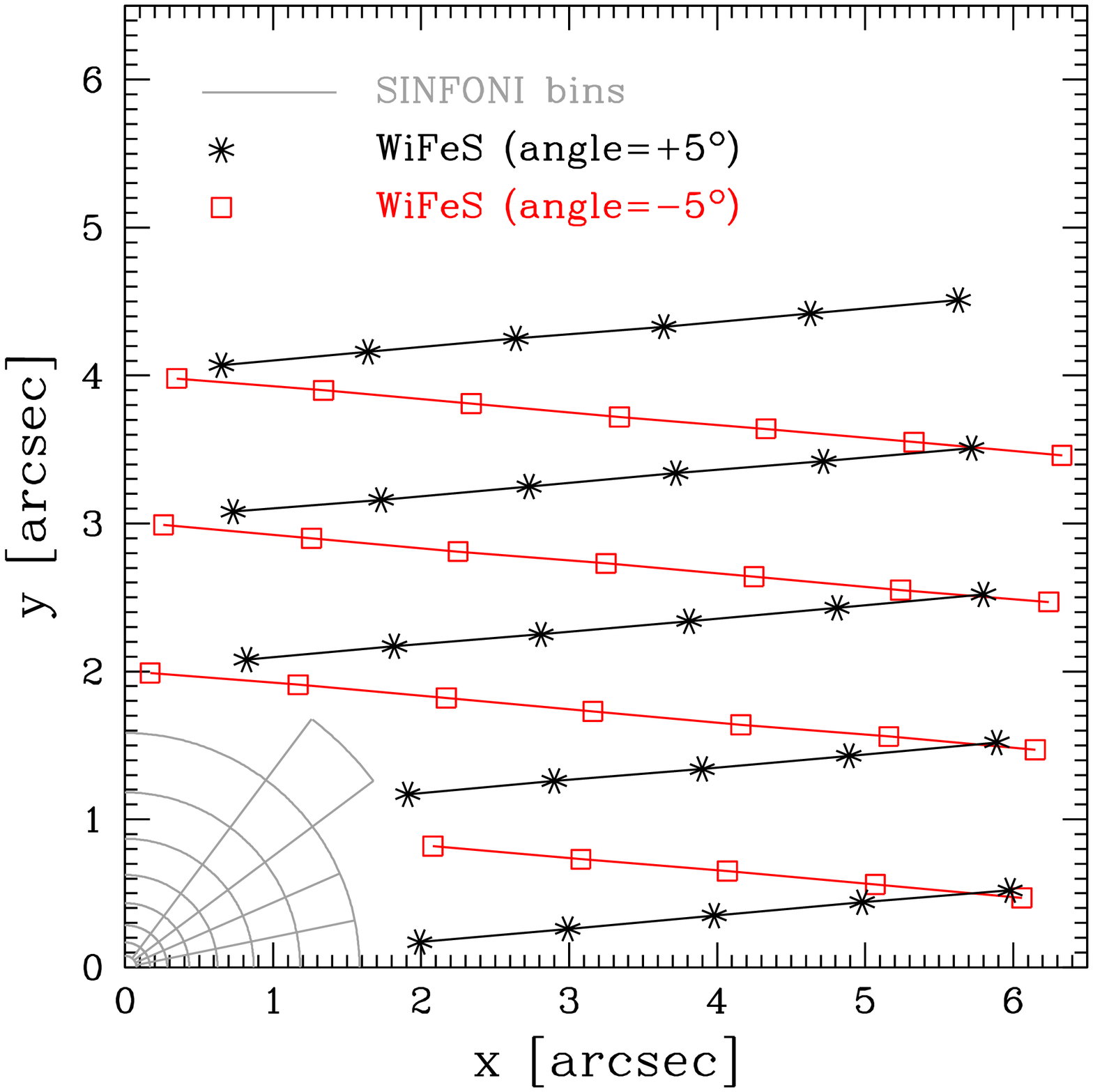} 
\caption{Spatial orientation of SINFONI and WiFeS datapoints in the
   folded quadrant of NGC\,4751. The x and y-axes correspond to the
   abscissa and ordinate in Fig. \ref{kinmap3}. The grey lines outline
   the SINFONI bins. WiFeS datapoints are plotted as asterisks and
   rectangles, connected by solid lines with the corresponding colors
   to depict the ``pseudoslits". The red lines show the pseudoslits
   positioned at an angle of -5\degree and the black lines outline
   those the angle of 5\degree.}
\label{ori4751}
\end{figure}

{\bf NGC\,4751}.  As complementary data, we used an IFU dataset
obtained using the Wide-Field Spectrograph WiFeS at the Siding Spring
ANU 2.3m telescope (see Appendix \ref{appendix47515516}). In the
modeling, the WiFeS datapoints were treated as multiple slits across
the central part of the galaxy. They were parallel to each other and
did not align with the major or minor axis. Due to the axisymmetry
assumption we folded the four quadrants of the galaxy; LOSVDs with
identical spatial position in the folded quadrant were averaged. This
resulted in two sets of slit data which were different in spatial
orientations. The first set of slits was positioned at an angle of
5\degree\ and the second one at -5\degree, measured counter-clockwise
(positive sign) from the major axis, each having 5 and 4 pseudo-slits,
respectively (see Fig. \ref{ori4751} for clarity). WiFeS data
  overlapping with the SINFONI data were excluded from the orbit
  modelling.

\begin{figure}
 \centering
 \includegraphics[scale=0.75]{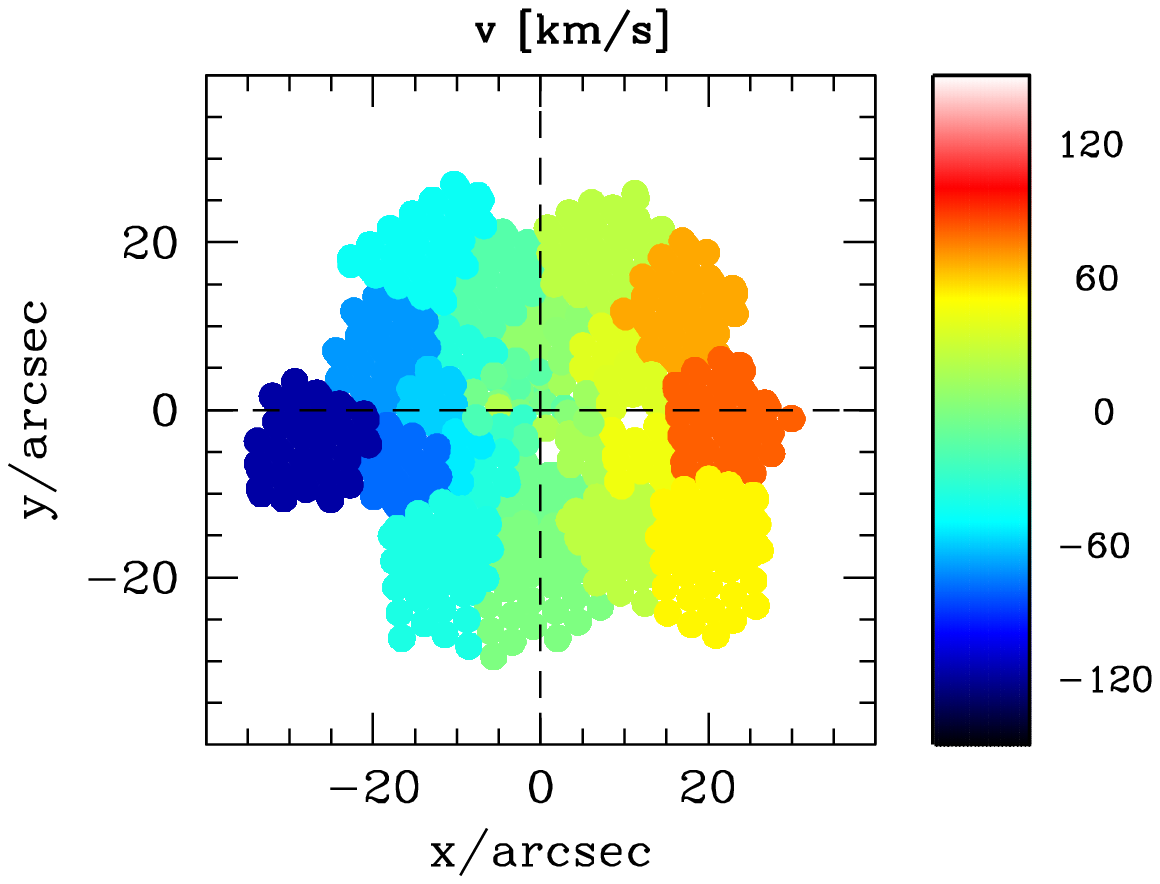}
 \caption{The velocity field of NGC\,5328 as observed with VIRUS-W.}
\label{ori5328}
\end{figure}

{\bf NGC\,5328}. As for NGC\,3091 we obtained new IFU data using
  VIRUS-W (see Appendix \ref{appendix3091}). But in this case 
  we employed a Voronoi binning scheme which yields a median S/N per
  \AA~ of 25. The bins cover a region out to about 30 arsec along the
  major and out to 20 arcsec along the minor axis. The binning is
  illustrated in Fig. \ref{ori5328}.

\begin{figure}
 \centering
 \includegraphics[scale=0.5]{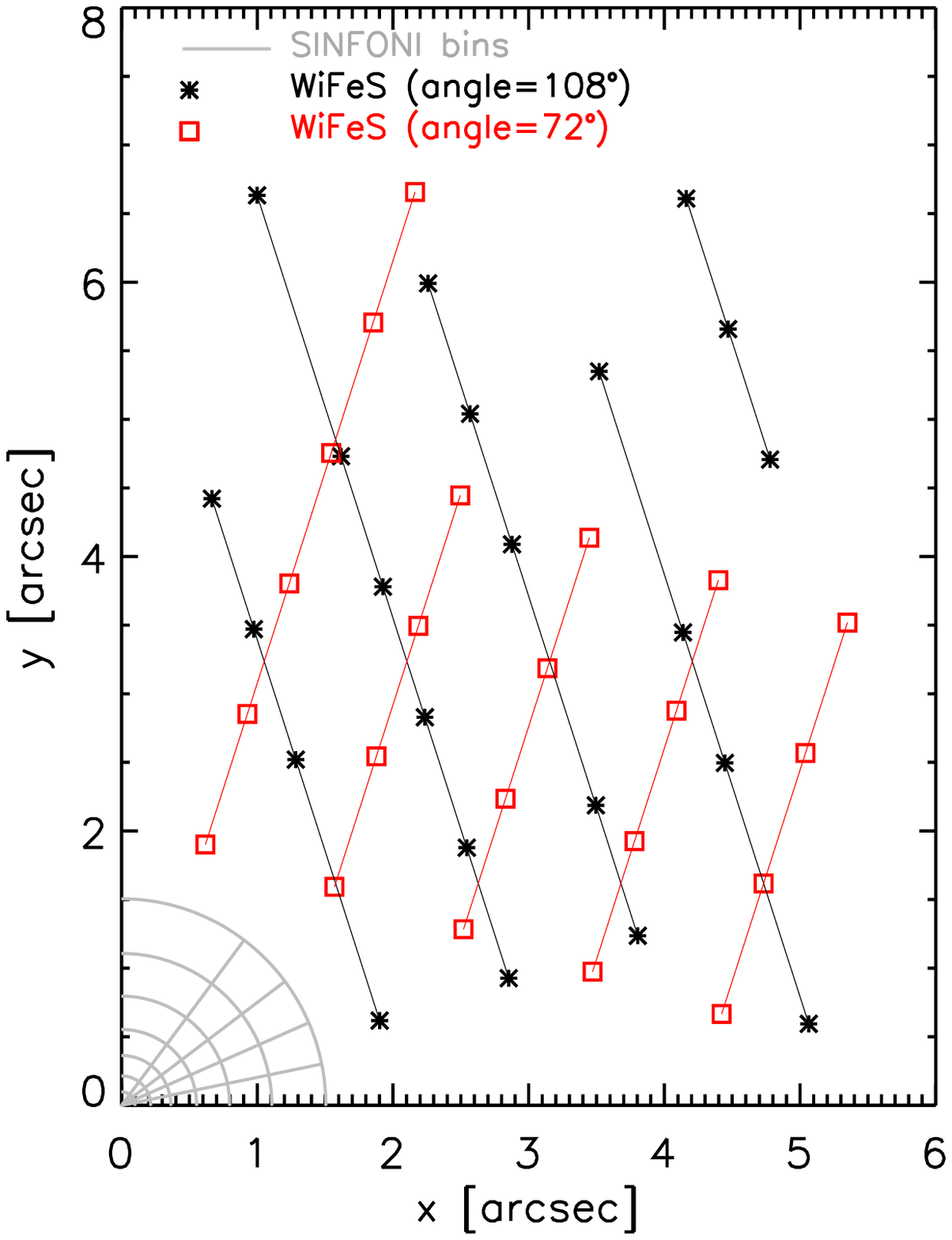} 
 \caption{The same as Fig. \ref{ori4751} for NGC\,5516. The red lines show the pseudoslits positioned at an angle of 72\degree and the black lines outline those at the angle of 108\degree.}
\label{ori5516}
\end{figure}

{\bf NGC\,5516}. The extended kinematic data for this galaxy were also
observed using WiFeS (Appendix \ref{appendix47515516}). The data used
in the model were limited up to 7 arcsec. Outside this radius the S/N
was low, giving unreasonable kinematic moments or uncertainties. The
WiFeS data were treated and folded in the same way as for
NGC\,4751. The two sets of ``pseudo-''slit data were at angles of
72\degree\ and 108\degree, measured counter-clockwise from the major
axis. Each set consists of five parallel slits (see
Fig. \ref{ori5516}). We did not include WiFeS datapoints located
  inside the SINFONI FoV in the dynamical modelling.

{\bf NGC\,6861}. {The additional dataset for this galaxy is long slit data along the major axis (up to $\sim$10 arcsec) taken using the EMMI spectrograph. For the details on the observation, reduction and the kinematic analysis, see Appendix \ref{appendix13746861}.}

{\bf NGC\,7619}. For this galaxy, we used the kinematic data from
\citet{Pu-10} who provided $V$, \sig, $h_3$ and $h_4$ out to 60
arcsec, almost twice the $R_e$, along both the major and minor
axes. The spatial distribution of the original datapoints was dense, slowing
down the computation while not increasing the significance of the
results. We therefore spatially rebinned the data to create a dataset with fewer
data points and used this in the modeling.

\section{Light distribution}
\label{lightdistrib}
To put constraints on the spatial distribution of stars we performed surface photometry and deprojected this 2D information into a light density distribution. The deprojection was done axisymetrically using the code of \citet{Magorrian-99}. The seeing correction was included in the deprojection as described in \citet{Rusli-11}, using a multi-gaussian profile as a representation of the actual PSF. We adopted an inclination of 90\degree\ for each galaxy, unless stated otherwise (see Fig. \ref{dchidm})

For two galaxies, we used the photometry that was already available in
the literature: the photometry for NGC\,7619 was taken from
\citet{Pu-10}, while for NGC\,4472 we used the photometry published in
\citet{Kormendy-09a}. We describe the photometry of the other galaxies
below\footnote{The photometry profile for each of these galaxies will
  be available in electronic form at:
  http://www2011.mpe.mpg.de/opinas/newresearch/blackholes/SINFONILOSVD/}.

\subsection{The photometry of NGC\,1374}
We used an HST image taken with ACS-WFC and the F475W filter (Proposal ID: 10911, PI: Blakeslee), and also an NTT EMMI image available from the ESO archive and taken with the filter "SPEC RS" (Program ID: 56.A-0430, PI: Macchetto/Giavalisco). We calibrated the final profile to the B band, using the aperture photometry of \citet{Caon-94}. The isophotal analysis was carried out using the software of \citet{Bender-87}. The matching between the HST and the ground-based profiles was performed in the region $5<a<20$ arcsec, where $a$ is the semi-major distance from the center, and we used the ground-based profile for $a>20$ arcsec.  The matching procedure (used also for all the following galaxies) consists in determining the scaling and sky value that minimize the magnitude square differences between the two profiles, assuming that the ground-based profile has the correct sky subtraction performed.  (Determining the sky background value for the HST image in this fashion is necessary when the galaxy fills the HST FoV.) The rms residuals in the region of the matching are small, typically 0.01-0.02 mag. The appropriate PSF to use for deprojecting the photometry is that of the ACS/WFC. We used the TinyTim program, which for the ACS/WFC generates an ``observed'' (i.e., optically distorted) PSF. Since the isophotal analysis was performed on the distortion-corrected ``drizzled'' image, we applied the same process to the PSF image, using custom-written Python code. The PSF image output by TinyTim was copied into the flat-fielded (but still distorted) ACS/WFC image, replacing the center of the galaxy. The full image was then run through the same \texttt{multidrizzle} processing to produce a distortion-corrected WFC mosaic. The processed PSF was then cut out from the mosaic. To generate the PSF image with the TinyTim software, we used the position of the galaxy center on the detector and a K giant spectrum as input (also for the galaxies below when TinyTim is used).

\subsection{The photometry of NGC\,1407}
We used an HST image taken with ACS/WFC and the F435W filter (Proposal
ID: 9427, PI: William Harris) and a ground-based B-band image taken at
the 40-inch telescope in Siding Spring equipped with the wide field
imager. On 2001 October 19 we took 6 pointings, each with a 10-minute
exposure. The images were reduced and combined using IRAF.  The
isophotal analysis was performed using the software of
\citet{Bender-87}. The ground-based surface brightness profile was
calibrated using photoelectric photometry of
  \citet{Burstein-87}. We matched the HST profile to the ground-based
data in the region $5<a<20$ arcsec, where $a$ is the semi-major axis
of the galaxy, and merged the two dataset using the matched HST
profile for $a<20$ arcsec.  The PSF used in the deprojection was
generated using the TinyTim and passed through the {\it drizzle}
software.

\subsection{The photometry of NGC\,1550}
We used an image from the Isaac Newton Telescope's Wide Field Camera (INT-WFC), originally observed by Michael Pohlen. The observation was a 300 second exposure in the Sloan $r$ filter, taken on 2003 December 15; the pixel scale was 0.331 arcsec/pixel and the median seeing FWHM was 1.4 arcsec.  We calibrated it to the Cousins $R$ using aperture photometry provided by \citet{Djorgovski-85}. Since there are no HST images of this galaxy, we used our $K$-band SINFONI image (generated by collapsing the reduced SINFONI datacube along the wavelength direction) for higher resolution in the central region of the galaxy. We perfomed the isophotal analysis on both images using the software of \citet{Bender-87} and matched the SINFONI surface brightness profile to the ground-based profile in the region $0.8 < a < 1.6$ arcsec. The combined profile uses the SINFONI data at $a < 1$ arcsec and the ground-based data for $a > 1$ arcsec. The SINFONI PSF was used in the deprojection.

\subsection{The photometry of NGC\,3091}
We used an HST ASC/WFC images taken using the F814W filter (Proposal
ID: 10787, PI: Charlton).  We calibrated them to the Cousins $I$ using
the approach of \citet{Sirianni-05} with the updated zero points
available at the STSCI
website\footnote{http://www.stsci.edu/hst/acs/analysis/zeropoints/},
along with colors from the galaxy photometry compilation of
\citet{Prugniel-98}.  Comparison with the aperture photometry of
\citet{Reid-94} shows that there is a -0.03 mag difference to the
Cousins I system. The isophotal analysis was performed using the
software of \citet{Bender-87}. The PSF used in the deprojection was
generated using the TinyTim software \citep{Krist-11} and passed through the {\it
  drizzle} software.

\subsection{The photometry of NGC\,4751}
We used an HST image taken with the NICMOS2 imager and the F160W filter (Proposal ID: 11219, PI: Alessandro Capetti) and a ground-based R-band image taken at the ESO-MPG 2.2m telescope in La Silla equipped with the wide field imager (WFI). The ground-based data (4 exposures of 5 minutes each) were taken on 2010 July 8. They were reduced (as all the WFI images described below) using the {\it mupipe} software \footnote{http://www.usm.lmu.de/$\mathtt{\sim}$arri/mupipe/} developed at the University Observatory in Munich \citep{Goessl-02}. After the initial bias and flat-field corrections, cosmic rays and bad pixels were masked, and the images were resampled to a common grid and stacked. A constant sky value was estimated from empty regions distant from the galaxy and subtracted from the stacked image. The calibration was performed using the photoelectric aperture photometry in the R band of \citet{Poulain-94}. The HST profile was matched to the ground-based one in the region $2''<a<5''$, and used for $a<5''$. The PSF was generated using TinyTim.

\subsection{The photometry of NGC\,5328}
We used the 60sec exposure EFOSC2 image taken at the 3.6m ESO
  telescope with the R filter \citep{gbauch-05} that we downloaded
  from the archive.  After the standard reduction, the calibration was
  performed using the photoelectic photometry in the V band from
  Hyperleda.  Since there are no HST images of this galaxy, we used
  our $K$-band SINFONI image (generated by collapsing the reduced
  SINFONI datacube along the wavelength direction) for higher
  resolution in the central region of the galaxy. We perfomed the
  isophotal analysis on both images using the software of
  \citet{Bender-87} and matched the SINFONI surface brightness profile
  to the ground-based profile in the region $0.8\arcsec < a <
  1.9\arcsec$. The combined profile uses the SINFONI data at $a <
  1.2\arcsec$ and the ground-based profile ground-based data for $a >
  1.2\arcsec$. The SINFONI PSF is used in the deprojection.

\subsection{The photometry of NGC\,5516}
We used an HST image taken with the PC of WFPC2 and the F814W filter
(Proposal ID: 6579, PI: John Tonry) and a ground-based R-band image
taken at the ESO-MPG 2.2m telescope in La Silla equipped with the Wide
Field Imager. The ground-based data (4 exposures of 230 seconds each)
were taken on 2010 July 10 and were reduced as for NGC\,4751. The
calibration was performed using the photoelectric photometry in the R
band from \citet{Poulain-94}. The HST profile was matched to the
ground-based one in the region 3 arcsec $<a<$ 10 arcsec, and used for
$a<8.9$ arcsec. The PSF was generated using TinyTim.

\subsection{The photometry of NGC\,6861}
We used an HST image taken with WFPC2 and the F814W filter (Proposal ID: 5999, PI: Andrew Philip) and a ground-based R-band image taken at the ESO-MPG 2.2m telescope in La Silla equipped with the wide field imager. The ground-based data (4 exposures of 230 seconds each) were taken on 2010 July 13 and were reduced as for NGC\,4751 and NGC\,5516.  We calibrated the HST profile of the Cousins I-band using the measured (F555W-F814W) colour and the equations of \citet{Holtzman-95}. We matched the ground-based profile to the HST one in the region $10''<a<85''$ and used the profile at $a>21.3''$. The PSF was generated using TinyTim.

\section{Dynamical models}
\label{dynmodels}
The dynamical modeling was performed using the axisymmetric orbit
superposition technique \citep{Schwarzschild-79}, described in
\citet{Gebhardt-00}, \citet{Thomas-04} and \citet{Siopis-09}. This
technique was implemented as follows. First, the gravitational
potential was calculated from the prescribed total mass distribution
using the Poisson equation. This total mass distribution is defined as
$\rho=\Upsilon \times \nu+ M_\mathrm{BH} \times \delta(r)+\rho_{\rm
  DM}$ where \ml\ is the mass-to-light ratio of the stars, $\nu$ is
the luminosity density distribution of the stars and $\rho_{\rm DM}$
is the dark halo density. Then, thousands of time-averaged stellar
orbits were generated in this potential. Each of their weights was
calculated such that the orbit superposition reproduced the luminosity
density and fitted the kinematics as good as possible. For each
potential we calculated about 24,000 orbits. We derived the
best-fitting set of parameters by setting up a parameter grid in the
modeling, with each gridpoint representing a trial potential. The
best-fit model was chosen based on the $\chi^2$ analysis.  For all
  galaxies except NGC\,5328 the $\chi^2$ is computed by comparing the
  LOSVDs derived by the model to the measured non-parametric LOSVDs
  and their uncertainties. 

As described above (Sec.~\ref{sinfonikinematics}), the shapes of
  the SINFONI LOSVDs for NGC\,5328 might be systematically affected by
  a faint sky emission line and we consider only the measured $V$ and
  $\sigma$ to be reliable. When we compute the $\chi^2$ difference
  between the models and the SINFONI data, we therefore do not use the
  full LOSVDs. Instead, for each model bin with SINFONI data we
  compute the Gauss-Hermite moments $h_1,\ldots,h_4$ of the respective
  model LOSVD. The series expansion is done based on the measured $V$
  and $\sigma$, rather than using the values of the corresponding
  best-fitting Gaussian to the model LOSVD. Hence, $h_1$ and $h_2$
  will be non-zero unless a Gaussian fit to the model LOSVD yields the
  measured $V$ and $\sigma$. For the comparison between model and
  SINFONI data we therefore calculate the $\chi^2$-difference in $h_1$
  and $h_2$, using $h_1=h_2=0$ as ``data'' (e.g. \citealt{Cretton-99}). 
The corresponding errors
  $dh_1$ and $dh_2$ were derived from the measured $dV$ and $d\sigma$
  through MC simulations. Leaving all the $h_n$ unconstrained for
  $n>2$ results in unrealistically noisy model LOSVDs. To reduce the
  noise, we extend the $\chi^2$-sum over $h_3$ and $h_4$ as
  well. However, the measured $h_3$ and $h_4$ are likely biased
  (Sec.~\ref{sinfonikinematics}) and instead of trying to reproduce
  them by a model fit, we use $h_3=h_4=0$ as ``data''. We artificially
  increase the uncertainties to $dh_3=dh_4=0.1$, in order not to bias
  the models too strongly towards any particular values, while still
  keeping them within a plausible range. In early-type galaxies, $\pm
  0.1$ is the typical range of observed $h_3$ and $h_4$.  Using $h_1$
  to $h_4$ for the $\chi^2$ as described above allows us to constrain
  the models towards the observed $V$ and $\sigma$ with reasonably
  smooth LOSVDs, but without a strong bias towards any particular
  LOSVD shape, as parameterised by $h_3$ and $h_4$.  The best-fitting
  $V$, $\sigma$, $h_3$ and $h_4$ are shown in Fig.~\ref{kin5328_all}.
  For the outer VIRUS-W data, we compare model and data using the full
  LOSVDs as for the other galaxies.

Along with the axisymmetry assumption in the modeling, we folded
  the four quadrants of each galaxy into one. For the SINFONI and
  VIRUS-W data, the folding was done by averaging the LOSVDs of four
  bins at the same angular and radial position. For the slit
  data, we fitted both sides of the slit simultaneously. The Voronoi
  binning of NGC\,5328 is not axisymmetric and we fit the four
  quadrants independently, using the sum of the four $\chi^2$ in the
  final analysis.

\subsection{The Importance of DM in the modeling}
\label{dmimportance}
The degeneracy between \mbh\ and \ml\ is often a problem in \mbh\
measurements. It is evident when the two-dimensional $\chi^2$
distribution (as a function of \mbh\ and \ml) appears diagonal
(e.g. \citealt{Gueltekin-09b}; \citealt{Nowak-10},
\citealt{Schulze-11}). It is thought that placing more stringent
constraints on \ml\ will help to pin down the black hole mass more
accurately. To do this, the naive approach would be to provide as
much/extended data as possible to constrain \ml. We show here that
this strategy is not advisable when DM is neglected in the models.

\begin{figure*}
\centering
  \includegraphics[scale=0.58, trim=0mm 0mm 0mm 0mm, clip]{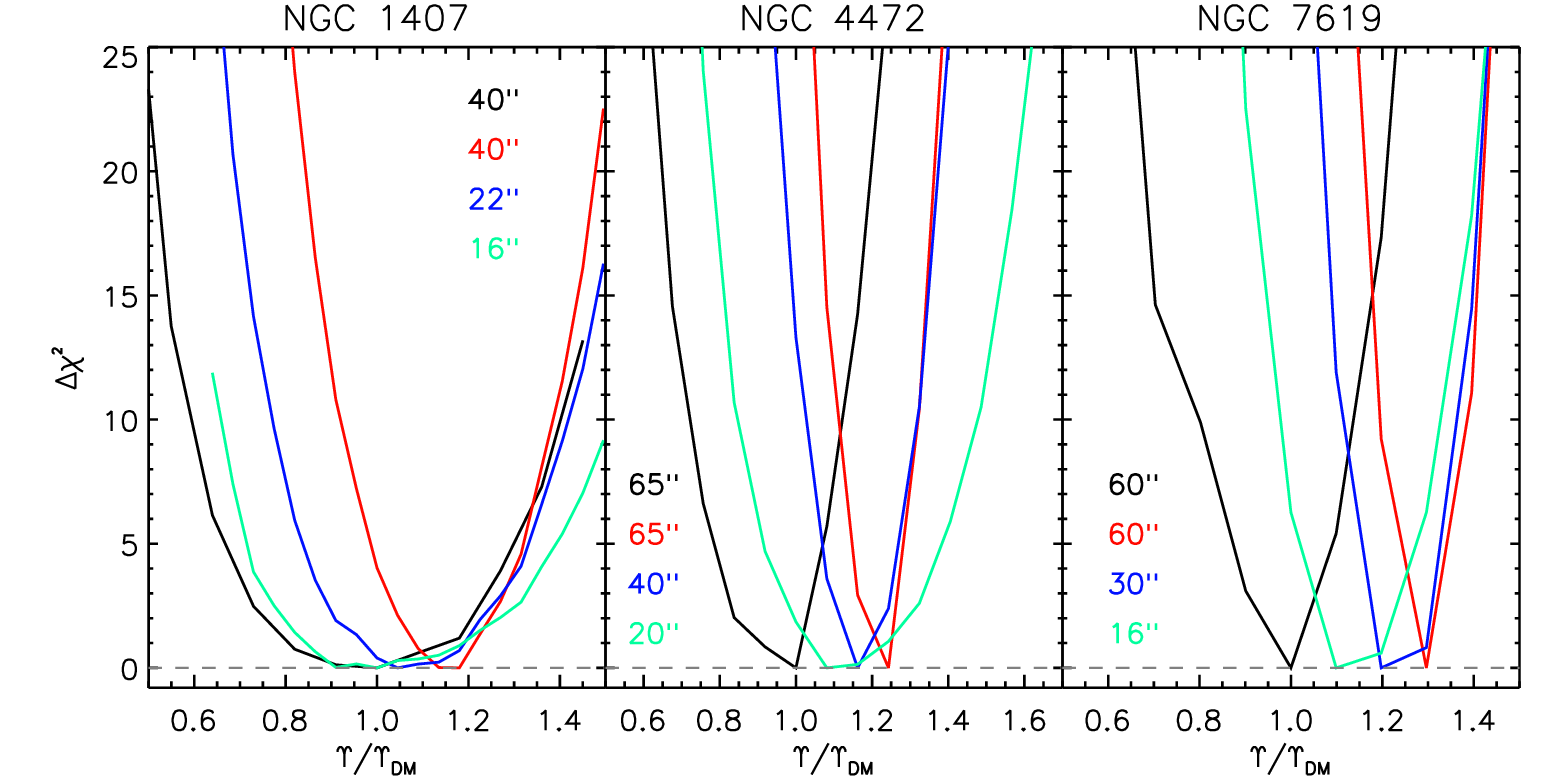}
  \caption[]{The $\Delta\chi^2$ distribution as a function of \ml. The \ml\ values are all normalized by the best-fit \ml\ obtained from modeling with DM (black line). The red, blue and green lines are models without DM, with different $r_{\rm trunc}$. The value of $r_{\rm trunc}$ for each run is written in the plot.\\}
\label{mldm}
\end{figure*}

From the slit data of NGC\,1407, NGC\,4472 and NGC\,7619, we created
multiple sets of data for each galaxy by truncating the slit data at
different outer radii ($r_{\rm trunc}$). For each galaxy, we ran models
using these kinematic datasets and also the full dataset to determine
\ml\ without having DM present in the model ($\rho_{\rm DM}$ is
zero). \mbh\ was set to zero. Fig. \ref{mldm} plots the $\Delta\chi^2$
vs. \ml\ obtained from these runs for each of the three galaxies,
$\Delta\chi^2$ being the difference between $\chi^2$ of each model and
the minimum $\chi^2$ of the run. As a comparison, we also show the
$\Delta\chi^2$ distribution when DM is included in the models for the
run with the full dataset (described in Section
\ref{dminclusion}). The \ml\ values along the x-axis are normalized by
the best-fit \ml\ obtained from the run with DM (\ml\ with
$\Delta\chi^2=0$ along the black line). The red line shows the run
without DM, but using the full slit data, i.e. the same kinematic
dataset as the run with DM. The blue and green represent the runs
without DM, with decreasing $r_{\rm trunc}$.

It can be seen for each of the three galaxies in Fig. \ref{mldm}
  that the highest and lowest values of the best-fit \ml\ were found
  when the full kinematic dataset was used: the \ml\ is lowest when DM
  is included and is highest when DM is omitted. For the case without
DM, \ml\ increases as $r_{\rm trunc}$ gets larger. This is expected
because more spatially extended data go farther into the region where DM is
dominant. In the modeling, omitting the DM component requires \ml\ to
increase in order to compensate for the missing dark mass. The larger
$r_{\rm trunc}$ is, the more dark mass there is which must be compensated for,
thereby increasing \ml\ further. The situation is worsened by the
fact that the 1\sig\ error bars of $\Upsilon$ ($\Delta\chi^2=1$)
decrease with increasing $r_{\rm trunc}$. When more extensive data are
used, the $\chi^2$ curve becomes steeper, excluding the ``true'' \ml\
(with DM) with a higher confidence. For the three galaxies, this
systematic bias appears to be as large as $\sim$20-30 percent.

It is clear that omitting DM in the modeling biases \ml. In principle,
this bias in \ml\ could also affect the black hole mass due to the
degeneracy. It is therefore important to consider DM in the modeling
to investigate how much effect this bias has on \mbh. We discuss this
further in Section \ref{bhduetodm}. In the following Section, we
describe how we incorporate DM in the modeling to fit \mbh.

\subsection{Inclusion of DM in the Model}
\label{dminclusion}

We used a spherical cored logarithmic (LOG) dark halo profile
\citep{Binney-87}. As found in \citet{Gebhardt-09} and
\citet{McConnell-11a}, the exact shape of the dark halo is of little
importance to the black hole mass and in most cases the LOG halo gives
a better fit compared to the other commonly used profiles. Since our
aim is to constrain \mbh\ and not the dark halo, we avoid a detailed
study of the halo parameters. The LOG halo is given by
\begin{equation}
  \rho_{\rm LOG}(r)=\frac{V_c^2}{4\pi G}\frac{3r_c^2+r^2}{(r_c^2+r^2)^2}
\end{equation}
where $r_c$ is the core radius, within which the density slope of the
DM is constant, and $v_c$ is the asymptotic circular velocity of the
DM. When the dark halo is present in the model, there is a total of
four free parameters ($\Upsilon$,\mbh, $V_c$ and $r_c$) to be
determined.

Calculating models to determine all four parameters
  simultaneously is computationally expensive. It requires about
  $\sim$ 100,000 models per galaxy and until now such an analysis has
  only been carried out for four galaxies: M\,87
  \citep{Gebhardt-09}, NGC\,4649 \citep{Shen-10}, NGC\,4594
  \citep{Jardel-11} and NGC\,1277 \citep{vandenBosch-12}. Most of the
  recent studies considering DM for dynamical \mbh\ determinations
  used a fixed (or a handfull of fixed) DM halos
  (e.g. \citealt{Schulze-11},
  \citealt{McConnell-11a,McConnell-12}). This lowers the computational
  effort since only \mbh\ and \ml\ are varied. However, to a certain
  extent a lower (higher) \ml\ can be compensated for by a more (less)
  massive halo. A fixed halo may therefore artificially narrow down
  the range of \ml\ consistent with the data, and, moreover, bias it
  towards a certain value. This, in turn, can affect the derived \mbh\
  and its uncertainties. In order to avoid such a possible bias, but
  still keep the computational effort feasible for our sample of ten
  galaxies, we here take an intermediate approach.

We first calculate models without the central, high-resolution
  SINFONI data. In this run we set \mbh\ to zero and vary $V_c$, $r_c$
  and \ml. We do not use the central SINFONI data since it probes the
  black hole. From this first set of models we determine the
  best-fitting $V_c$ and $r_c$ for each \ml. We then proceed to
  determine the best-fit \mbh\ (Section 6) by varying \mbh\ and \ml\,
  using the best-fit $r_c(\Upsilon)$ and $V_c(\Upsilon)$ for each \ml\
  as calculated in the first step. In this way, we reduce the
  computational load and time quite significantly, but still allow the DM 
  distribution to adapt to different \ml.  

Fitting the dark halo can be done only when the kinematic data are
sufficiently extensive, which is not always the case. For galaxies
with limited data, we skipped the first step and we directly
determined \mbh\ and \ml\ by fixing $r_c$ and $V_c$ to values from
dark matter scaling relations which were derived from a sample of Coma
galaxies \citep{Thomas-09}. The relations are written as follows:

\begin{equation}
  {\rm log}\,r_c = 1.54 + 0.63({\rm log}(L_B/L_\odot)-11.0)
\label{eq1}
\end{equation}
\begin{equation}
  {\rm log}\,V_c = 2.78 + 0.21({\rm log}(L_B/L_\odot)-11.0)\\
\label{eq2}
\end{equation}
with $r_c$ and $V_c$ stated in kpc and \kms\ respectively. The galaxy
luminosity $L_B$ is calculated from the B-band absolute magnitude
given in Table \ref{sample}. These equations are used to assign
  $r_c$ and $V_c$ values for NGC\,1374, NGC\,1550 and NGC\,5516.

  The best-fitting $r_c$ and $V_c$ for all galaxies with fitted halo
  are presented in Table \ref{tab:res}. They are not the exact values
  expected from the dark matter scaling relations, but they fall
  within the scatter. For galaxies with long slit data, we included
  all the slit datapoints to fit the dark halo. For NGC\,3091, we used
  the VIRUS-W data and the outer part of SINFONI 250-mas data because
  the VIRUS-W data started only at 10 arcsec. These VIRUS-W data were
  advantageous because of the larger spatial coverage, but they did
  not deliver useful results when used by themselves. Modeling \ml\
  and DM by using only the VIRUS-W data resulted in a rather flat
  $\chi^2$ distribution for \ml. We therefore included the 250-mas
  SINFONI data as well, but only used datapoints outside 1 arcsec.

\section{Black hole masses}
\label{bhmasses}
\begin{table*}
\caption{The best-fit \mbh\ and \ml\ with and without DM \label{tab:res}}
{\tiny
\begin{tabular}{lllllcccc}
Galaxy               & $M_{\rm BH,NoDM}$            & $\Upsilon_{\rm NoDM}$  & $M_{\rm BH,DM}$           & $\Upsilon_{\rm DM}$  & $r_c$  & $V_c$ & ${\rm D_{SoI}}$ & band\\
                     &   (\msun)                 &                     &     (\msun)            &                     & (kpc) & (\kms) & (\arcsec) & \\         
\hline
NGC\,1374\,$^\dagger$   & $5.8(5.3, 6.3)\times10^8$  &  $ 5.7( 5.3, 6.1) $ &  $5.8(5.3,6.3)\times10^8$ &  $ 5.3( 4.7, 5.9) $& 6.0  & 336  & 1.54& B       \\
NGC\,1407             &  $3.8(2.4, 4.4)\times10^9$  &  $10.7( 9.7,11.7) $ &  $4.5(4.1,5.4)\times10^9$ &  $ 6.6( 5.8, 7.5) $& 10.9 & 340  &3.89& B       \\
NGC\,1550\,$^\dagger$   &  $3.7(3.3, 4.2)\times10^9$ &  $ 4.3( 3.9, 4.7) $ &  $3.7(3.3,4.1)\times10^9$ &  $ 4.0( 3.4, 4.5) $& 20.7 & 507 & 1.34& R       \\
NGC\,3091             &  $9.7(8.1,11.0)\times10^8$  &  $ 5.4( 5.2, 5.7) $ &  $3.6(3.4,3.7)\times10^9$ &  $ 3.8( 3.6, 4.1) $& 29.8 & 809  &1.21& F814W   \\ 
NGC\,4472             &  $1.7(1.0, 2.0)\times10^9$  &  $ 7.3( 6.9, 7.7) $ &  $2.5(2.4,2.8)\times10^9$ &  $ 4.9( 4.5, 5.3) $& 13.6 & 780  &3.00& V       \\
NGC\,4751             &  $1.4(1.3, 1.5)\times10^9$  &  $13.1(12.7,13.4) $ &  $1.4(1.3,1.5)\times10^9$ &  $12.2(11.5,12.8) $& 5.2  & 300  &0.70& R       \\
NGC\,5328             &  $4.8(1.7, 8.0)\times10^7$  &  $10.1( 9.9,10.4) $ &  $4.7(2.8,5.6)\times10^9$ &  $ 4.9( 4.3, 5.5) $& 18.6 & 400  &1.34& V       \\
NGC\,5516\,$^\dagger$   &  $3.2(3.1, 3.4)\times10^9$ &  $ 5.6( 5.5, 5.8) $ &  $3.3(3.0,3.5)\times10^9$ &  $ 5.2( 5.1, 5.5) $& 31.8 & 585 & 1.06& R       \\
NGC\,6861             &  $2.2(2.1, 2.4)\times10^9$ &  $ 5.9( 5.7, 6.1) $ &  $2.0(1.8,2.2)\times10^9$ &  $ 6.1( 6.0, 6.3) $& 21.8 & 200  &0.76& I       \\
NGC\,7619             &  $4.2(3.0, 5.5)\times10^8$  &  $ 5.2( 4.9, 5.6) $ &  $2.5(2.2,3.3)\times10^9$ &  $ 3.0( 2.6, 3.3) $& 39.2 & 700  &0.83& I       \\                
\hline
\end{tabular}
}
\tablecomments{The quoted uncertainties for \mbh\ and \ml\ are 1\sig\ errors. The values of \ml\ are given after applying the extinction correction \citep{Schlegel-98}; for NGC\,3091, we adopt the Landolt I-band value of extinction. ${\rm D_{SoI}}$ is the diameter of the sphere of influence, calculated based on \mbh\ from models with DM ($M_{\rm BH,DM}$) and central velocity dispersions from HyperLeda, which are listed in Table \ref{sample}. The daggers ($\dagger$) mark galaxies for which DM parameters are not fitted, but derived from equations \ref{eq1} and \ref{eq2}. The other galaxies have fitted halos and $r_c$ and $V_c$ represent the best-fitting DM core radius and circular velocity (see Section \ref{dminclusion}). \\}
\end{table*}

In this Section we present the best-fitting black hole masses along
with the \ml\ values that are derived from the dynamical modeling. In
addition to the high-resolution SINFONI data, we also included the
other ground-based data at larger radii (Section
\ref{additionalkinematics}). When the DM was included in the model, we
used either the best-fit LOG halos as a function of \ml\ or a fixed
halo according to the B absolute magnitude taken from HyperLeda
(equations \ref{eq1} and \ref{eq2}). We also ran models without DM,
fitting only \mbh\ and \ml\ and using the same kinematic dataset as
the runs with DM. Table \ref{tab:res} shows the results with the
corresponding 1$\sigma$ uncertainties. The size of the sphere of
influence, calculated based on \mbh\ with DM and the central $\sigma$
taken from HyperLeda, is also listed in that table. We include $r_c$
and $V_c$ for the galaxies in which we use a fixed halo. The
Gauss-Hermite fit of the best-fitting model to the data for each
galaxy is shown in Appendix \ref{appendixc}
(Figs. \ref{kin1374}-\ref{kin7619}).

\begin{figure*}
\centering
  \includegraphics[scale=1.0]{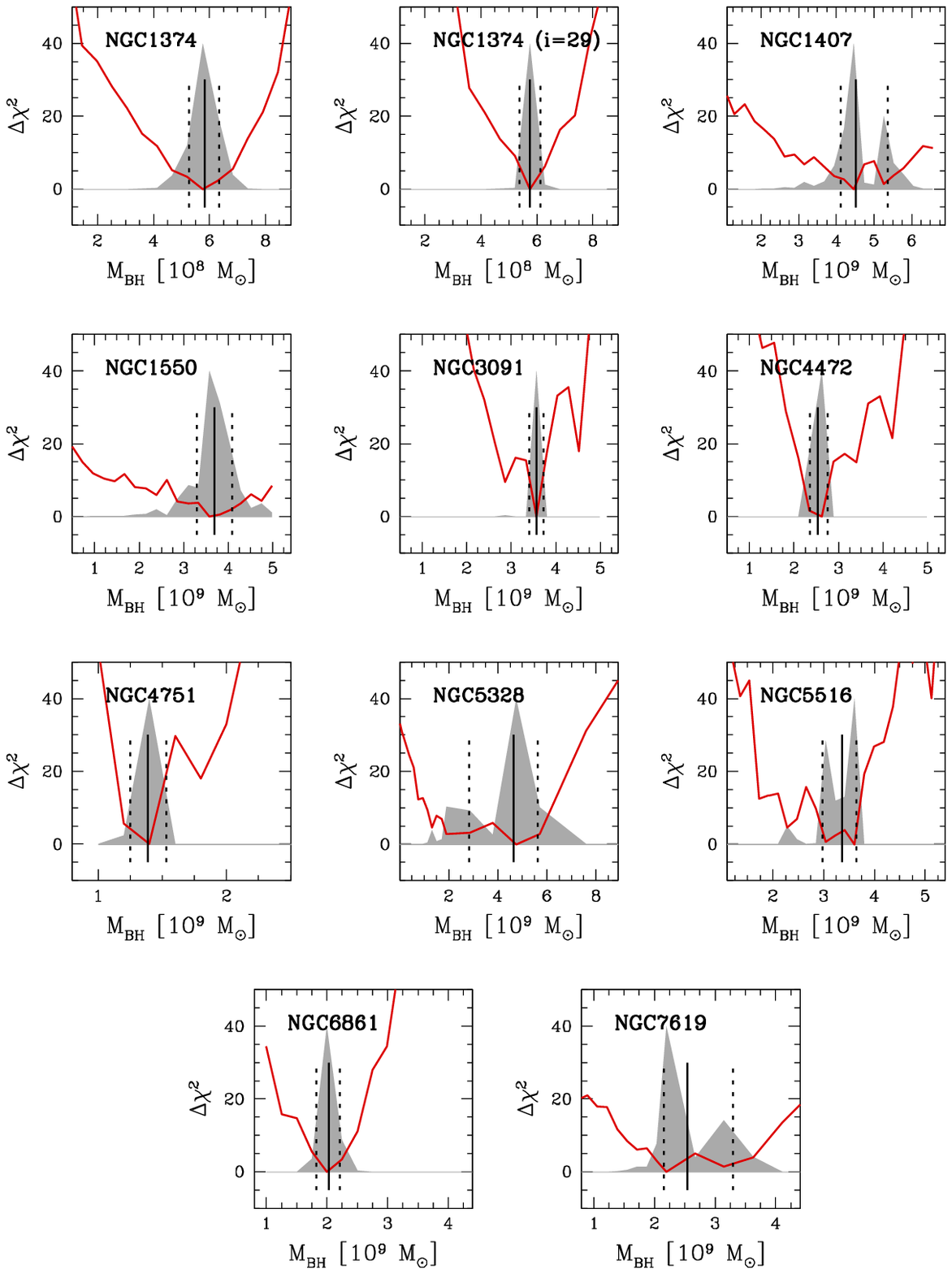} 
\caption[]{$\Delta\chi^2$ (red lines) and marginalized likelihood
    $P$ (shaded, cf. equation \ref{eq:likelihood}) vs \mbh\ for models
    with DM. The likelihood is scaled arbitrarily to a maximum value
    of $40$ in each panel. For NGC\,1374, we show the plot for two
    inclinations. The vertical full lines show the derived black hole masses; 
the vertical dashed lines the 1$-\sigma$ errors (see Table \ref{tab:res}). \\}
\label{dchidm}
\end{figure*}

In Fig. \ref{dchidm} we plot the $\chi^2$ distribution as a function
of \mbh\ (marginalized over \ml\ and also over $r_c$ and $V_c$ in some
cases) for each galaxy, when DM is included. For some galaxies, these
profiles look noisier than for others and the noise does not
necessarily decrease with increasing grid resolution. The profiles of
galaxies with a larger rotation over velocity dispersion ratio, like
NGC\,1374 and NGC\,7619, seem to be smoother. In spite of the noise,
each profile does show a clear minimum.

We obtain the \mbh\ and their uncertainties from the cumulative of
  the marginalized likelihood distribution
\begin{equation}
  P(M_\mathrm{BH})   \propto    \sum 
  \exp \left[ -\frac{1}{2} \Delta \chi^2(M_\mathrm{BH},\Upsilon) \right] 
  \delta \Upsilon,
\label{eq:likelihood}
\end{equation}
where the sum extends over all sampled \ml\
(e.g. \citealt{McConnell-11a}).  Another common approach is to derive
\mbh\ from the minimum of $\chi^2$ and the uncertainties from $\Delta
\chi^2 = 1$. In our case, either approach gives similar results.  
Fig. \ref{dchidm} shows also $P$ for all
galaxies.

\citet{Longo-94} and \citet{Donofrio-95} suggest that NGC\,1374 might be a face-on misclassified lenticular, based on the kinematic profile derived from their long-slit data. For this reason, we additionally ran models with an inclination of 29\degree; with this inclination, the projected flattening ($b/a$; $a$ and $b$ are the semi major and semi minor axes radius) of 0.88 translates into an intrinsic flattening of 0.25 (disk-like). We show the results in Fig. \ref{dchidm}. We see that changing the inclination to i=29 hardly changes \mbh\ and \ml. The same is true when the galaxy is modeled without DM. \mbh\ and \ml\ in this galaxy seem to be robust against DM inclusion and a change of inclination. 

Despite the steep change in \ml between models with and without DM,
the \mbh\ in NGC\,1407 stays almost the same. The black hole mass
measured for this galaxy is $4.3\times10^9$ \msun, comparable
to \mbh\ in NGC\,4649 which has a higher velocity dispersion of around
350 \kms\ \citep{Shen-10}. NGC\,1407 has the highest \mbh\ in the
sample, although the velocity dispersion is rather low. In
  NGC\,5328, the \mbh\ without DM is too small to be formally
  resolved; without DM, only an upper limit for \mbh\ can be derived.

There are two sets of SINFONI data for NGC\,3091, which overlap in the
inner 1.5 arcsec radius. The 100-mas dataset excels in spatial
resolution while 250-mas data is better in S/N. To fit \mbh\, we use
the 100-mas data out to a radius of 0.65 arcsec, which corresponds to
the 6th radial bin in the kinematic maps. With \mbh\ of
$3.6\times10^9$\msun, the radius of the sphere of influence is 0.68
arcsec, meaning that the region within the sphere of influence is
fully constrained by the high-resolution data (100-mas). Outside this
radius (out to $\sim 4$ arcsec) the kinematics is provided by the 250-mas
data.

\section{The change of \mbh\ due to DM}
\label{bhduetodm}

In three out of our ten galaxies, \mbh\ stays the same or changes only
within the $1\sigma$ errors when a DM halo is included. In one galaxy,
NGC\,6861, \mbh\ decreases (but still being slightly above the upper limit
of \citealt{Beifiori2009}). However, the confidence intervals for
models with and without DM are nearly identical for this galaxy. There
are three galaxies whose black hole masses change significantly
outside their $1\sigma$ errors, i.e. NGC\,3091, NGC\,5328 and
NGC\,7619. Without considering NGC\,5328 (see above), \mbh\ increases
by a factor of 2 on average when a DM halo is included. This is
similar to the increase in \mbh\ found for M87 \citep{Gebhardt-09},
but larger than that found by \citet{Schulze-11} using their sample of
12 galaxies.

\citet{Gebhardt-11} suggest two strategies for deriving an accurate
\mbh. These are to get \ml\ right in the first place or to have high
resolution data that cover the central region of the galaxy. Both
strategies have been tested on M87 by \citet{Gebhardt-09} and
\citet{Gebhardt-11}. Below we address how these two strategies apply
to our sample galaxies.

\subsection{The influence of data resolution}
\label{subsec:resolution}
\begin{figure}
\centering
  \includegraphics[scale=0.5]{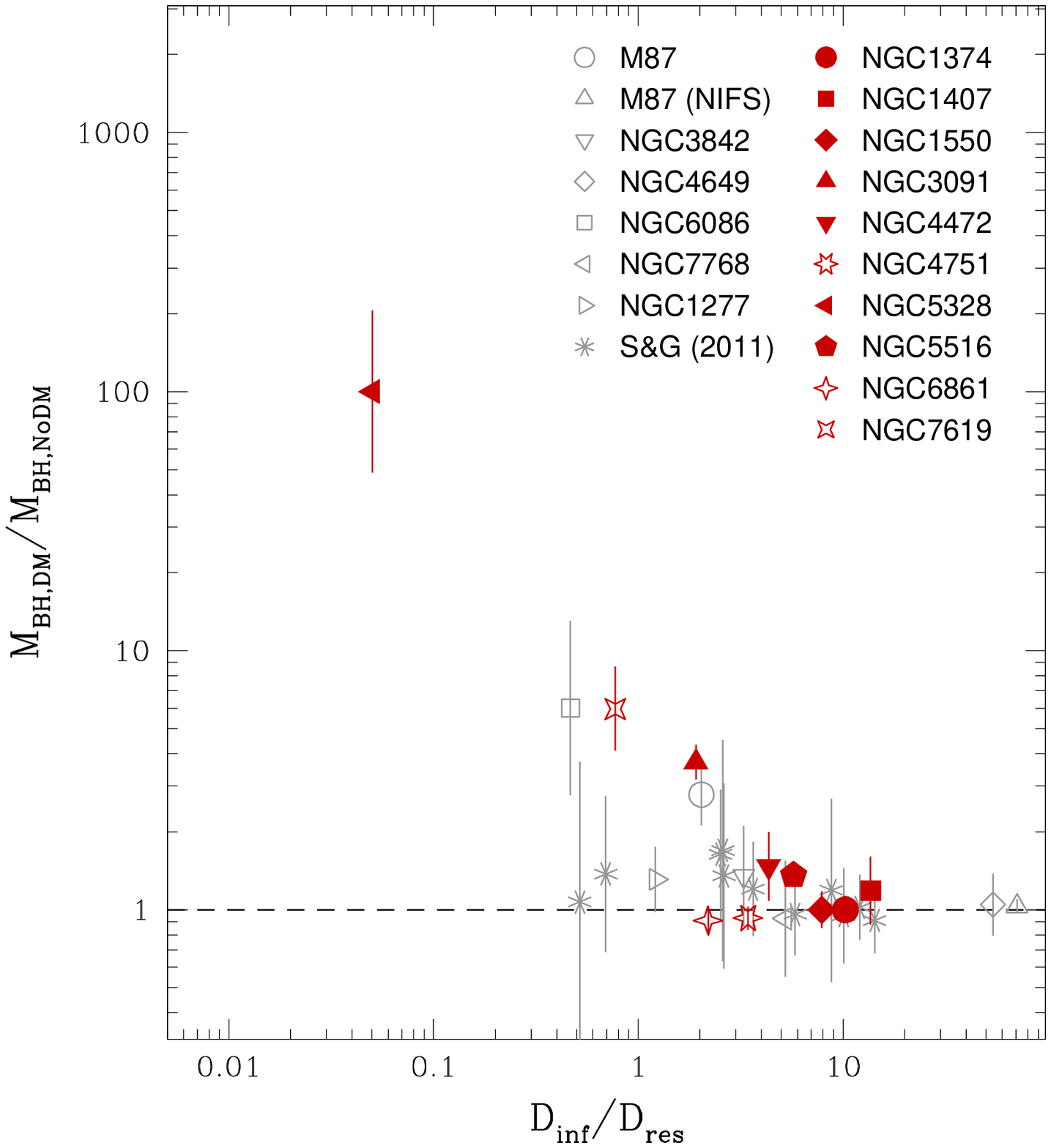}
\caption[]{The ratio of \mbh\ from models with and without DM, versus relative spatial
    resolution. ${\rm D_{inf}}$ is the diameter of the sphere of
    influence of the black hole {\it before} taking into account DM
    and using the central $\sigma$ from HyperLeda. ${\rm D_{res}}$ is
    the diameter of the resolution element (see text). We include
    galaxies outside our sample for comparison purposes (M87,
    NGC\,3842, NGC\,4649, NGC\,6086, NGC\,7768, NGC\,1277 and the sample
    of \citealt{Schulze-11}). There are
    two datapoints for M87, corresponding to the two measurements by
    \citet{Gebhardt-09} and \citet{Gebhardt-11} (marked as 'M87
    (NIFS)'). \\}
\label{ratmbh_res}
\end{figure}

We show a plot of spatial resolution vs the change in \mbh\ in
Fig. \ref{ratmbh_res}. $D_{\rm res}$ is the diameter of the resolution
element, which is largest of the aperture size, seeing, the PSF FWHM
and the model bin size. For our galaxies, $D_{\rm res}$ represents the
PSF FWHM (last column in Table \ref{obsruns}), except for NGC\,1407
where we use the average size of the model bins inside the SINFONI
FOV, i.e. 0.24 arcsec (larger than the PSF FWHM of 0.19 arcsec). The
velocity dispersion values from HyperLeda ($\sigma_0$ in Table
\ref{sample}) were used to derive $D_{\rm inf}$, the diameter of the
sphere of influence.  We have checked that for the ten galaxies in our
sample, the velocity dispersions from spatially-averaged SINFONI
spectra are similar to these. Using these SINFONI $\sigma$ values, or
the $\sigma_e$ values from Table \ref{sample}, does not change the
appearance of the plot. $D_{\rm inf}$ is calculated using \mbh\
without dark matter. We include M87 \citep{Gebhardt-09,Gebhardt-11},
NGC\,4649 \citep{Shen-10}, NGC\,6086 \citep{McConnell-11a}, NGC\,3842
\& NGC\,7768 \citep{McConnell-12}, NGC\,1277 \citep{vandenBosch-12}
and the galaxies of \citet{Schulze-11} in the plot to complete the
sample of black holes in massive galaxies that have been studied for
the DM effect.

The plot confirms the importance of good data resolution for \mbh\
measurements, in the context of DM influence. When $D_{\rm inf}$ --
without DM -- is not or only marginally resolved ($D_{\rm inf}/D_{\rm
  res} < 5$), there is a large scatter in the black hole mass ratio
and therefore a large risk of getting a wrong \mbh\ when DM is not
included. However, when the BH -- without DM -- was already resolved
by more than $D_{\rm inf} \sim 10 \times D_{\rm res}$, then its mass
is reliable. For values of $D_{\rm inf}/D_{\rm res}$ between 5 and 10,
\mbh\ could be wrong by $\sim30$ percent, but still on the safe side
given that this number is similar to a typical \mbh\ measurement
error. Based on this plot, one can, to some degree, assess individual
galaxies to see whether it is necessary to have DM present in the
models without first having to go through the time-consuming process
of modeling with DM.

\subsection{The influence of \ml}
\begin{figure}
\centering
  \includegraphics[scale=0.5]{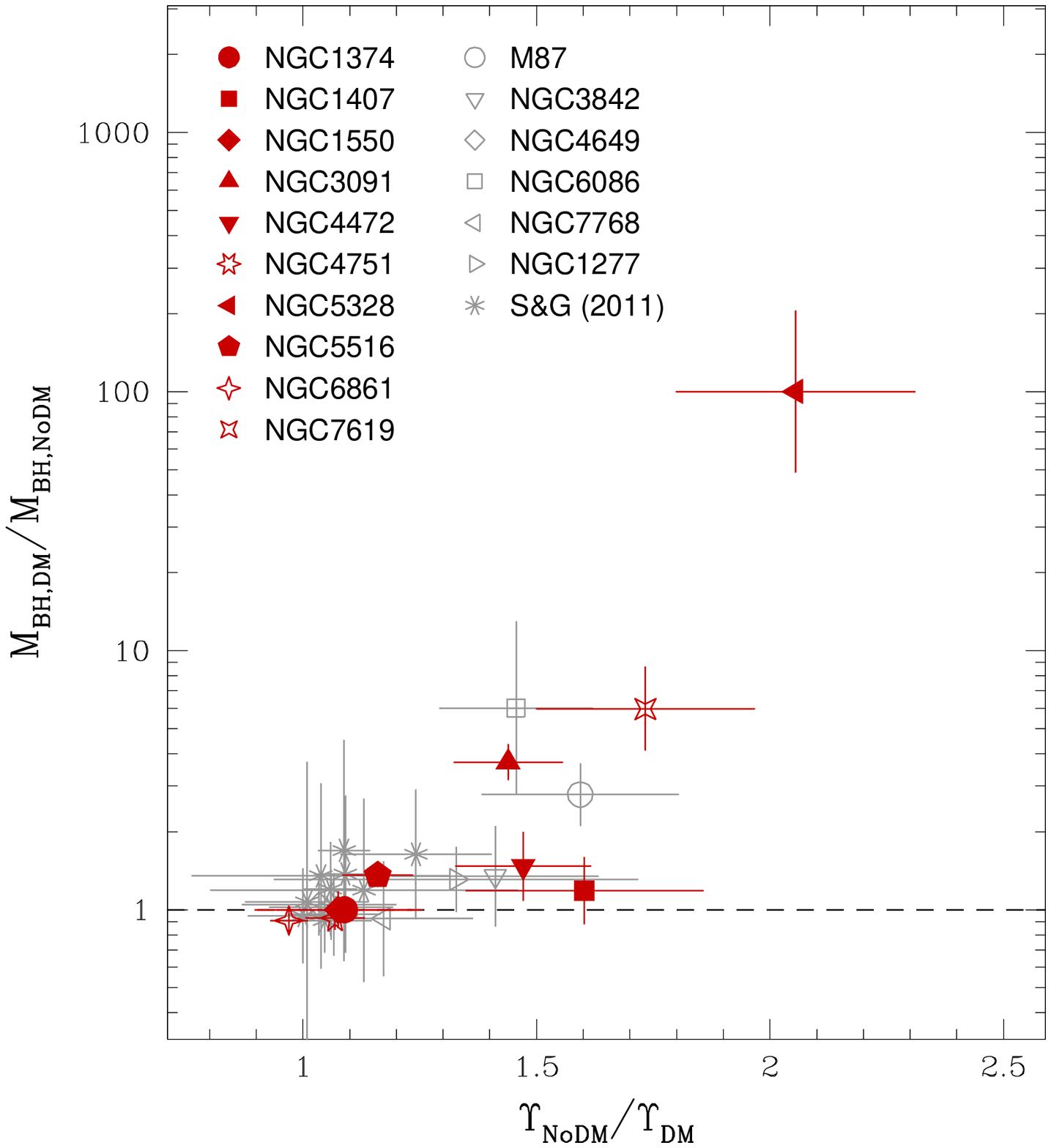}
\caption[]{The ratio of \mbh\ with and without DM against the corresponding ratio for \ml. Note the
    different numerator and denominator in both ratios. Galaxies outside
    our sample are included for comparison purposes and shown in grey (M87,
    NGC\,3842, NGC\,4649, NGC\,6086, NGC\,7768, NGC\,1277 and the galaxies
    from \citealt{Schulze-11}).\\}
\label{ratmbh_ratml}
\end{figure}

Nine of the ten galaxies see a nominal change in \ml\ (Table
  \ref{tab:res}). In all galaxies but NGC\,6861 it systematically
  decreases from models without DM to models with DM.  We show how the
  change in \ml\ is related to the change in \mbh\ in
  Fig. \ref{ratmbh_ratml}. We again include galaxies outside our
  sample for comparison. 

The figure shows that the effect of DM on \mbh\ measurements is
  indirect, mediated by the \ml. In galaxies where models with
  or without DM result in the same \ml, the black-hole masses
  \mbh\ with DM and without DM are similar. The DM halo has no direct
  influence on the measured \mbh. In most galaxies, however, the
  inclusion of DM leads to a reduction of the \ml. Significant
  differences between \mbh\ determined with and without DM appear only
  in galaxies where the \ml\ changes significantly. Despite
  the different modelling approaches this behaviour is seen in all
  galaxies that have been analysed for DM so far.

The scatter in Fig. \ref{ratmbh_ratml} is large. As already
  discussed above (Section \ref{subsec:resolution}) one source of
  scatter is the varying data resolution in the centre. In some
  galaxies the resolution is so good that despite a strong change in
  \ml, \mbh\ stays constant because its determination is decoupled
  from \ml.  However, with decreasing resolution, \mbh\ and \ml\
  become more and more degenerate. If the total central mass, i.e. the
  sum of \mbh\ and the stellar mass inside some small radius, would be
  entirely constrained by the observed kinematics, independently of
  whether DM is included in the outer parts or not, then the changes
  in \mbh\ and \ml\ should compensate for each other.

\begin{figure}
\centering
  \includegraphics[scale=0.5]{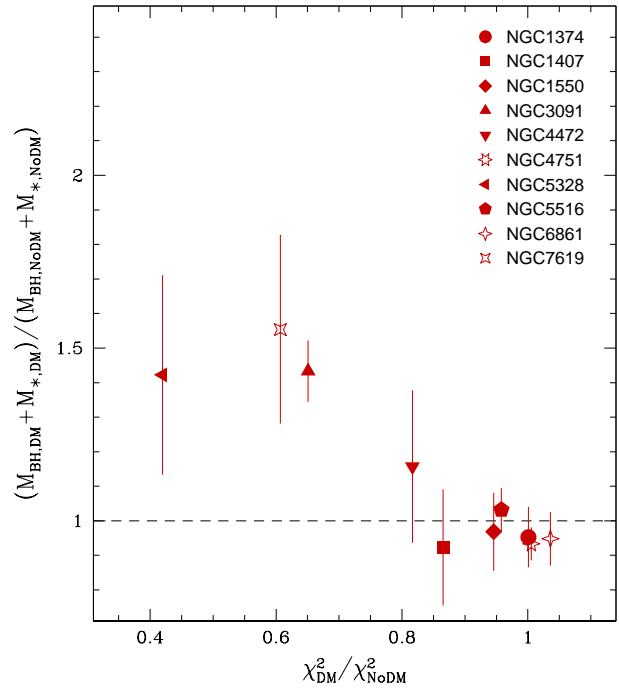}
\caption[]{The relative change of the total mass (BH plus stars) inside
    the sphere of influence ${\rm D_{inf,DM}}$ of the best-fit BH (with DM) against
    the change in $\chi^2$.\\}
\label{dm_chi}
\end{figure}

In Fig.~\ref{dm_chi} we test explicitly for this degeneracy by
  comparing the total enclosed mass of the best-fit model with DM and
  the best-fit model without DM, inside the same radius. As a
  benchmark we take $D_{\rm inf, DM}$ (calculated from the best-fit
  \mbh\ with DM). Comparison with Fig.~\ref{ratmbh_ratml} shows that
  the total mass in the centre is indeed better conserved than the
  masses of the stars and the BH individually. In most galaxies it is
  conserved to about $\sim10$ percent, implying that the change in
  \mbh\ almost exactly compensates the change in the central stellar
  mass. When the $\chi^2$ changes drastically after the inclusion of
  dark matter (e.g. in NGC\,5328), the central mass is not conserved,
  however. The $\Delta \chi^2$ between the corresponding models is
  sometimes over $\sim 1000$. Consequently, the fit without DM does
  not represent the data well and not even its total central mass is
  reliable. When this happens, the changes in \mbh\ and \ml\ from
  models without DM to models with DM are no longer mutually related
  to each other and can not be predicted. In fact, in some of our
  galaxies the change in \mbh\ overcompensates the change in \ml\ and
  the total mass in the centre increases. In addition to the varying
  data resolution, this adds to the scatter in Fig.~\ref{ratmbh_ratml}
  as well.

\begin{figure}
\centering
  \includegraphics[scale=0.5]{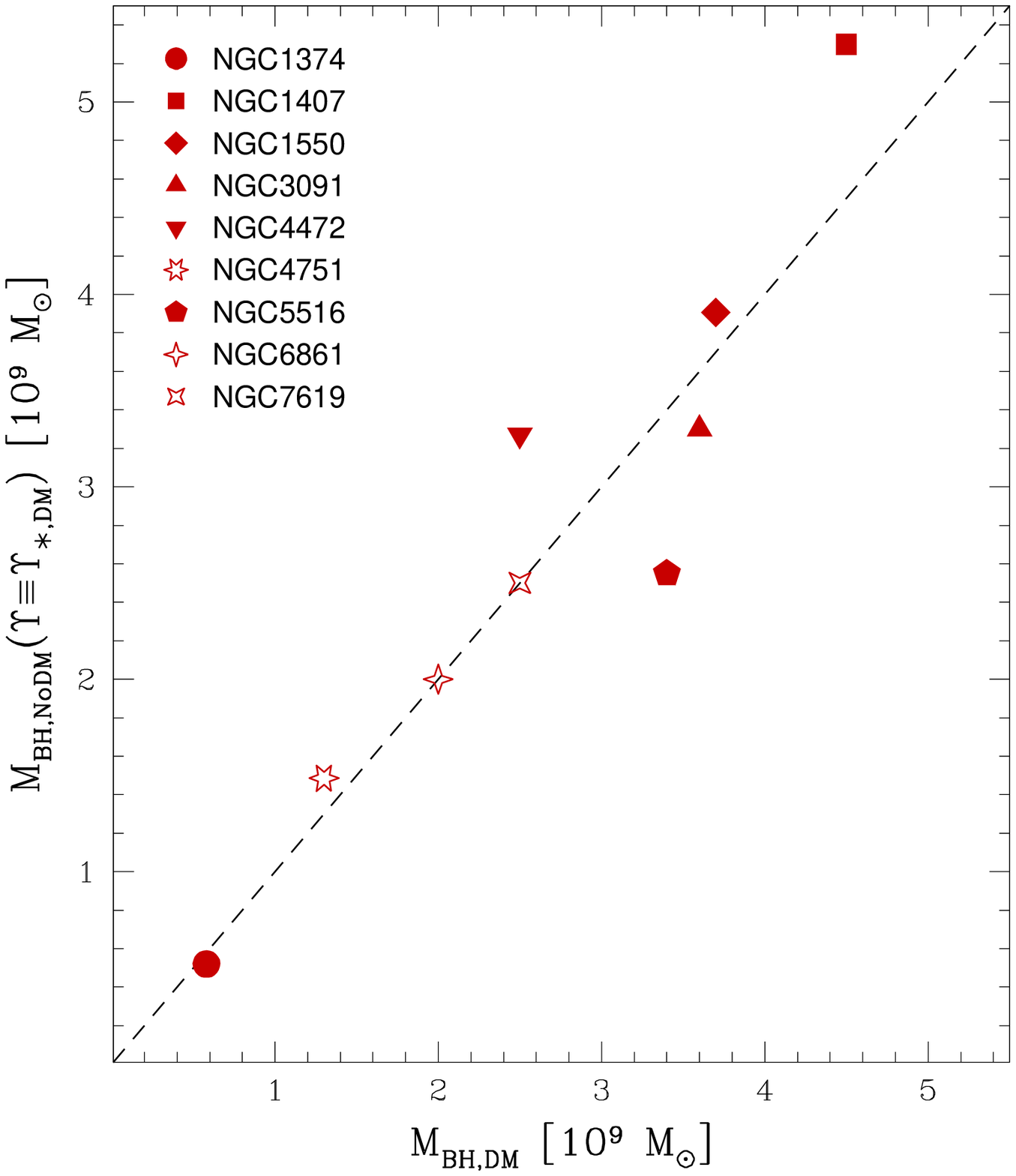}
\caption[]{Comparison of the best-fit \mbh\ with DM (x-axis) and the
    best-fit \mbh\ one would get without DM in the models, but still using
    the correct (DM-based) mass-to-light ratio (y-axis; $\Upsilon \equiv
    \Upsilon_{\ast,\mathrm{DM}}$).  NGC\,5328 has been omitted because
    $\Upsilon_{\ast,\mathrm{DM}}$ is outside the grid of \ml\ probed
    without DM.\\}
\label{recover}
\end{figure}

\subsection{Modelling strategies}
Our findings have straightforward
  implications for the modelling. As long as the sphere of influence is well sampled by
  the data, a biased \ml\ is not a problem in recovering the correct
  \mbh\ because \ml\ and \mbh\ are decoupled and, thus, including DM
  is not a necessity. The example for this would be NGC\,1407. When
  the data resolution is not sufficiently high, it is important to
  have an unbiased \ml\ to derive an accurate \mbh, as in the case of
  NGC\,3091, NGC\,5328 and NGC\,7619.

Based on our results in Section \ref{dmimportance}, it is clear
  that including the DM is necessary to derive an unbiased dynamical \ml. When DM
  is not included, the \ml\ obtained with less extended kinematic data
  is more reliable, although this does not completely remove the
  bias. It is however not clear, in advance of the modeling, where to
  spatially truncate the kinematic data. $R_e$ is not necessarily a
  good criterion. In the case of NGC\,1407 and NGC\,4472, the data go
  out to only a fraction of $R_e$ (around 0.5 and 0.25 $R_e$
  respectively) but still contribute to an appreciable change in \ml,
  while in NGC\,1550 the change in \ml\ is within the error although
  the data extend out to $R_e$ (Table \ref{tab:res}).  

In Fig.~\ref{recover} we compare the best-fit \mbh\ with DM to the
  best-fit \mbh\ that we get without DM, when we constrain \ml\ to its
  ``correct'' value (i.e. the value obtained from models with DM). Since
  the average ratio of these two \mbh\ determinations is $1.03$ with
  an rms-scatter of only $0.17$, one could recover \mbh\ accurately
  without DM if the correct \ml\ was known.

One could perhaps rely on the mass-to-light ratio derived from a
  single stellar population analysis (\mlssp). However, a major
  uncertainty in this approach is our ignorance related to the stellar
  IMF. Recent dynamical and lensing studies of early-type galaxies
  show that the amount of mass following the light differs from the
  prediction of SSP stellar masses based on a constant IMF. Galaxies
  with velocity dispersions around 200 \kms\ are consistent with a
  Kroupa or Milky-Way like IMF, while at dispersions of 300 \kms\
  dynamical masses are a factor of 1.6-2.0 times larger than predicted
  by the same IMF (\citealt{Cappellari-06}; \citealt{Napolitano-10};
  \citealt{Treu-10}; \citealt{Thomas-11}; \citealt{Cappellari-12};
  \citealt{Wegner-12}). The interpretation of this trend as a
  systematic variation of the IMF with velocity dispersion depends on
  assumptions about the dark matter distribution.  Independent support
  comes from the recent stellar-population models of near-infrared
  spectra by \citet{Conroy-12}, which can reproduce the trend with
  velocity dispersion in terms of a variable stellar IMF.  

At least two galaxies in our sample do not follow this trend,
  however. For NGC\,1407, \citet{Spolaor-08b} give a central age of
  $12.0 \pm 1.1$ Gyr and a metallicity of $0.29 \pm 0.08$. Using the
  SSP models of Maraston (2005) and the galaxy's velocity dispersion
  ($\sigma=276$ \kms) to estimate the IMF, we derive a
  stellar-population \mlssp\ $= 12.5 \pm 3 \, M_\odot/L_\odot$. Our
  dynamical models with DM yield instead a much lower value, \ml\ 
$=5.8-7.5\, M_\odot/L_\odot$. For NGC\,7619, using the SSP analysis of
  \citet{Pu-10} and estimating again the IMF from the galaxy's
  velocity dispersion ($\sigma=292$ \kms), we get \mlssp\ $= 5.9 \pm 1
  \, M_\odot/L_\odot$. As in NGC\,1407, this is much larger than the
  dynamical \ml\ $= 2.6-3.3\, M_\odot/L_\odot$. A similar result was
  found by \citet{vandenBosch-12} when modelling the BH, stars and
  dark matter in NGC\,1277 simultaneously. Checks on a much larger
  sample of galaxies would be necessary to justify the use of \mlssp\
  as a surrogate.

\section{Black hole-bulge relation}
\label{bhbulgerel}
\begin{figure*}[t!]
\centering
  \includegraphics[scale=0.80, trim=5mm 0mm 0mm 0mm, clip]{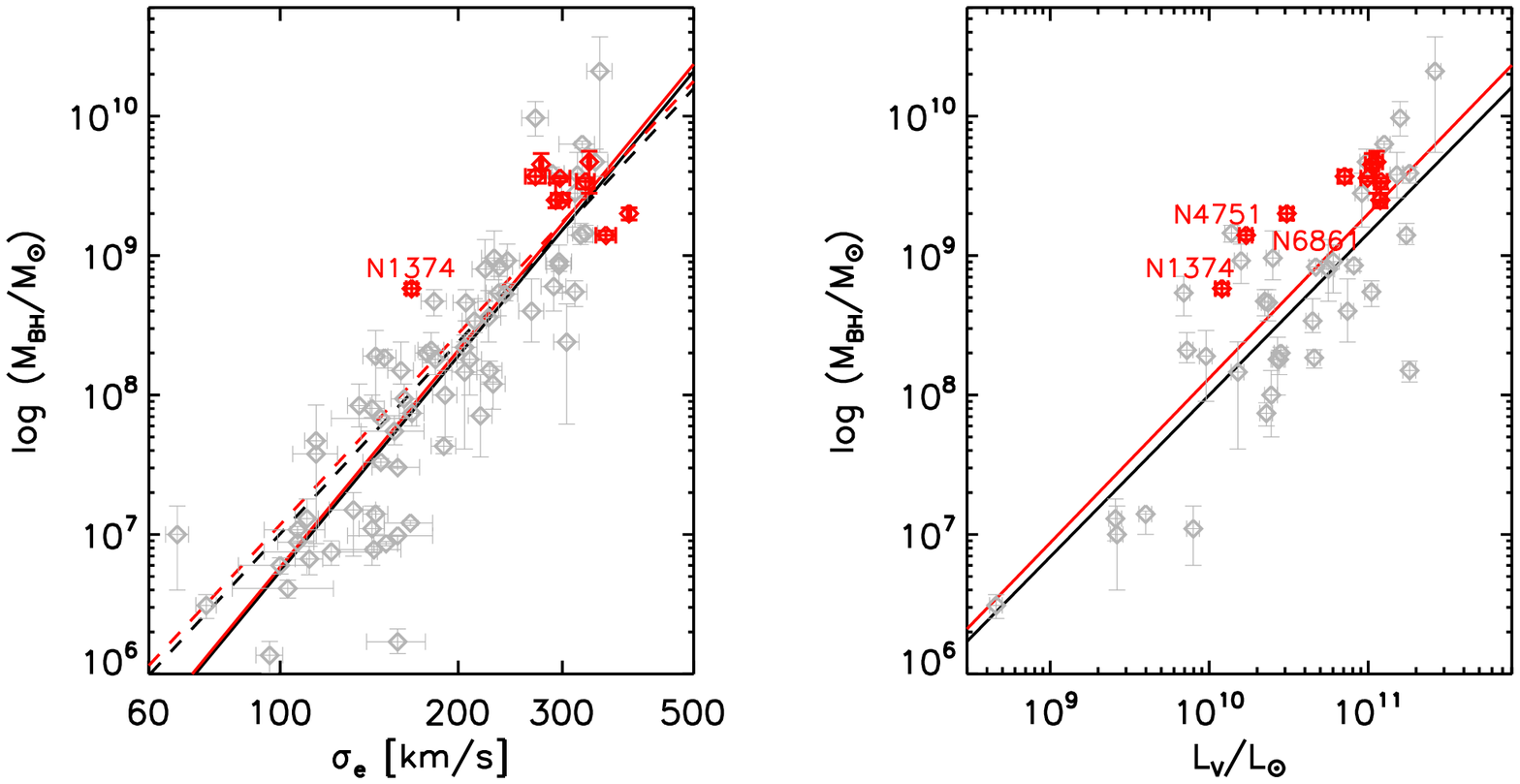}   
\caption[]{The \msig\ (left) and \mlum\ (right) diagrams with our
    updated relations. Grey diamonds are the galaxies in the sample of
    \citet{McConnell-11b}. The relations that they derive are
    indicated by black lines. The solid and dashed black lines in the
    \msig\ relation show the fit to all galaxies and to only the
    early-type galaxies in their sample, respectively. The red
    diamonds represent the ten galaxies in our sample, where we
    plot \mbh\ obtained by taking into account the DM. Using the
    McConnell et~al.\ sample and these ten galaxies, we updated the
    relations, shown by the red lines. The solid and dashed red lines
    in the \msig\ relation use the 'all1' and 'early-type' sample,
    respectively (see Table \ref{msigmlum}). \\}
\label{updatedrel}
\end{figure*}

We consider the latest version of the \msig\ and \mlum\ relations
derived by \citet{McConnell-11b} in Fig. \ref{updatedrel}. We do
  not consider the results of \citet{McConnell-13}, because they are
  based on a sample that lists preliminary values of a subset of the
  galaxies analysed here. The galaxies in the sample of
\citet{McConnell-11b} are plotted in grey and their relations are
shown by the black lines. In the \msig\ diagram, the solid and dashed
lines are the fit to all galaxies in the sample and to only early-type
galaxies, respectively. We overplot our new measurements as red
diamonds; we calculate the velocity dispersion of the ten
galaxies as in \citet{Gueltekin-09}, i.e. $\sigma_e^2 =
\int_0^{R_e}{(\sigma^2+V^2)I(r)dr/\int_0^{R_e}{I(r)dr}}$ and the \mbh\
values are the ones obtained with DM. All galaxies lie well above the
\mlum\ relation of \citet{McConnell-11b}. In the \msig\ diagram, two
galaxies (NGC\,6861 and NGC\,4751) fall below their relation.

The standard parametrization of the \msig\ and \mlum\ relationships
are ${\rm log_{10}} (M_{\rm BH}/M_\odot) = \alpha + \beta\, {\rm
  log_{10}} (\sigma/200\, \kmseq)$ and ${\rm log_{10}} (M_{\rm
  BH}/M_\odot) = \alpha + \beta\, {\rm log_{10}}
(L_V/10^{11}L_\odot)$, respectively, with some intrinsic scatter
$\epsilon_0$. This functional form is a single-index power law with
$\alpha$ as the zero-point and $\beta$ as the slope. We refitted the
relationships after incorporating our black hole measurements by using
a Bayesian method with Gaussian errors and intrinsic scatter
$\epsilon_0$ as described in \citet{Kelly-07}, assuming that the
uncertainties in the luminosities or velocity dispersions are not
correlated with those in the black hole masses. We were able to
recover the \msig\ and \mlum\ relations and their intrinsic scatters
as fitted by \citet{Gueltekin-09} and \citet{McConnell-11b}, using
their corresponding samples.  For our new fits, we define three
  samples: The sample ``all1" refers to all galaxies in the sample of
  \citet{McConnell-11b} plus our ten galaxies. The ``early-type"
  sample is the early-type subset of ``all1''. "all2'' contains the
  ``ML'' sample of \citet{McConnell-11b} plus the ten galaxies
  from this work.  The fit results are given in Table
\ref{msigmlum} for each relationship and shown in
Fig. \ref{updatedrel} as red lines. In the \mlum\ fit we adopt an
uncertainty for each $M_V$ of 0.1 magnitude, compatible with the
extrapolation errors in our integrated luminosity profiles. Note that
the fit parameters and the intrinsic scatter are not very sensitive to
a precise ${\rm M_V}$ uncertainty. Collectively, our ten galaxies
introduce a negligible change in the \msig\ relation of
\citet{McConnell-11b} and a slight increase in the zeropoint of their
\mlum\ relation, such that a given luminosity now corresponds to a
higher \mbh.

The average deviation of our galaxies from the new \msig\
relation is 0.47 dex in \mbh\ for the ``all1'' sample and 0.42 dex for
the early-type sample. Our galaxies are located 0.43 dex away
from the derived \mlum\ relation, on average. If we use the relations
of McConnell et al., then the average deviation would be higher,
i.e. 0.48, 0.44 and 0.53 dex from the \msig\ (all), \msig\
(early-type) and the \mlum\ relations, respectively. The above
  mean deviations are equal to the intrinsic scatter in the respective
  relations, except for our \mlum\ relation, which has a larger
  intrinsic scatter.

The galaxy that is farthest from the \msig\ relation is NGC\,1374, for
which we do not see a change in \mbh\ due to DM. If NGC\,1374 is
  intrinsically flattened and seen close to face-on, then we have
  observed the smallest projected velocity dispersion possible.
NGC\,4751 deviates the most in the \mlum\ relation: the galaxy
luminosity is approximately a factor 4 too small for its black
hole mass.  

\begin{table}[h!]
\caption{Parametric fits for the \msig\ and \mlum\ relations \label{msigmlum}}
\begin{tabular}{lllll}
Diagram &Sample & $\alpha$ & $\beta$&$\epsilon_0$ \\
\hline
\msig\           & all1           & $8.32\pm0.05$    & $5.16\pm0.32$    & 0.44  \\
\msig\           & early-type     & $8.44\pm0.06$    & $4.55\pm0.34$    & 0.39 \\ 
\mlum\           & all2           & $9.30\pm0.10$    & $1.18\pm0.13$    & 0.50\\
\hline
\end{tabular}
\tablecomments{The sample ``all1" refers to all galaxies in the sample of \citet{McConnell-11b} plus our ten galaxies. "early-type" sample is the early-type subset of ``all1''. "all2'' contains the ``ML'' sample of \citet{McConnell-11b} plus the ten galaxies from this work. $\epsilon_0$ is the intrinsic scatter of the relations. \\}
\end{table}

\begin{figure}[t!]
\centering
  \includegraphics[scale=0.70, trim=5mm 0mm 0mm 0mm, clip]{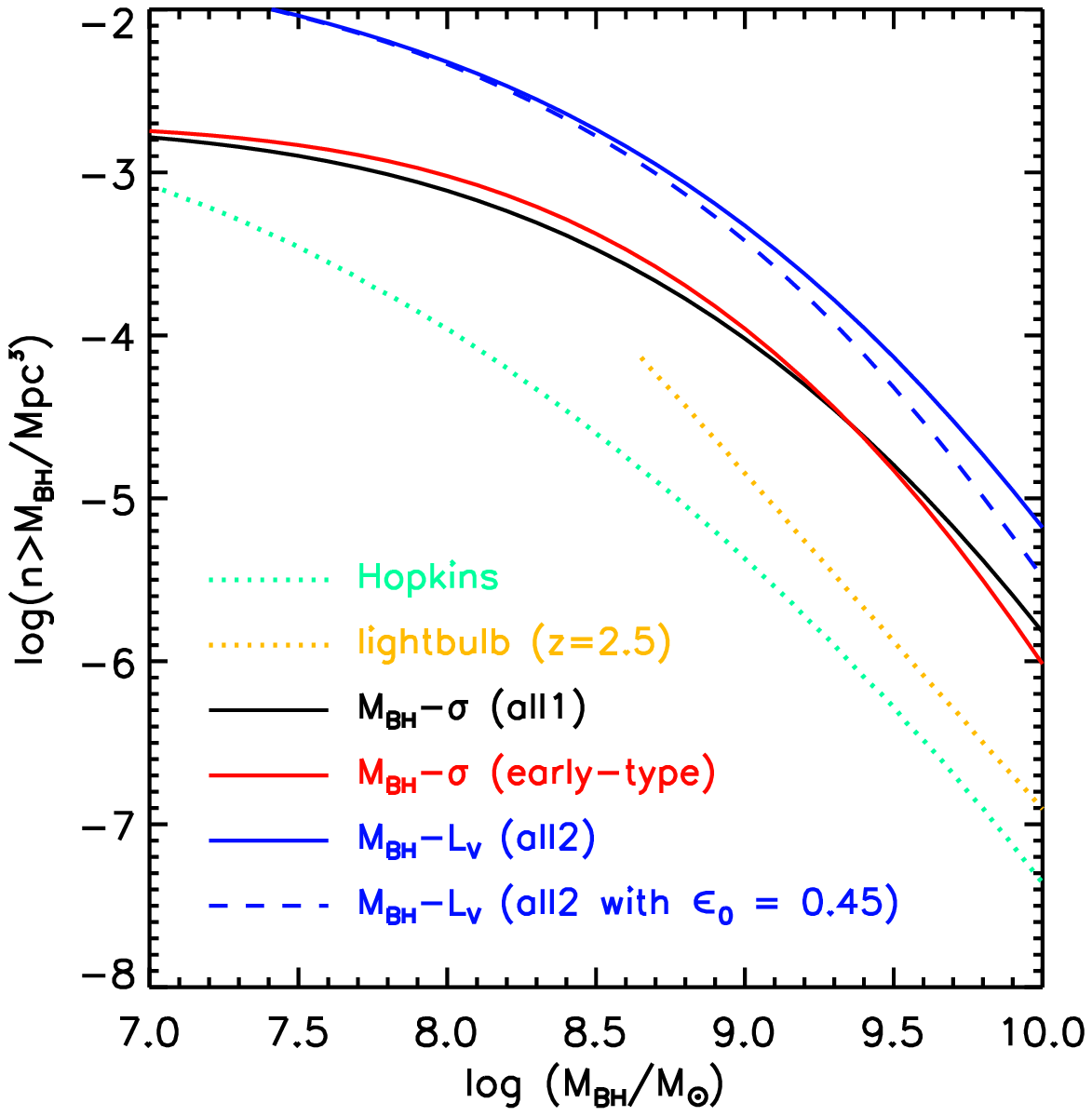}   
\caption[]{Black hole cumulative space density derived from the \msig\ and \mlum\ relations in Table \ref{msigmlum} (solid lines). If the intrinsic scatter of the \mlum\ relation is the same as that of the \msig\ relation (all1 sample), the resulting space density is shown by the dashed blue line. The lightbulb model is based on the luminosity function from SDSS at a redshift of z=2.5 \citep{Richards-05} and the Hopkins model is based on the quasar model of \citet{Hopkins-06}. Both quasar models are identical to those in Fig. 11 of  \citet{Lauer-07}}
\label{mbhdens}
\end{figure}

Based on the black hole-bulge relationships, combined with a velocity
dispersion or luminosity function, one can construct an estimate of
the local black hole space density. \citet{Lauer-07} and
\citet{Gueltekin-09} find a marked discrepancy between the $L$-based
and $\sigma$-based local density of the largest black holes ($M_{\rm
  BH} \gtrsim 10^9$\msun), i.e. the former is significantly larger
than the latter. Using our results from Table \ref{msigmlum}, we
derive the cumulative density distribution of black holes (see
Fig. \ref{mbhdens}) following the same approach and assumptions as in
\citet{Lauer-07}, which were also used by
\citet{Gueltekin-09}. The dispersion-based density is computed
  using the velocity dispersion function of \citet{Sheth-03} in the
  form of Schechter function fit. To this, a correction to the number
  of high-\sig\ galaxies ($\sigma > 350$ \kms) is applied based on the
  work of \citet{Bernardi-06}. For the L-based space density, we
  use the SDSS g' luminosity function of \citet{Blanton-03},
  converted to the V-band. The high-L part of this function is
  corrected by adding an estimate of the BCG luminosity function of
  \citet{Postman-95}. We refer the reader to Section 6.1 of
  \citet{Lauer-07} for a more thorough description.

The \msig\ relation fitted here has a considerably higher zeropoint,
slope and intrinsic scatter than those derived in
\citet{Gueltekin-09}, leading to a higher \sig-based BH density which
should therefore place the latter closer to the prediction based on the \mlum\
relation. However, the intrinsic scatter and the zeropoint in our
\mlum\ fit are also larger compared to those in \citet{Gueltekin-09},
which results in an even higher luminosity-based BH density. Since
2009, a number of \mbh\ measurements have been
added to the high-\mbh\ part of the relationships, which have
changed the correlations quite significantly. As shown in
Fig. \ref{mbhdens}, the difference between the two mass functions is
not reconciled using our updated \msig\ and \mlum\
relations. Black holes more massive than $10^9$ \msun\ now
  correspond to $\sigma \gtrsim 280$ \kms\ or $L_{\rm V} \gtrsim
  7\times10^{10} L_\odot$ (Fig. \ref{updatedrel}) and there are fewer
  galaxies with $\sigma > 280$ \kms\ than there are galaxies with $L > 7\times10^{10}
  L_\odot$. If the \mlum\ relation were to have the same scatter as
  the \msig\ relation, the predicted space density of BH for the
  largest BHs would be in a better agreement: a lower intrinsic
  scatter results in fewer high-mass black holes (see Fig. 3 of
  \citealt{Lauer-07c}).

\citet{Lauer-07} make a comparison between the local BH
  cumulative space density estimated from the BH--host-galaxy
  relations and the density estimated from luminous QSOs; they argue
  that this comparison favors the the \mlum\ relation, since the
  \msig\ relation seems to underpredict the density of the most
  massive BHs compared to the QSO-based models (their Fig.~11). In
  contrast to their results, we find that the updated \mlum\ and
  \msig\ relations both \textit{overpredict} the density of the most
  massive BHs compared to the same QSO models (Fig.~\ref{mbhdens}).
  Since the \mlum\ relation predicts the highest density of BHs at the
  upper-mass end, it is now the prediction from the \msig\ relation
  which is the better match to the same QSO models (represented by the
  ``Hopkins'' and ``lightbulb'' lines in Fig.~\ref{mbhdens}) is now
  that from the \msig\ relation. Nonetheless, even this relation
  appears to predict a BH density about an order of magnitude larger
  than that from the QSO models for $M_{\rm BH} \gtrsim 10^{9}$
  \msun. We note, however, that there is considerable uncertainty and
  freedom in QSO-based models of BH space density, depending on
  factors such as how Eddington vs.\ sub-Eddington accretion duty
  cycles are handled, overall radiative efficiency, etc.\ \citep[see,
  e.g.,][]{Kelly-12}.

\section{Summary}
\label{summ}
This paper presents new AO-assisted SINFONI observations of ten
  nearby early-type galaxies, nine of which have velocity dispersions
  greater than 250 \kms. The SINFONI data are complemented by other
  ground-based data with larger spatial scales, which are found in the
  literature or come from new observations. We use the data to measure
  the masses \mbh\ of supermassive black holes in the centers of these
  galaxies using three-integral axisymmetric orbit-superposition
  models. The black hole masses reported here represent the first such
  dynamical measurements for each of the sample galaxies.  

We pay particular attention to the question of how including or
  ignoring a dark matter (DM) halo in the dynamical models affects the
  derived \mbh. We adopt a cored logarithmic profile for the DM
  halo. For seven galaxies, where the kinematic data extend to large
  radii, the parameters of the halo are determined by the fitting
  process; for the others, we set the halo parameters to fixed values
  based on the $L_{B}$--DM scaling relation of \citet{Thomas-09}.  

We first study models without black holes; this shows that
  omitting DM in the modeling results in a bias in the stellar M/L
  ratio \ml, which is systematically higher when DM is not
  included. This bias becomes stronger when the spatial extent of the
  kinematic data is maximised. This suggests that it is necessary to
  consider DM in the dynamical modeling in order to derive a reliable
  value of \ml.  

We then explore models with and without DM which include black
  holes, and we assess how changes in \ml\ or data resolution affect
  the black hole mass determination. For completeness, we include
  \mbh\ measurements, with and without DM, for 17 massive galaxies
  from the literature. We find that changes in \mbh\ are triggered by
  changes in \ml\ due to DM inclusion. The underlying relationship
  between \mbh\ and the DM halo is indirect: including the DM halo
  affects \ml, which in turn affects \mbh. If we fix \ml\ to the
  ``correct'' value (i.e., that obtained by including DM in the
  modeling), then the ``correct'' value of \mbh\ is obtained from the
  modeling even if DM is excluded.  

The ratio between the black hole mass determined with a DM halo
  model and the mass determined without one, $M_{\rm BH, DM} / M_{\rm
    BH, NoDM}$, shows scatter and bias which increases as the spatial
  resolution of the central kinematic data decreases. If we define the
  resolution relative to the BH's sphere of influence as $D_{\rm
    inf}/D_{\rm res}$, where $D_{\rm inf}$ is the size of the sphere
  of influence (using \mbh\ from models without DM) and $D_{\rm res}$
  is the best resolution of the observations, then \mbh\ is
  significantly underestimated for $D_{\rm inf}/D_{\rm res} < 5$ when
  DM is \textit{not} included. For $D_{\rm inf}/D_{\rm res} > 10$,
  \mbh\ is negligibly affected by the presence or absence of DM,
  despite the large change in \ml; \mbh\ and \ml\ are effectively
  decoupled when the resolution is this good. For values of $D_{\rm
    inf}/D_{\rm res}$ between 5 and 10, \mbh\ can be underestimated by
  about 30 percent if DM is not included, a level similar to typical
  measurement errors for \mbh.  

All of the measured BH masses are located above the \mlum\
  relation of \citet{McConnell-11b}; seven of the ten galaxies have BH
  masses above their \msig\ relation. Including these ten galaxies in
  the relationships introduces a negligible change in the \msig\
  relation and slightly increases the zeropoint of the \mlum\
  relation, such that a given luminosity now corresponds to a higher
  \mbh. Using our updated relations, we find that the cumulative space
  density of the most massive black holes predicted by the \mlum\
  relation is about one order of magnitude higher than that predicted
  by the \msig\ relation. The latter predictions, in turn, are about
  an order of magnitude higher than the quasar-count-based density
  models from \citet{Lauer-07}, which inverts the claim made in that
  study (that the \msig\ relation predicts too few high-mass black
  holes to match the quasar-count models).  

\section*{Acknowledgements}
We thank the anonymous referee for comments that helped us
  improving the paper.  We thank the Paranal Observatory Team for
support during the observations. SPR acknowledges support from the DFG
Cluster of Excellence Origin and Structure of the Universe. PE was
supported by the Deutsche Forschungsgemeinschaft through the Priority
Programme 1177 'Galaxy Evolution'.

This research has made use of the NASA/IPAC Extragalactic Database (NED) which is operated by the Jet Propulsion Laboratory, California Institute of Technology, under contract with the National Aeronautics and Space Administration.

\appendix

\section{The long-slit kinematics of NGC\,1374 and NGC\,6861}
\label{appendix13746861}
The long-slit observations of NGC\,1374 and NGC\,6861 were part of the program described by \citet{Saglia-02}. NGC\,1374 was observed during the period 2001 November 10--14 at the La Silla ESO NTT telescope, equipped with the EMMI spectrograph.  We used the red arm (REMD mode with longslit, 6 arcmin long) with the 13.5 $\,$\AA\ mm$^{-1}$ grating (\#6), the OG530 order sorting filter and the Tektronix $2048\times2047$ 24-$\mu$m pixels CCD. The slit width was 3\arcsec\ and the scale 0.27 arcsec/pixel, giving a resolution of 70 km s$^{-1}$ and covering the wavelength range $\lambda= 8298-8893$\AA. We observed the galaxy at PA=120 with 90 minutes integration time.

NGC\,6861 was observed during the period 2001 May 8--14 at the 2.3
Siding Spring telescope, where we used the red arm of the Double Beam
Spectrograph \citep{Rodgers-88} in longslit (6.7
arcmin) mode with the 600 R grating (73 $\,$\AA\ mm$^{-1}$) grating,
no beamsplitter and the Site $1752\times532$ 15-$\mu$m pixels CCD. The
slit width was 3.5\arcsec\ and the scale was 0.91 arcsec/pixel, giving
a resolution of 75 km s$^{-1}$ and covering the wavelength range
$\lambda=7645-9573$. We observed the galaxy at PA=142 with 142 min
integration time.

The standard CCD data reduction was carried out with the image
processing package MIDAS provided by ESO. After bias subtraction and
flatfielding, the fringing disappeared, resulting in CCD spectra flat
to better than 0.5\% and slit illumination uniform to better than
1\%. No correction for dark current was applied, since it was always
negligible. Hot pixels and cosmic ray events were removed with a
$\kappa-\sigma$-clipping procedure. The wavelength calibration was
performed using calibration lamps (HeAr, FeNe, FeAr) for NGC\,6861,
and sky lines for NGC\,1374, where the calibration lamp exposures were
too weak or not enough calibration lines were present in the
wavelength range. A third order polynomial was used to perform the
calibration, achieving 0.1 \AA\ rms precision. The spectra were
rebinned to a logarithmic wavelength scale and the mean sky spectrum
during each exposure was derived by averaging several lines from the
edges of the CCD spectra. By applying the same procedure to the blank
sky observations, the systematic sky residuals were measured to be
less than 1\%. After subtraction of the sky spectra, spectra of the
same galaxy taken at identical slit positions were centred and added;
one-dimensional spectra were produced for the kinematic standard
stars. The galaxy spectra were rebinned along the slit in order to
guarantee a signal-to-noise ratio that allowed the derivation of the
kinematic parameters.

The galaxy kinematics were derived using the Fourier Correlation
Quotient (FCQ) method (Bender 1990). Following Bender, Saglia and
Gerhard (1994, hereafter BSG94), the line-of-sight velocity
distributions (LOSVDs) were measured and fitted to provide the stellar
rotational velocities $V$, the velocity dispersions $\sigma$ and the
first orders of asymmetric ($h_3$) and symmetric ($h_4$) deviations of
the LOSVDs from real Gaussian profiles. Monte Carlo simulations were
performed to establish that a fourth order polynomial in the
wavelength range $\approx 8250-8950$\AA\ provided good continuum fits,
resulting in systematic residuals always smaller than the estimated
statistical errors. These were calibrated testing a grid of input S/N,
taking into account the noise contributions of the galaxy and the sky
signals. The errors from systematics in the sky subtraction (at the
1\% level, see above) were less than the statistical errors. Tables
\ref{Tab_CaT_NGC1374} and \ref{Tab_CaT_NGC6861} give the resulting
kinematics.

\begin{table*}[h!]
\caption{The long-slit kinematics of NGC\,1374 \label{Tab_CaT_NGC1374}}

\begin{tabular}{lllllllll}
R (\arcsec)  & V (\kms) & dV (\kms) & $\sigma$ (\kms) & $d\sigma$ (\kms) &$h_3$ 
& $dh_3$ & $h_4$ & $dh_4$\\
\hline
  -15.16 & -46.18   &  9.42 & 106.82 & 14.21 & -0.087 &  0.069 &  0.045 &  0.067\\ 
   -4.62 & -50.65   &  3.14 & 134.27 &  4.11 & -0.090 &  0.024 &  0.042 &  0.028\\ 
   -1.76 &  -51.9   &  2.00 & 144.72 &  2.28 & -0.003 &  0.010 &  0.001 &  0.016\\ 
   -0.63 & -28.74   &  2.23 & 171.29 &  3.00 & -0.025 &  0.012 &  0.015 &  0.014\\ 
    0.14 &     10   &  2.15 & 180.22 &  2.92 &  0.013 &  0.011 & -0.013 &  0.009\\ 
    1.04 &  40.45   &  1.95 & 168.26 &  3.22 &  0.041 &  0.012 &  0.016 &  0.012\\ 
    2.66 &  49.13   &  2.49 & 146.97 &  2.89 &  0.054 &  0.012 &  0.033 &  0.017\\ 
    7.29 &  42.56   &  2.70 & 139.42 &  3.63 &  0.050 &  0.027 & -0.012 &  0.023\\ 
   19.83 &  46.67   & 11.75 & 118.26 & 11.72 & -0.036 &  0.121 & -0.065 &  0.090\\
\hline
\end{tabular}
\end{table*}

\begin{table*}[h!]
\caption{The long-slit kinematics of NGC\,6861 \label{Tab_CaT_NGC6861}}
\begin{tabular}{lllllllll}
R (\arcsec)  & V (\kms) & dV (\kms) & $\sigma$ (\kms) & $d\sigma$ (\kms) &$h_3$ & $dh_3$ & $h_4$ & $dh_4$\\
\hline
  -10.01 & -257.90  &   13.73 & 279.18&  11.94 & 0.195 &  0.033 &-0.107 &  0.039\\  
   -7.78 & -273.20  &    9.91 & 275.60&   8.89 & 0.201 &  0.025 &-0.107 &  0.029\\  
   -5.96 & -263.80  &    9.36 & 299.99&  10.24 & 0.173 &  0.021 &-0.029 &  0.026\\  
   -4.14 & -235.30  &    7.84 & 307.07&  12.17 & 0.128 &  0.022 & 0.001 &  0.026\\  
   -2.81 & -196.60  &   10.48 & 323.65&  13.82 & 0.116 &  0.022 & 0.067 &  0.032\\  
   -1.90 & -126.20  &    8.38 & 362.01&   8.09 & 0.139 &  0.022 & 0.050 &  0.021\\  
   -0.99 &  -81.17  &    8.37 & 365.61&   6.58 & 0.065 &  0.017 & 0.050 &  0.021\\  
   -0.08 &    0.89  &    9.07 & 403.95&  11.68 & 0.083 &  0.016 & 0.034 &  0.021\\  
    0.84 &   70.00  &   10.72 & 380.21&  10.71 & 0.019 &  0.018 &-0.015 &  0.022\\  
    1.75 &  128.90  &    7.97 & 361.96&  10.57 &-0.012 &  0.017 & 0.019 &  0.024\\  
    2.66 &  196.60  &    8.82 & 333.02&   7.74 &-0.065 &  0.020 &-0.031 &  0.022\\  
    3.57 &  258.80  &    9.17 & 311.83&   9.65 &-0.092 &  0.017 & 0.044 &  0.032\\  
    4.89 &  273.30  &    8.16 & 299.81&  10.54 &-0.084 &  0.020 &-0.003 &  0.025\\  
    6.72 &  284.70  &   10.21 & 275.83&  11.75 &-0.100 &  0.029 &-0.040 &  0.036\\  
    8.94 &  260.80  &   12.40 & 292.37&  15.25 &-0.046 &  0.035 &-0.043 &  0.040\\
\hline
\end{tabular}
\end{table*}

\section{The VIRUS-W kinematics of NGC\,3091 and NGC\,5328}
\label{appendix3091}
The observations of NGC\,3091 and NGC\,5328 were carried out using the new
VIRUS-W spectrograph \citep{Fabricius-12b} at the 2.7\,m Harlan J. Smith
telescope at the McDonald Observatory in Texas. NGC\,3091 was observed during
the commissioning of VIRUS-W in December 2010 while for NGC\,5328 we were
granted with two nights of observing time in May 2011.

VIRUS-W offers a fiber-based Integral Field Unit (IFU) with a
  field of view (FoV) of 105 arcsec$\times$55 arcsec. The 267
  individual fibers have a core diameter of 3.14 arcsec on sky. The
  fibers are arranged in a rectangular matrix in a densepack scheme
  with a fill factor of 1/3, such that three dithered observations are
  necessary in order to achieve 100\,\% coverage of the FoV. The
  instrument offers further two different modes of spectral
  resolution.  For both galaxies we used the lower resolution mode (R
  = 3300, translating to $\sigma_{instr}=39$ \kms) with a spectral
  coverage of 4320\,\AA~to 6042\,\AA.

\begin{figure}
 \centering
 \includegraphics[width=.48\textwidth]{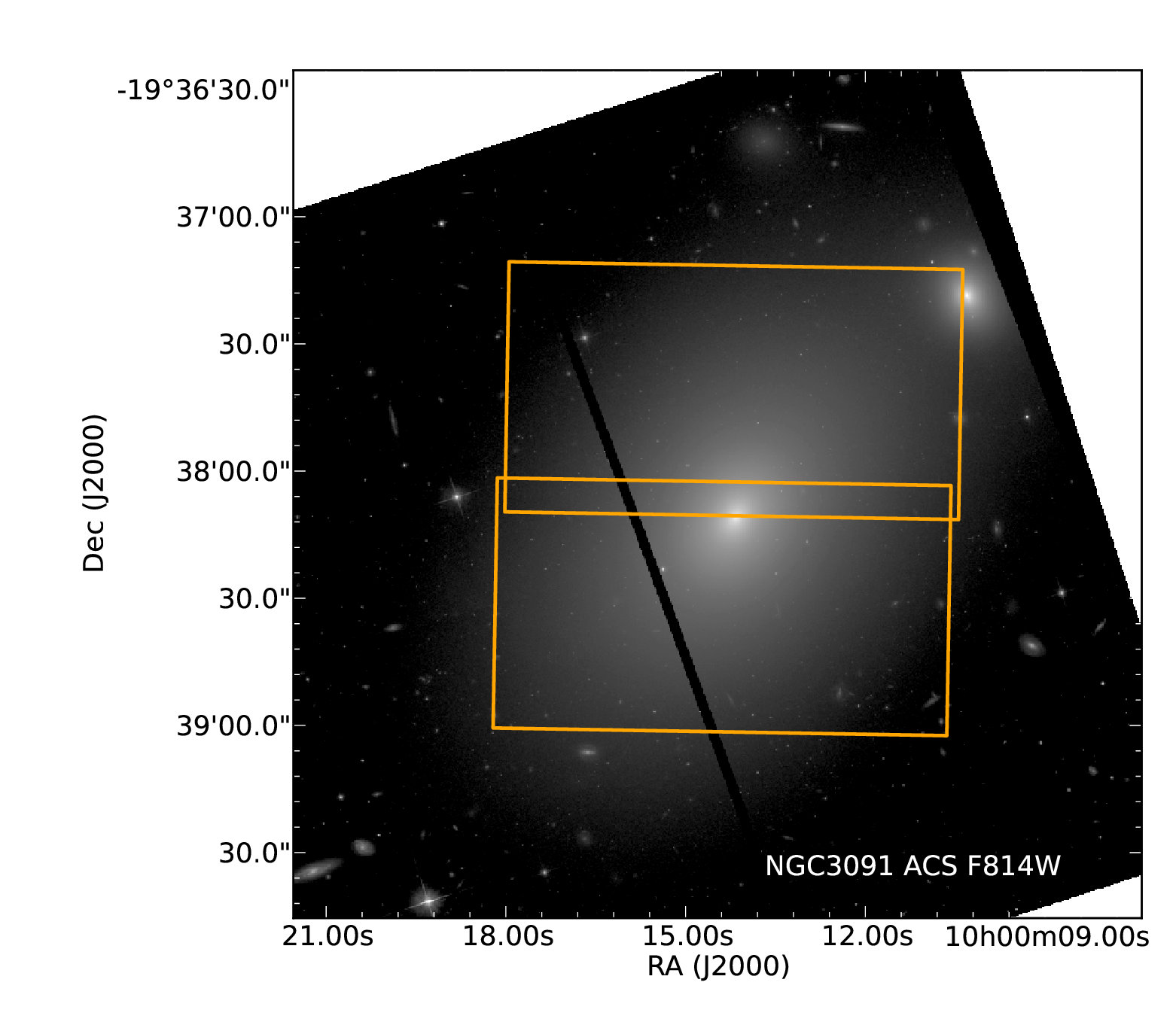} 
 \includegraphics[width=.48\textwidth]{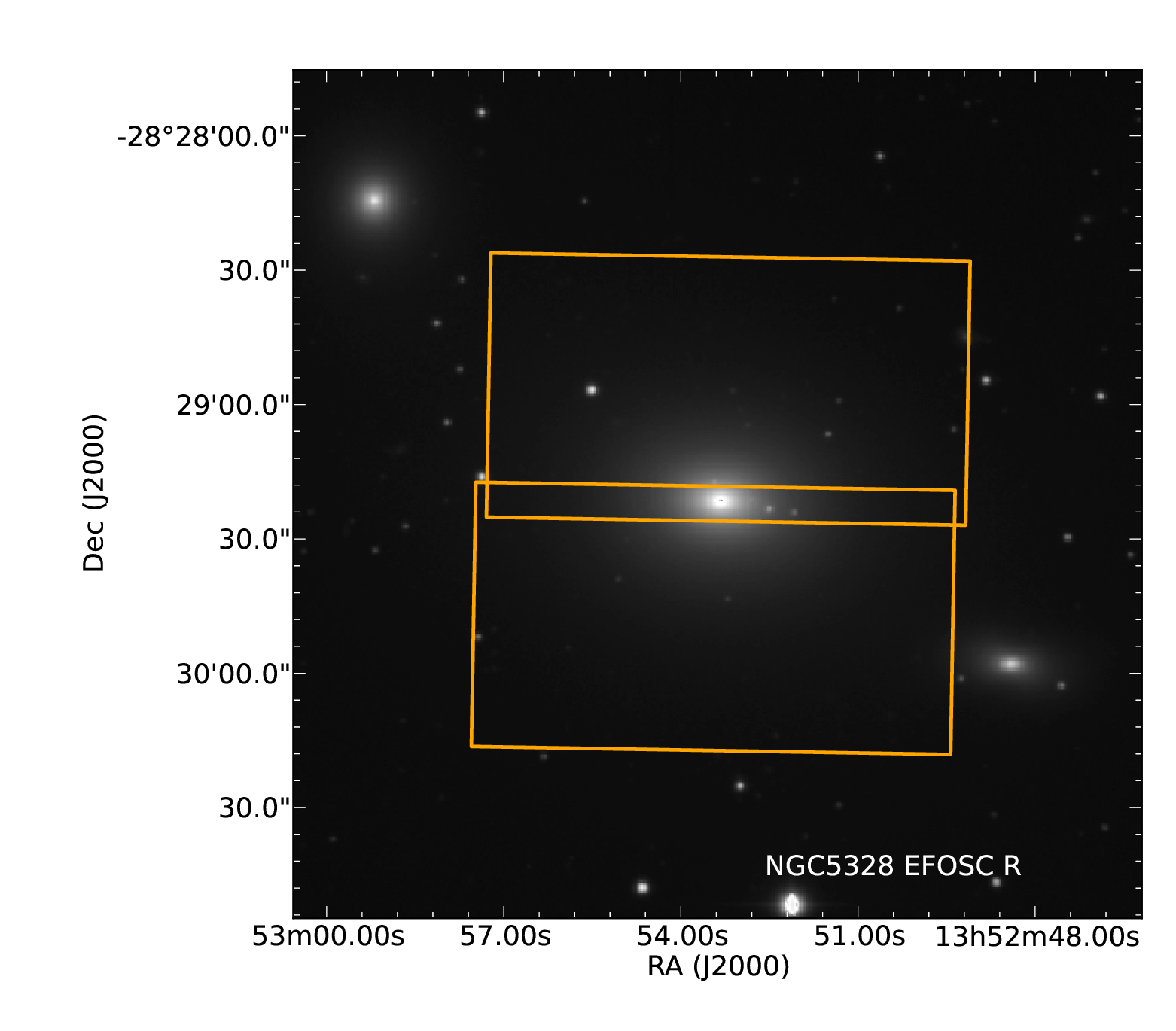}
 \caption[]{Left panel: HST ACS F814W image of NGC\,3091. The boxes
   outline the instrumental FoV of our two different VIRUS-W
   fields. Right panel: VIRUS-W FoV for NGC\,5328 overplotted on a
   EFOSC2 R band \citep{gbauch-05} image.}
\label{fig:N3091_N5328_FoV}
\end{figure}

We obtained dithered observations of two off-centred and slightly overlapping
regions of NGC\,3091 and NGC\,5328 in the night of December 5 and the two
nights of May 24 and 25 respectively.  The two plots in
Fig.\,\ref{fig:N3091_N5328_FoV} show images of the two galaxies with boxes that
outline the field positions.

The exposure time of the individual on-object pointings was 600\,s for
NGC\,3091 and 700\,s for NGC\,5328. We repeated each pointing once for cosmic
ray rejection, resulting in a total on-object integration time of 2\,h and
2.3\,h for NGC\,3091 and NGC\,5328 respectively. All on-object observations
were bracketed and interleaved with 300\,s sky nods that were offset by 9
arcmin to the north.  During the observations of NGC\,3091 the seeing varied
from 2.0 to 2.2 arcsec (FWHM) while for NGC\,5328 it varied from 2.0 to 2.5
arcsec, which is small compared to the fiber size.  In addition to the galaxies,
we also observed the two giant stars HR\,7576 and HR\,2600 which serve as
templates for the kinematic extraction. 

We extracted the individual fiber spectra using the {\tt cure} pipeline that
was originally developed by Ralf Koehler, Niv Drory and Jan Snigula for the
HETDEX experiment. The basic image reduction uses the {\tt fitstools} package
\citep{Goessl-02} and follows standard recipes for creation and subtraction of
master bias frames and the averaging of flat fields and calibration lamp
frames. {\tt cure} traces the positions of the fiber spectra on the chip using
the twilight or dome flat frames and then searches for the calibration lamp
line positions along those traces. Trace and wavelength positions are modelled
as a two-dimensional Chebyshev polynomial of 7-th degree across the CCD
surface.  The standard deviation of this wavelength/distortion solution is
0.2\, pixels which corresponds to 0.1\,\AA~at the linear dispersion of
0.52\,\AA/pixel.  In the next step, {\tt cure} calculates transformations
between fiber number/wavelength and x/y positions and corresponding inverse and
cross transformations. Given these models for fiber position and wavelength
calibration, the fiber extraction walks along the trace positions across the
CCD and extracts the spectra at given wavelength steps. We used a 7-pixel-wide
extraction aperture around the centroid position of the individual fiber
spectra. Tests showed that this results in less than 1\,\% loss of flux and
less than 0.1\,\% crosstalk between neighbouring spectra.

The extraction produces a total of $2 \times 3 \times 267 = 1062$ wavelength
calibrated spectra, corresponding to the two fields with three dithers and 267
fiber apertures. We extracted spectra binned linearly in wavelength space.

Preliminary kinematic maps for NGC\,3091 showed unrealistic north-south
gradients of the dispersion which we identified to be a consequence of an
imperfect sky subtraction when the data of the interleaved sky nods were used.
We rejected the data from the sky nods and identified fibers around the edges
of the combined FoV to estimate the background. We selected them to be located
further than 50 arcsec from the galaxy center and visually rejected fibers that
fall on the neighbouring galaxy MCG\,-03-26-006.  We averaged the skyfiber
spectra for each dither position separately while rejecting spurious events
using a standard, iterative kappa-sigma-clipping algorithm. The resulting
average sky spectrum is subtracted from the corresponding object spectra.

For NGC\,5328 the use of the sky nods did not result in similar gradients.
We averaged the bracketing sky exposures for each science exposure while
rejecting cosmics.  The sky signal is then calculated from the extracted sky
spectra by averaging the flux of 20 neighbouring fibers in a moving window
approach. The resulting spectra are scaled by exposure time and subtracted from
the science data.

The modelling employs a radial binning scheme for the SINFONI data
described in Section \ref{kinematicdata}.  For NGC\,3091 we binned the
VIRUS-W data following the same procedure. The binning step again
rejects residual events.  Bins that are located inside a radius of 10
arcsec around the center are rejected because there the size of the
fiber aperture is comparable or larger than the bin size of the radial
binning scheme.  The median S/N per \AA~in the final bins is 19.

For NGC\,5328 we instead employed a Voronoi binning scheme. The bin
sizes were calculated using the code of \citet{Cappellari-03} and
then combined in the same manner as for NGC\,3091. The minimum S/N per
\AA~in the used bins is 25 with a median of 40.

After rebinning of the spectra into log-wavelength space, we used the
Fourier Correlation Quotient (FCQ) algorithm (FCQ;
\citealp{Bender-90}; BSG94) to extract the line of sight velocity
distributions. For this work, we use a single template spectrum of the K3III
star HR\,7576. We tested fits with the additional K2III star HR\,2600,
which yielded consistent results. We fit the kinematics in the rest
frame spectral range from 4537\,\AA\ to 5443\,\AA. The continuum is
removed using an 8th degree polynomial and the first and last three
channels in the Fourier space are filtered out. We finally used Monte
Carlo simulations as described in \citet{Mehlert-00} for the
derivation of statistical errors.

Table \ref{Tab_NGC3091} provides the resulting kinematics for
  NGC\,3091, using the same spatial binning scheme as for the SINFONI
  data shown in Fig.  \ref{kinmap1}-\ref{kinmap4}. Table
  \ref{Tab_NGC3091_folded} shows the kinematics for the folded
  quadrant used in the model. The velocities $V$ are given relative to
  the average velocity.  Similarly, Table \ref{Tab_NGC5328} gives the
  resulting kinematics for NGC\,5328. The Voronoi bin centers are
  listed in the first two columns.  The velocities $V$ are given
  relative to the average velocity.  

\begin{table*}[h!]
\caption{The VIRUS-W Kinematics of NGC\,3091: four quadrants \label{Tab_NGC3091}}
\begin{tabular}{lrrrrrrrrr}
Radius (\arcsec)  & iv & $V$ (\kms) & $dV$ (\kms) & $\sigma$ (\kms) & $d\sigma$ (\kms) &$h_3$ & $dh_3$ & $h_4$ & $dh_4$\\
\hline
-42.29 &-3 &-156.03 &20.43 &235.67 & 18.02 & 0.130 &0.045 &-0.059 &0.032  \\
-42.29 &-1 &-150.63 &18.92 &272.33 & 25.56 & 0.043 &0.037 &-0.008 &0.031 \\
-42.29 & 1 &-207.45 &15.69 &221.72 & 21.44 & 0.004 &0.044 &-0.086 &0.031 \\
-42.29 & 3 &-134.08 &39.99 &322.73 & 56.95 & 0.034 &0.050 &-0.034 &0.053 \\
-32.75 &-5 &  -8.43 &30.24 &323.64 & 35.80 & 0.088 &0.043 & 0.034 &0.035 \\
\hline
\end{tabular}
\tablecomments{The binning is the same as for the SINFONI data. Radius is the radial midpoint of each bin. iv is the numbering for the five angular bins (see Section \ref{kinematicdata}). The sign of the radius and iv indicate the quadrant in which the bin is located. Positive radii and iv correspond to the quadrant with positive x-axis and y-axis values, respectively (see the kinematic map in Fig. \ref{kinmap2}). This table is published in its entirety in the electronic edition of AJ; a portion is shown here for guidance regarding its form and content.\\}
\end{table*}

\begin{table*}[h!]
\caption{The VIRUS-W Kinematics of NGC\,3091: the folded quadrant  \label{Tab_NGC3091_folded}}
\begin{tabular}{lrrrrrrrrr}
Radius (\arcsec)  & iv & $V$ (\kms) & $dV$ (\kms) & $\sigma$ (\kms) & $d\sigma$ (\kms) &$h_3$ & $dh_3$ & $h_4$ & $dh_4$\\
\hline
  11.73  & 1  &   88.63  &    6.67  &  292.87  &    6.59  &   -0.016  &    0.016  &   -0.008  &    0.016  \\
  15.17  & 1  &  100.14  &    6.50  &  284.19  &    6.52  &   -0.020  &    0.016  &   -0.004  &    0.016  \\
  19.61  & 1  &  116.13  &    7.21  &  296.92  &    7.00  &   -0.020  &    0.017  &   -0.018  &    0.017  \\
  25.35  & 1  &  130.88  &    7.30  &  286.45  &    7.40  &   -0.022  &    0.018  &   -0.001  &    0.017  \\
  32.75  & 1  &  108.75  &    7.81  &  303.42  &    7.50  &   -0.015  &    0.018  &   -0.025  &    0.017  \\
  42.29  & 1  &  157.37  &    7.90  &  268.64  &    7.54  &   -0.015  &    0.020  &   -0.028  &    0.019  \\
  11.72  & 2  &   72.27  &    6.56  &  284.45  &    6.59  &   -0.002  &    0.016  &   -0.005  &    0.016  \\
  25.35  & 2  &  112.73  &    6.88  &  275.51  &    6.78  &   -0.013  &    0.018  &   -0.015  &    0.017  \\
  32.75  & 2  &  132.37  &    7.78  &  294.77  &    7.28  &   -0.032  &    0.018  &   -0.038  &    0.018  \\
  11.72  & 3  &   73.21  &    6.75  &  290.16  &    6.76  &    0.015  &    0.016  &   -0.002  &    0.016  \\
  15.17  & 3  &   50.97  &    6.88  &  290.66  &    6.82  &   -0.004  &    0.017  &   -0.010  &    0.016  \\
  19.61  & 3  &   78.73  &    6.89  &  298.15  &    6.24  &   -0.021  &    0.016  &   -0.045  &    0.016  \\
  25.35  & 3  &   95.69  &    6.94  &  283.83  &    6.49  &   -0.025  &    0.017  &   -0.034  &    0.017  \\
  32.75  & 3  &  110.00  &    8.17  &  303.03  &    7.88  &    0.004  &    0.019  &   -0.024  &    0.018  \\
  42.29  & 3  &  102.34  &   10.27  &  300.84  &    9.90  &   -0.022  &    0.023  &   -0.025  &    0.021  \\
  11.72  & 4  &   42.38  &    6.54  &  277.24  &    6.56  &   -0.007  &    0.016  &   -0.002  &    0.016  \\
  15.17  & 4  &   59.26  &    6.68  &  277.34  &    6.65  &    0.014  &    0.017  &   -0.004  &    0.017  \\
  19.61  & 4  &   58.89  &    7.12  &  311.81  &    6.55  &   -0.008  &    0.016  &   -0.039  &    0.016  \\
  25.35  & 4  &   73.76  &    7.53  &  291.71  &    7.63  &   -0.031  &    0.018  &   -0.001  &    0.018  \\
  32.75  & 4  &   69.99  &    7.97  &  280.78  &    8.15  &   -0.031  &    0.020  &   -0.007  &    0.019  \\
  11.72  & 5  &   23.11  &    6.38  &  278.03  &    6.71  &    0.019  &    0.016  &    0.019  &    0.016  \\
  15.17  & 5  &   20.00  &    6.23  &  266.46  &    6.15  &    0.001  &    0.016  &   -0.007  &    0.016  \\
  19.61  & 5  &   15.97  &    6.44  &  260.06  &    6.20  &    0.023  &    0.017  &   -0.017  &    0.017  \\
  25.35  & 5  &   22.20  &    6.86  &  263.70  &    6.68  &    0.007  &    0.018  &   -0.018  &    0.018  \\
  32.75  & 5  &   23.16  &    7.65  &  264.44  &    7.89  &   -0.031  &    0.020  &    0.005  &    0.019 \\
\hline
\end{tabular}
\tablecomments{The binning is the same as for the SINFONI data. Radius is the radial midpoint of each bin in the folded quadrant (see the SINFONI kinematic map in Fig. \ref{kinmap2}). iv is the numbering for the five angular bins (see Section \ref{kinematicdata}).\\}
\end{table*}

\begin{table*}[h!]
\caption{The VIRUS-W Kinematics of NGC\,5328 \label{Tab_NGC5328}}
{\tiny
\begin{tabular}{rrrrrrrrrr}
x (\arcsec)  & y (\arcsec) & $V$ (\kms) & $dV$ (\kms) & $\sigma$ (\kms) & $d\sigma$ (\kms) &$h_3$ & $dh_3$ & $h_4$ & $dh_4$\\
\hline
  0.14 &   1.44 &   11.86 &  5.45 & 365.91 &  5.64 & -0.011 & 0.009 &  0.032 & 0.010 \\
  0.32 &   0.84 &    4.47 &  8.31 & 380.59 &  8.66 & -0.039 & 0.013 &  0.020 & 0.017 \\
  0.27 &   3.98 &   -4.47 & 13.86 & 364.20 & 16.18 & -0.027 & 0.023 &  0.045 & 0.023 \\
  2.53 &   3.20 &  -14.69 &  8.75 & 344.91 &  9.57 & -0.020 & 0.014 &  0.025 & 0.016 \\
 -0.59 &  10.12 &   -7.79 &  6.76 & 319.64 &  9.62 & -0.011 & 0.017 &  0.003 & 0.017 \\
 -0.20 &   4.55 &   14.76 & 11.47 & 347.10 & 13.51 & -0.029 & 0.020 &  0.024 & 0.018 \\
  2.82 &   0.19 &   12.83 & 12.73 & 359.68 & 14.04 & -0.023 & 0.019 &  0.012 & 0.023 \\
  3.25 &  -0.39 &   -3.25 & 14.54 & 348.74 & 16.36 & -0.051 & 0.021 &  0.004 & 0.024 \\
 -2.41 &   2.23 &    1.93 & 13.92 & 367.98 & 13.59 &  0.006 & 0.018 &  0.001 & 0.020 \\
 -2.80 &   2.70 &   15.40 &  6.94 & 370.44 &  7.95 &  0.000 & 0.012 &  0.053 & 0.013 \\
 -2.54 &  -0.31 &   -1.29 &  8.47 & 369.81 & 11.09 & -0.012 & 0.014 &  0.030 & 0.017 \\
 -2.10 &  -0.93 &   29.23 &  9.59 & 366.87 & 11.88 & -0.015 & 0.019 &  0.020 & 0.018 \\
  0.58 &  -2.17 &  -18.73 & 10.68 & 333.10 & 12.94 &  0.001 & 0.022 &  0.004 & 0.023 \\
 -2.85 &   5.66 &   12.72 & 10.59 & 328.00 &  9.96 &  0.000 & 0.015 &  0.005 & 0.017 \\
 -5.07 &  15.42 &   16.40 &  7.16 & 298.97 & 10.68 &  0.014 & 0.015 &  0.035 & 0.019 \\
  3.35 &   8.70 &   -6.00 &  7.51 & 317.11 &  8.94 & -0.013 & 0.015 &  0.015 & 0.015 \\
  7.12 &  19.26 &  -24.02 &  6.37 & 251.77 & 10.02 &  0.025 & 0.017 &  0.034 & 0.020 \\
 -5.12 &   3.51 &   13.76 & 12.76 & 351.44 & 15.83 &  0.017 & 0.023 &  0.016 & 0.023 \\
 -5.08 &   0.84 &    0.54 &  8.48 & 339.17 &  8.22 & -0.001 & 0.013 &  0.014 & 0.014 \\
 -5.04 &   0.33 &  -19.94 & 10.50 & 355.19 & 12.10 & -0.040 & 0.020 & -0.011 & 0.019 \\
 -8.93 &   7.33 &   36.24 &  8.35 & 318.68 &  9.06 &  0.018 & 0.013 &  0.015 & 0.016 \\
 -5.56 &   4.08 &    8.15 & 12.29 & 357.87 & 12.71 & -0.019 & 0.020 & -0.006 & 0.018 \\
 -0.20 &  -9.84 &    5.40 &  5.84 & 280.98 &  9.85 & -0.020 & 0.016 &  0.025 & 0.016 \\
 -1.86 &  -3.47 &   16.22 &  9.84 & 346.22 & 12.14 &  0.040 & 0.018 &  0.040 & 0.016 \\
 -4.81 &  -2.22 &   28.34 & 10.29 & 346.75 & 11.43 &  0.042 & 0.017 &  0.003 & 0.022 \\
 -5.82 &  -7.84 &   36.81 &  9.65 & 301.21 & 11.98 &  0.011 & 0.018 &  0.026 & 0.020 \\
 -0.58 & -20.40 &    0.50 &  6.38 & 270.63 &  8.52 & -0.052 & 0.015 &  0.034 & 0.016 \\
 -7.94 &   1.98 &    4.90 &  8.04 & 303.80 &  9.52 &  0.010 & 0.016 &  0.011 & 0.017 \\
-19.08 &   7.40 &   70.12 &  6.99 & 278.10 &  7.54 & -0.005 & 0.015 & -0.019 & 0.015 \\
-12.81 &   1.48 &   59.15 &  5.67 & 298.24 &  8.20 &  0.030 & 0.013 &  0.016 & 0.012 \\
-13.78 &  19.79 &   45.78 & 10.24 & 299.89 & 12.51 &  0.005 & 0.020 &  0.004 & 0.021 \\
 -9.10 &  -5.04 &   53.46 &  9.79 & 313.76 &  9.75 &  0.010 & 0.017 &  0.000 & 0.015 \\
-16.18 &  -6.96 &   78.87 &  8.42 & 302.84 &  9.32 &  0.030 & 0.014 &  0.009 & 0.016 \\
 -7.61 &  -1.11 &   20.66 & 10.91 & 345.02 & 11.26 & -0.022 & 0.020 &  0.004 & 0.020 \\
-27.52 &  -4.08 &  117.83 &  7.54 & 283.45 &  9.55 &  0.065 & 0.019 & -0.005 & 0.017 \\
-12.11 & -18.81 &   39.38 &  6.42 & 287.55 &  7.50 & -0.017 & 0.016 &  0.004 & 0.015 \\
  7.42 &   7.45 &  -37.88 &  8.37 & 304.66 & 10.30 & -0.012 & 0.018 &  0.006 & 0.018 \\
 16.96 &  12.47 &  -67.12 &  6.60 & 280.65 &  8.00 & -0.013 & 0.017 & -0.032 & 0.014 \\
  5.56 &   1.77 &  -12.65 &  9.24 & 334.85 & 13.65 & -0.035 & 0.018 &  0.035 & 0.018 \\
 10.80 &   2.82 &  -36.74 &  8.02 & 306.04 &  8.68 & -0.032 & 0.016 &  0.012 & 0.015 \\
  4.85 &  -5.29 &  -16.55 &  7.99 & 347.65 &  9.91 & -0.019 & 0.014 &  0.042 & 0.014 \\
  9.07 & -15.65 &  -24.16 &  6.16 & 271.87 &  7.25 &  0.000 & 0.015 &  0.000 & 0.014 \\
 11.62 &  -5.40 &  -44.78 &  8.04 & 299.28 &  9.48 & -0.040 & 0.020 &  0.027 & 0.018 \\
  5.67 &  -1.31 &   -8.51 &  9.38 & 341.20 &  9.63 & -0.020 & 0.016 &  0.006 & 0.017 \\
 21.69 &  -1.25 &  -81.45 &  6.42 & 251.71 &  6.31 & -0.027 & 0.014 & -0.041 & 0.015 \\
 20.24 & -16.77 &  -51.68 &  6.68 & 277.99 &  7.65 & -0.007 & 0.015 & -0.025 & 0.014 \\
\hline
\end{tabular}
}
\tablecomments{
The x and y coordinates correspond to the centers of the Voronoi bins. The
x-axis is aligned with the major axis. Negative x-values lie eastwards of the
galaxy center while positive y-values point to north.
\\}
\end{table*}

\section{The WiFeS Kinematics of NGC\,4751 and NGC\,5516}
\label{appendix47515516}
NGC\,4751 and NGC\,5516 were observed in May 2010 at the ANU 2.3m telescope at Siding Spring Observatory. The instrument was WiFeS Integral Field spectrograph \citep{Dopita-07,Dopita-10}. The dual-beam system was used with the B3000 and R7000 VPH grating combination with Dichroic $\#$3. For NGC\,4751 we collected 6x1200 sec exposures in May 2010 and a 1200 sec sky exposure. For NGC\,5516 3x1200 sec object exposures plus a 1200 sec sky exposure were taken on 2010 May 19. The data reduction was performed using the dedicated IRAF pipeline software of the instrument. We derived the stellar kinematics from the blue frames, focusing on the wavelength region $\lambda=4900-5400$\AA, where we achieved a spectroscopic resolution of $\sigma_{\rm inst} = 58$ \kms\ measured from the width of the sky lines. The stellar kinematics was measured with the software described in \citet{Saglia-10} and \citet{Fabricius-12}. This uses the FCQ method of BSG94 in combination with a stellar library convolved to the instrumental resolution and can be applied iteratively to detect and subtract possible emission lines.  However, no detectable emission was found. Table \ref{Tab_NGC4751} and \ref{Tab_NGC5516} give the resulting kinematics for NGC\,4751 and NGC\,5516. We give the kinematics along NS slits (with positive radii indicating distances in arcsec from the center of the galaxy towards the North and negative radii towards the South), shifted with respect to the center of the galaxy along the East (positive shifts in arcsec) or West (negative shifts) directions. The velocities of NGC\,4751 are given relative to the one of the center pixel and those of NGC\,5516 are given relative to the average velocity. For both galaxies, we also provide the kinematics of the folded quadrant in Table \ref{Tab_NGC4751_folded} for NGC\,4751 and Table \ref{Tab_NGC5516_folded} for NGC\,5516. The spatial positions $x$ and $y$ follow Fig. \ref{ori4751} and Fig. \ref{ori5516}.

\begin{table*}[h!]
\caption{The Kinematics of NGC\,4751 Derived From WiFeS Data} \label{Tab_NGC4751}
\begin{tabular}{lcrrrrrrrr}
$R$ (\arcsec)  & Shift (\arcsec) & $V$ (\kms) & $dV$ (\kms) & $\sigma$ (\kms) & $d\sigma$ (\kms) &$h_3$ & $dh_3$ & $h_4$ & $dh_4$\\
\hline
 2.0 & 0.0 & 191.00 & 18.98 & 345.10 & 19.46 & -0.075 & 0.050 & -0.053 & 0.050\\
 3.0 & 0.0 & 226.50 & 18.57 & 319.10 & 20.01 & -0.092 & 0.053 & -0.035 & 0.053\\
 4.0 & 0.0 & 243.10 & 26.29 & 313.00 & 28.41 & -0.012 & 0.076 & -0.034 & 0.076\\
 5.0 & 0.0 & 286.80 & 24.75 & 230.40 & 23.62 &  0.009 & 0.098 & -0.077 & 0.098\\
 6.0 & 0.0 & 288.20 & 32.61 & 200.30 & 27.90 &  0.015 & 0.148 & -0.110 & 0.148\\
\hline
\end{tabular}
\tablecomments{$R$ are positive North, shifts are positive East. This table is published in its entirety in the electronic edition of AJ; a portion is shown here for guidance regarding its form and content.\\}
\end{table*}

\begin{table*}[h!]
\caption{The Kinematics of NGC\,5516 Derived From WiFeS Data \label{Tab_NGC5516}}
\begin{tabular}{lcrrrrrrrr}
$R$ (\arcsec)   & Shift (\arcsec) & $V$ (\kms) & $dV$ (\kms) & $\sigma$ (\kms) & $d\sigma$ (\kms) &$h_3$ & $dh_3$ & $h_4$ & $dh_4$\\
\hline
-6.0& -3.0&   10.10&  65.17& 373.09&  76.84&  0.033& 0.159& -0.001& 0.159\\   
-6.0& -4.0&   44.16&  57.13& 331.90&  57.88&  0.027& 0.156& -0.057& 0.156\\   
-6.0& -5.0&   40.87&  94.19& 509.47&  65.92&  0.022& 0.168& -0.163& 0.168\\   
-5.0&  1.0&   70.82&  36.98& 319.39&  25.42& -0.078& 0.105& -0.167& 0.105\\   
-5.0& -1.0&   63.88&  29.72& 308.21&  35.26& -0.004& 0.088&  0.002& 0.088\\   
\hline
\end{tabular}
\tablecomments{North is positive $R$ and East is positive shift. This table is published in its entirety in the electronic edition of AJ; a portion is shown here for guidance regarding its form and content.}
\end{table*}

\begin{table*}[h!]
\caption{The Kinematics of NGC\,4751: the folded quadrant} \label{Tab_NGC4751_folded}
{\tiny
\begin{tabular}{lrrrrrrrrr}
$x$ (\arcsec)  & $y$ (\arcsec) & $V$ (\kms) & $dV$ (\kms) & $\sigma$ (\kms) & $d\sigma$ (\kms) &$h_3$ & $dh_3$ & $h_4$ & $dh_4$\\
\hline
 0.65  &  4.07 &       55.60  &   19.15  &  217.39  &   20.73 &   0.0192 &   0.0802 &  -0.0355 &   0.0802\\
 1.64  &  4.16 &       62.61  &   17.88  &  158.15  &   13.06 &   0.1616 &   0.0974 &  -0.1429 &   0.0974\\
 2.64  &  4.25 &      138.20  &   22.51  &  188.10  &   19.81 &   0.0486 &   0.1098 &  -0.1056 &   0.1098\\
 3.64  &  4.33 &      185.71  &   29.82  &  197.98  &   24.41 &   0.0199 &   0.1430 &  -0.1542 &   0.1430\\
 4.63  &  4.42 &      199.36  &   43.41  &  231.75  &   35.55 &   0.0437 &   0.1640 &  -0.0648 &   0.1640\\
 5.63  &  4.51 &      211.42  &   61.72  &  202.26  &   57.83 &  -0.0701 &   0.2552 &  -0.0333 &   0.2552\\
 0.73  &  3.08 &       46.68  &   21.02  &  312.06  &   25.04 &  -0.0555 &   0.0604 &   0.0201 &   0.0604\\
 1.73  &  3.16 &       70.31  &   18.84  &  256.98  &   23.77 &   0.0072 &   0.0672 &   0.0298 &   0.0672\\
 2.73  &  3.25 &      171.41  &   20.31  &  234.63  &   18.42 &  -0.0142 &   0.0788 &  -0.0923 &   0.0788\\
 3.72  &  3.34 &      212.55  &   33.03  &  275.24  &   39.13 &   0.0763 &   0.1107 &  -0.0016 &   0.1107\\
 4.72  &  3.42 &      192.60  &   31.79  &  220.86  &   28.64 &   0.0668 &   0.1251 &  -0.0652 &   0.1251\\
 5.72  &  3.51 &      189.79  &   36.46  &  205.80  &   31.61 &   0.0721 &   0.1677 &  -0.1149 &   0.1677\\
 0.82  &  2.08 &       72.20  &   12.63  &  293.20  &   11.40 &   0.0107 &   0.0377 &  -0.0837 &   0.0377\\
 1.82  &  2.17 &      150.11  &   17.45  &  316.07  &   20.87 &   0.0259 &   0.0505 &   0.0032 &   0.0505\\
 2.81  &  2.25 &      190.80  &   17.30  &  261.40  &   19.87 &   0.0426 &   0.0601 &  -0.0111 &   0.0601\\
 3.81  &  2.34 &      185.47  &   18.11  &  231.61  &   18.21 &   0.0072 &   0.0742 &  -0.0498 &   0.0742\\
 4.81  &  2.43 &      206.05  &   26.10  &  215.86  &   22.25 &  -0.0350 &   0.1094 &  -0.1100 &   0.1094\\
 5.80  &  2.52 &      244.31  &   55.79  &  242.38  &   43.54 &  -0.0033 &   0.2093 &  -0.1379 &   0.2093\\
 1.91  &  1.17 &      180.28  &   11.49  &  318.19  &   12.98 &  -0.0465 &   0.0325 &  -0.0170 &   0.0325\\
 2.90  &  1.26 &      203.59  &   15.19  &  298.71  &   16.40 &   0.0307 &   0.0461 &  -0.0344 &   0.0461\\
 3.90  &  1.34 &      225.26  &   18.87  &  267.69  &   19.15 &   0.0225 &   0.0646 &  -0.0529 &   0.0646\\
 4.89  &  1.43 &      244.06  &   25.32  &  237.45  &   22.23 &  -0.0030 &   0.0948 &  -0.0935 &   0.0948\\
 5.89  &  1.52 &      275.37  &   35.85  &  233.09  &   30.17 &   0.0161 &   0.1361 &  -0.0965 &   0.1361\\
 1.99  &  0.17 &      147.60  &   13.47  &  345.68  &   13.44 &  -0.0370 &   0.0354 &  -0.0620 &   0.0354\\
 2.99  &  0.26 &      209.58  &   13.38  &  289.65  &   14.68 &  -0.0742 &   0.0420 &  -0.0298 &   0.0420\\
 3.98  &  0.35 &      221.17  &   19.79  &  262.47  &   19.40 &   0.0015 &   0.0649 &  -0.0521 &   0.0649\\
 4.98  &  0.44 &      268.58  &   20.07  &  216.06  &   15.38 &  -0.0280 &   0.0824 &  -0.1146 &   0.0824\\
 5.98  &  0.52 &      288.23  &   30.32  &  203.53  &   25.96 &   0.0163 &   0.1352 &  -0.1097 &   0.1352\\
 2.08  &  0.82 &      148.56  &   12.32  &  342.49  &   14.39 &  -0.0401 &   0.0325 &  -0.0018 &   0.0325\\
 3.08  &  0.73 &      199.69  &   18.06  &  336.42  &   21.41 &   0.0001 &   0.0466 &   0.0122 &   0.0466\\
 4.07  &  0.65 &      246.89  &   19.86  &  259.00  &   21.53 &  -0.1633 &   0.0613 &   0.0324 &   0.0613\\
 5.07  &  0.56 &      254.04  &   17.10  &  202.09  &   17.08 &  -0.0028 &   0.0771 &  -0.0610 &   0.0771\\
 6.06  &  0.47 &      269.76  &   31.74  &  229.37  &   33.49 &  -0.0019 &   0.1237 &  -0.0272 &   0.1237\\
 0.17  &  1.99 &      -11.38  &   13.44  &  305.63  &   10.86 &  -0.0364 &   0.0377 &  -0.0608 &   0.0377\\
 1.17  &  1.91 &       28.88  &   13.64  &  277.95  &   13.37 &  -0.0211 &   0.0412 &  -0.0212 &   0.0412\\
 2.17  &  1.82 &      109.12  &   20.05  &  290.62  &   22.25 &  -0.0210 &   0.0625 &  -0.0242 &   0.0625\\
 3.16  &  1.73 &      185.97  &   17.83  &  258.51  &   18.12 &   0.0283 &   0.0613 &  -0.0507 &   0.0613\\
 4.16  &  1.64 &      191.94  &   18.28  &  195.69  &   18.24 &  -0.0769 &   0.0850 &  -0.0627 &   0.0850\\
 5.16  &  1.56 &      218.44  &   19.97  &  173.68  &   18.03 &  -0.0321 &   0.1039 &  -0.0938 &   0.1039\\
 6.15  &  1.47 &      278.57  &   35.55  &  223.35  &   39.18 &   0.0101 &   0.1397 &  -0.0003 &   0.1397\\
 0.26  &  2.99 &        5.98  &   22.22  &  349.15  &   32.47 &  -0.0425 &   0.0549 &   0.1163 &   0.0549\\
 1.26  &  2.90 &       27.78  &   23.44  &  302.25  &   29.42 &  -0.0229 &   0.0675 &   0.0243 &   0.0675\\
 2.25  &  2.81 &      117.60  &   19.01  &  254.26  &   20.17 &   0.0519 &   0.0647 &  -0.0258 &   0.0647\\
 3.25  &  2.73 &      162.61  &   21.29  &  235.29  &   20.48 &   0.0076 &   0.0810 &  -0.0752 &   0.0810\\
 4.25  &  2.64 &      133.15  &   28.03  &  239.49  &   27.98 &   0.0242 &   0.1045 &  -0.0469 &   0.1045\\
 5.24  &  2.55 &      160.07  &   24.30  &  154.39  &   18.74 &   0.1511 &   0.1223 &  -0.0964 &   0.1223\\
 6.24  &  2.47 &      153.89  &   33.83  &  163.14  &   23.21 &   0.0303 &   0.1653 &  -0.1794 &   0.1653\\
 0.35  &  3.98 &       47.72  &   22.14  &  246.66  &   23.73 &   0.1088 &   0.0815 &  -0.0370 &   0.0815\\
 1.34  &  3.90 &       50.20  &   19.54  &  216.86  &   20.26 &  -0.0611 &   0.0826 &  -0.0498 &   0.0826\\
 2.34  &  3.81 &       98.76  &   28.14  &  244.92  &   32.55 &   0.0856 &   0.1034 &   0.0164 &   0.1034\\
 3.34  &  3.72 &      100.89  &   33.56  &  238.94  &   37.43 &  -0.0226 &   0.1278 &  -0.0280 &   0.1278\\
 4.33  &  3.64 &      204.88  &   40.01  &  257.43  &   36.70 &   0.0606 &   0.1397 &  -0.0836 &   0.1397\\
 5.33  &  3.55 &      186.09  &   31.74  &  159.29  &   26.43 &  -0.0175 &   0.1519 &  -0.0558 &   0.1519\\
 6.33  &  3.46 &      279.52  &   48.84  &  210.46  &   40.96 &  -0.1381 &   0.2133 &  -0.1137 &   0.2133\\
\hline
\end{tabular}
}
\tablecomments{The $x$ and $y$ values denote the spatial position of WiFeS datapoints according to Fig. \ref{ori4751}.}
\end{table*}

\begin{table*}[h!]
\caption{The WiFeS kinematics of NGC\,5516: the folded quadrant \label{Tab_NGC5516_folded}}
{\tiny
\begin{tabular}{lrrrrrrrrr}
$x$ (\arcsec)   & $y$ (\arcsec) & $V$ (\kms) & $dV$ (\kms) & $\sigma$ (\kms) & $d\sigma$ (\kms) &$h_3$ & $dh_3$ & $h_4$ & $dh_4$\\
\hline

   5.06  &    0.59  &   38.14  &   16.22  &  305.53  &   15.24  &   -0.048  &    0.038  &   -0.038  &    0.036  \\
   3.80  &    1.24  &   42.37  &   14.72  &  305.76  &   13.93  &   -0.025  &    0.035  &   -0.034  &    0.033  \\
   2.85  &    0.93  &   18.46  &   12.99  &  291.31  &   12.68  &   -0.035  &    0.032  &   -0.017  &    0.031  \\
   1.90  &    0.62  &    9.71  &   12.04  &  310.70  &   11.19  &   -0.018  &    0.028  &   -0.035  &    0.027  \\
   4.45  &    2.50  &   52.96  &   15.86  &  281.27  &   16.33  &    0.015  &    0.040  &    0.001  &    0.038  \\
   3.50  &    2.19  &   25.89  &   15.56  &  346.35  &   14.11  &   -0.014  &    0.032  &   -0.051  &    0.030  \\
   2.54  &    1.88  &   10.88  &   14.41  &  325.40  &   13.91  &    0.015  &    0.031  &   -0.020  &    0.030  \\
   4.14  &    3.45  &   73.87  &   18.02  &  360.91  &   17.66  &   -0.047  &    0.035  &   -0.016  &    0.033  \\
   2.24  &    2.83  &   34.81  &   13.01  &  285.49  &   13.29  &    0.026  &    0.032  &    0.002  &    0.031  \\
   1.28  &    2.52  &    9.20  &   12.47  &  316.55  &   12.60  &    0.009  &    0.028  &   -0.003  &    0.027  \\
   4.78  &    4.71  &    9.20  &   22.75  &  331.65  &   23.06  &   -0.005  &    0.047  &   -0.003  &    0.043  \\
   2.88  &    4.09  &   55.65  &   17.08  &  289.01  &   16.88  &   -0.026  &    0.041  &   -0.014  &    0.039  \\
   1.93  &    3.78  &   55.55  &   20.79  &  389.65  &   19.31  &    0.056  &    0.037  &   -0.037  &    0.035  \\
   0.98  &    3.47  &   56.85  &   18.97  &  358.69  &   18.38  &    0.008  &    0.037  &   -0.027  &    0.035  \\
   4.47  &    5.66  &   66.83  &   26.54  &  349.82  &   25.59  &    0.011  &    0.050  &   -0.019  &    0.046  \\
   3.52  &    5.35  &   45.58  &   20.66  &  334.85  &   20.51  &    0.001  &    0.041  &   -0.003  &    0.039  \\
   2.57  &    5.04  &   30.47  &   16.96  &  302.37  &   16.88  &   -0.045  &    0.039  &   -0.011  &    0.037  \\
   1.62  &    4.73  &   34.03  &   17.61  &  363.94  &   16.58  &    0.001  &    0.034  &   -0.036  &    0.032  \\
   0.67  &    4.42  &   -4.11  &   15.46  &  317.96  &   13.81  &   -0.010  &    0.035  &   -0.059  &    0.033  \\
   4.16  &    6.61  &   54.65  &   33.06  &  420.04  &   27.89  &    0.003  &    0.050  &   -0.050  &    0.047  \\
   2.26  &    5.99  &   28.34  &   17.75  &  232.11  &   17.13  &   -0.000  &    0.053  &   -0.025  &    0.049  \\
   1.00  &    6.63  &   15.22  &   22.93  &  310.71  &   21.17  &   -0.024  &    0.051  &   -0.042  &    0.046  \\
   0.62  &    1.90  &   -3.37  &   11.35  &  293.34  &   11.63  &    0.040  &    0.027  &    0.009  &    0.027  \\
   1.57  &    1.59  &    2.69  &   11.82  &  312.92  &   11.46  &   -0.002  &    0.027  &   -0.018  &    0.026  \\
   2.52  &    1.28  &   18.04  &   12.56  &  329.71  &   12.88  &   -0.005  &    0.027  &    0.004  &    0.027  \\
   3.47  &    0.98  &   27.05  &   14.88  &  320.47  &   13.79  &    0.037  &    0.033  &   -0.039  &    0.031  \\
   4.42  &    0.67  &   88.28  &   17.69  &  316.53  &   16.94  &    0.060  &    0.038  &   -0.020  &    0.036  \\
   0.93  &    2.85  &   -5.61  &   14.12  &  320.68  &   12.76  &    0.025  &    0.032  &   -0.052  &    0.030  \\
   1.88  &    2.54  &   11.21  &   13.28  &  316.55  &   13.69  &   -0.003  &    0.030  &    0.006  &    0.029  \\
   2.83  &    2.24  &    9.02  &   14.82  &  302.00  &   13.62  &   -0.005  &    0.035  &   -0.046  &    0.034  \\
   3.78  &    1.93  &   26.09  &   14.92  &  294.87  &   15.36  &   -0.032  &    0.035  &    0.005  &    0.034  \\
   4.73  &    1.62  &   31.84  &   17.95  &  311.08  &   17.18  &   -0.012  &    0.040  &   -0.025  &    0.038  \\
   1.24  &    3.80  &    3.38  &   15.04  &  323.55  &   15.28  &    0.045  &    0.033  &    0.001  &    0.031  \\
   2.19  &    3.50  &   -8.42  &   15.22  &  318.27  &   16.88  &   -0.041  &    0.034  &    0.042  &    0.032  \\
   3.14  &    3.19  &   11.10  &   16.58  &  313.03  &   16.50  &   -0.015  &    0.037  &   -0.011  &    0.035  \\
   4.09  &    2.88  &   -8.51  &   17.13  &  299.39  &   17.73  &    0.019  &    0.040  &    0.007  &    0.038  \\
   5.04  &    2.57  &   16.86  &   19.91  &  315.80  &   21.11  &    0.009  &    0.044  &    0.014  &    0.041  \\
   1.55  &    4.76  &  -35.67  &   17.36  &  319.34  &   17.85  &   -0.023  &    0.038  &    0.001  &    0.036  \\
   2.50  &    4.45  &    6.78  &   17.87  &  335.38  &   16.72  &    0.006  &    0.038  &   -0.040  &    0.035  \\
   3.45  &    4.14  &  -33.58  &   20.48  &  314.23  &   20.05  &    0.017  &    0.045  &   -0.016  &    0.041  \\
   4.40  &    3.83  &  -20.74  &   21.79  &  310.97  &   20.57  &   -0.005  &    0.048  &   -0.034  &    0.044  \\
   5.35  &    3.52  &  -28.60  &   25.26  &  336.72  &   23.52  &   -0.030  &    0.051  &   -0.038  &    0.046  \\
   1.85  &    5.71  &  -57.00  &   20.41  &  323.03  &   19.63  &    0.026  &    0.044  &   -0.028  &    0.041  \\
   2.16  &    6.66  &  -34.64  &   26.79  &  297.06  &   24.72  &   -0.015  &    0.060  &   -0.036  &    0.054 \\
\hline
\end{tabular}
}
\tablecomments{The $x$ and $y$ values denote the spatial position of WiFeS datapoints according to Fig. \ref{ori5516}.}
\end{table*}

\clearpage

\section[]{Kinematic fits of the best-fit models}
\label{appendixc}
\begin{figure*}[h!]
\centering
  \includegraphics[scale=0.555]{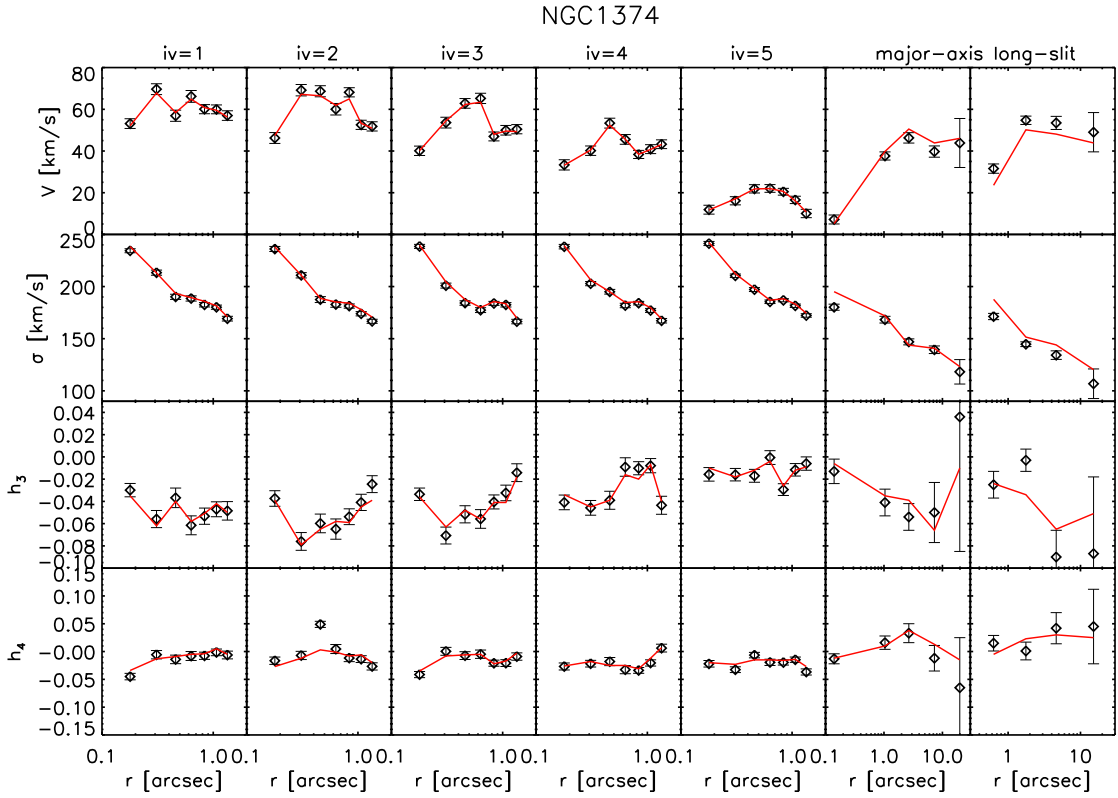}
  \caption[]{Kinematic data (black diamonds) and the best-fit model
    (red lines) for NGC\,1374. The first to fifth columns are for
    SINFONI data with increasing angle from major towards minor axis
    (see Fig. \ref{kinmap1}). The 6th and the 7th columns show both
    sides of the slit data along the major axis.\\}
\label{kin1374}
\end{figure*}

\begin{figure*}[h!]
\centering
  \includegraphics[scale=0.552]{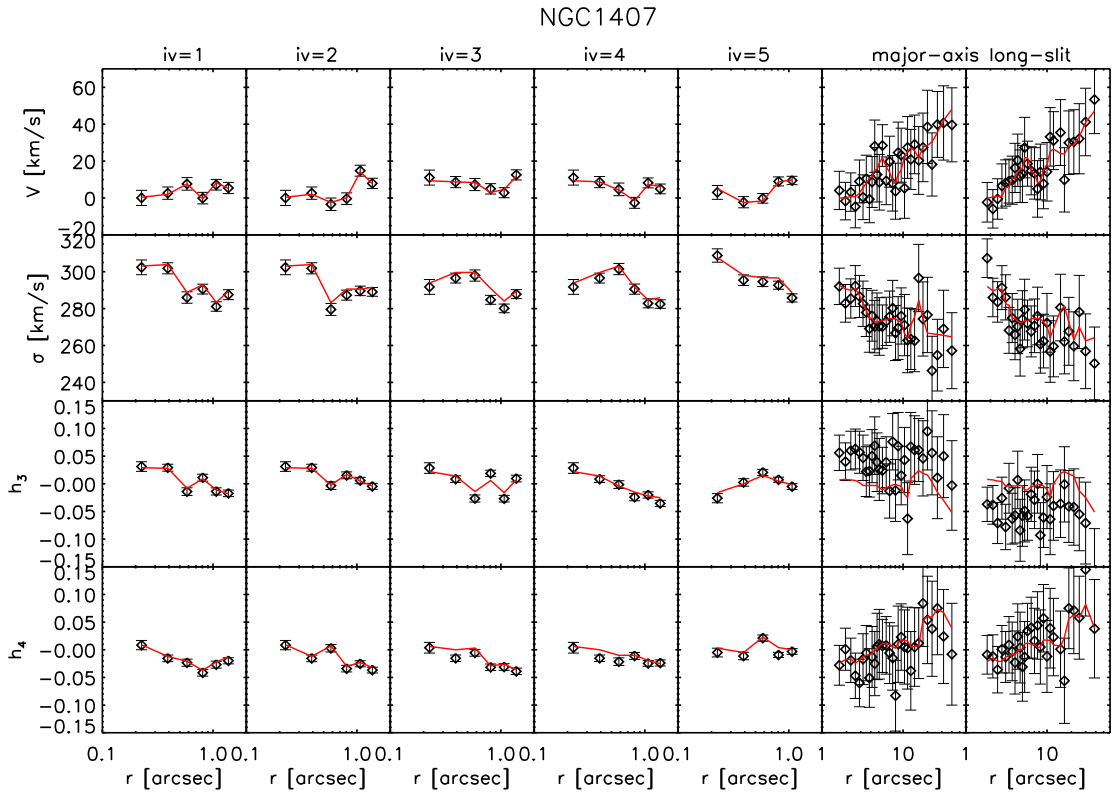}
  \caption[]{The same as Fig. \ref{kin1374} for NGC\,1407. The 6th and
    the 7th columns display both sides of the major-axis slit data presented in
    \citet{Spolaor-08a}.\\}
\label{kin1407}
\end{figure*}

\begin{figure*}[h!]
\centering
  \includegraphics[scale=0.552]{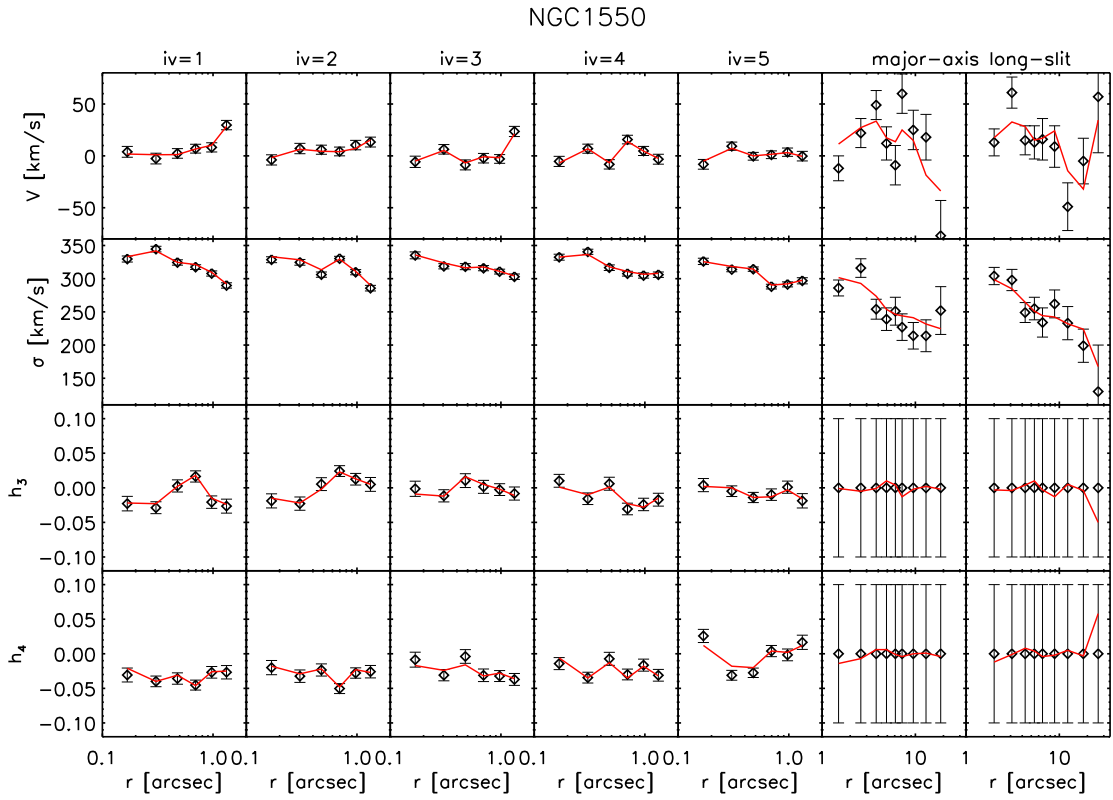}
  \caption[]{The same as Fig. \ref{kin1374} for NGC\,1550. The 6th and
    the 7th columns display both sides of the major-axis slit data from
    \citet{Simien-00}.\\}
\label{kin1550}
\end{figure*}

\begin{figure*}[h!]
\centering
  \includegraphics[scale=0.55]{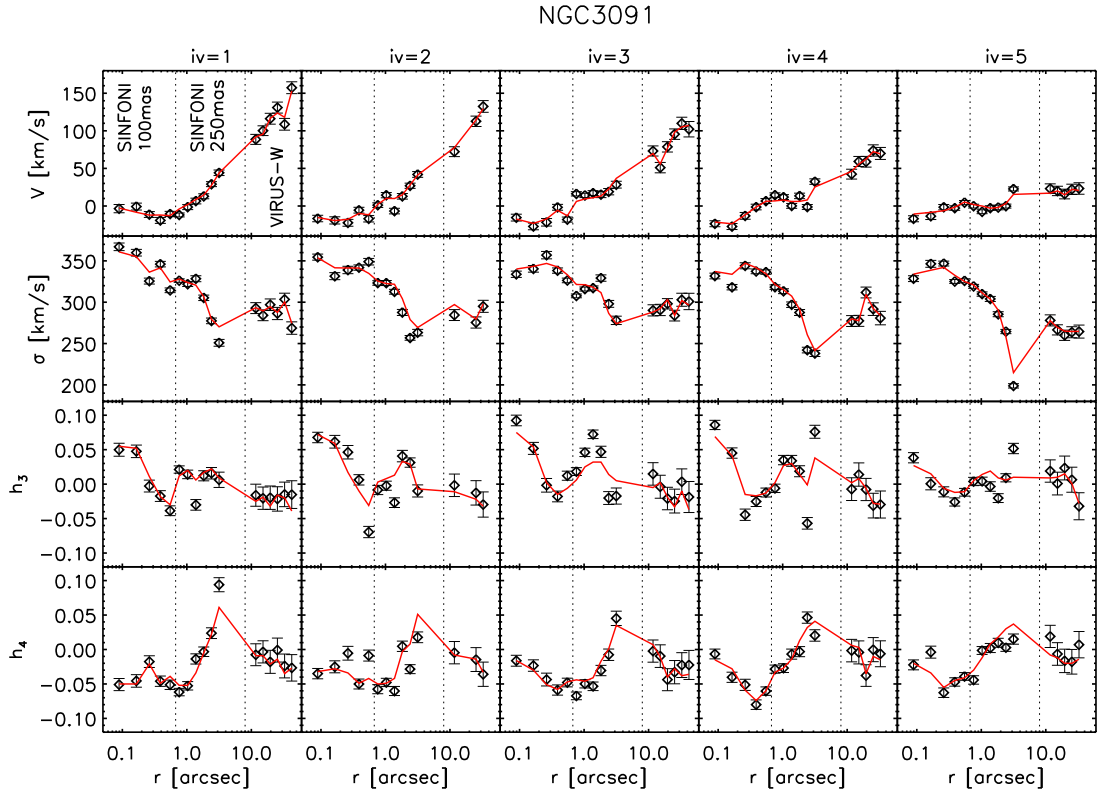}
  \caption[]{The same as Fig. \ref{kin1374} for NGC\,3091. The dotted vertical lines mark the region covered by SINFONI 100 mas data (leftmost), SINFONI 250mas data (middle) and VIRUS-W data (rightmost).}
\label{kin3091}
\end{figure*}

\begin{figure*}[h!]
\centering
  \includegraphics[scale=0.555]{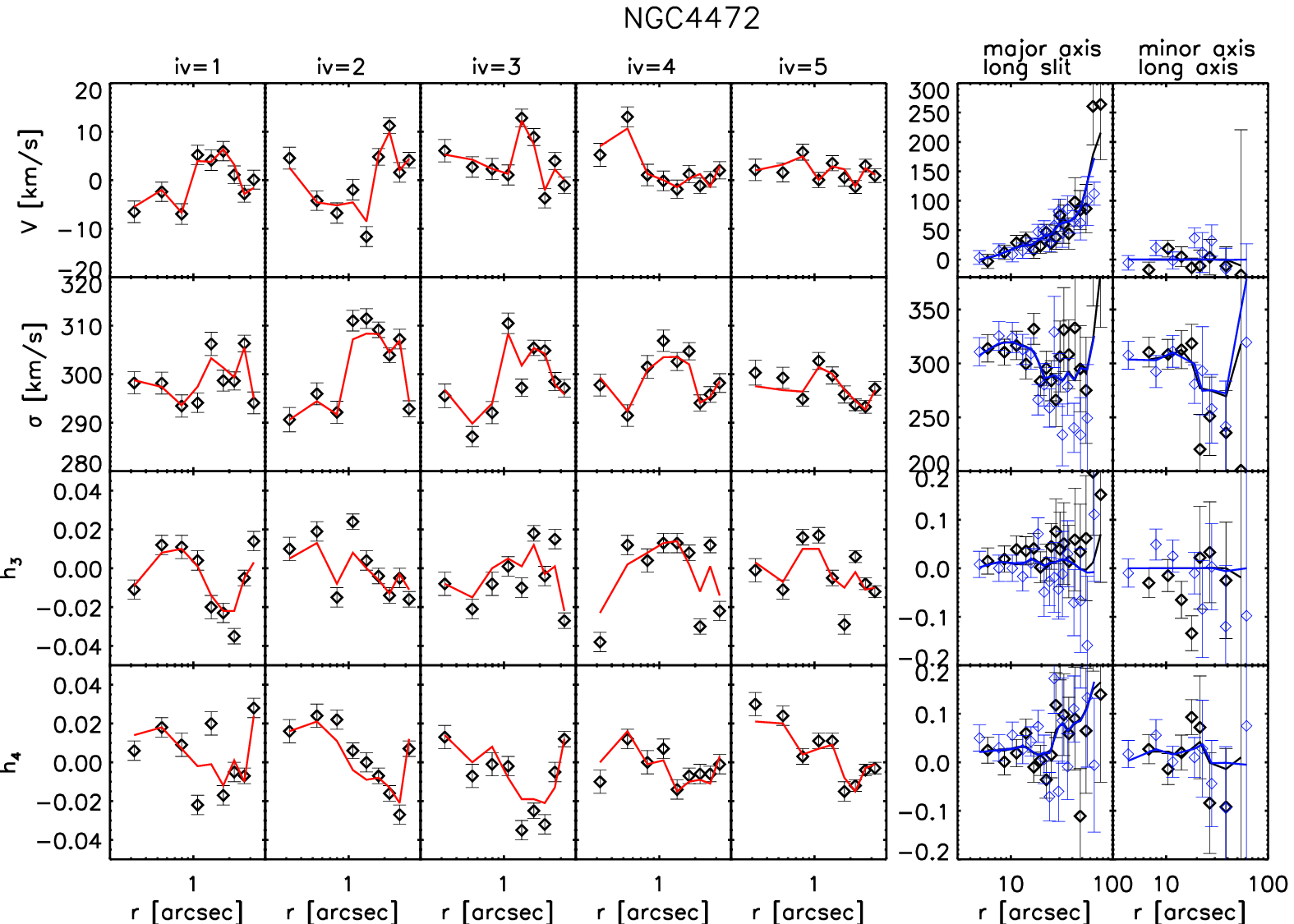}
  \caption[]{Kinematic data (diamonds) and the best-fit model (lines) for NGC\,4472. The first to fifth columns display SINFONI data with increasing angle from major towards minor axis. The 6th column shows both sides of the major axis data from \citet{Bender-94}. The 7th column shows both sides of minor axis data from the same paper. The different colors represent data from different sides.\\}
\label{kin4472}
\end{figure*}

\begin{figure*}[h!]
\centering
  \includegraphics[scale=0.54]{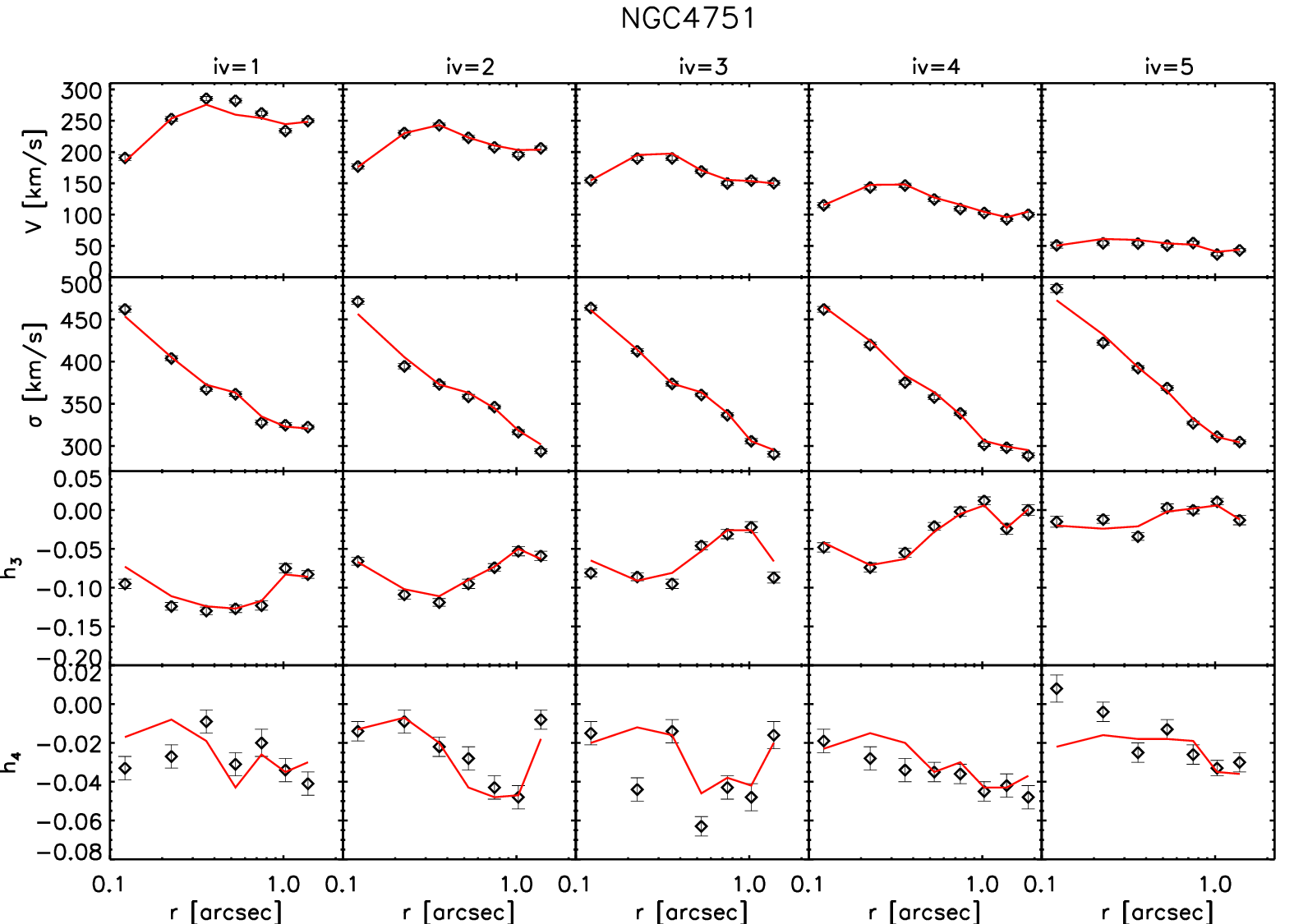}
  \caption[]{The same as Fig. \ref{kin1374} for NGC\,4751. The complementary data are shown in the next figure.\\}
\label{kin4751_sinfo}
\end{figure*}

\begin{figure*}[h!]
\centering
  \includegraphics[scale=0.54]{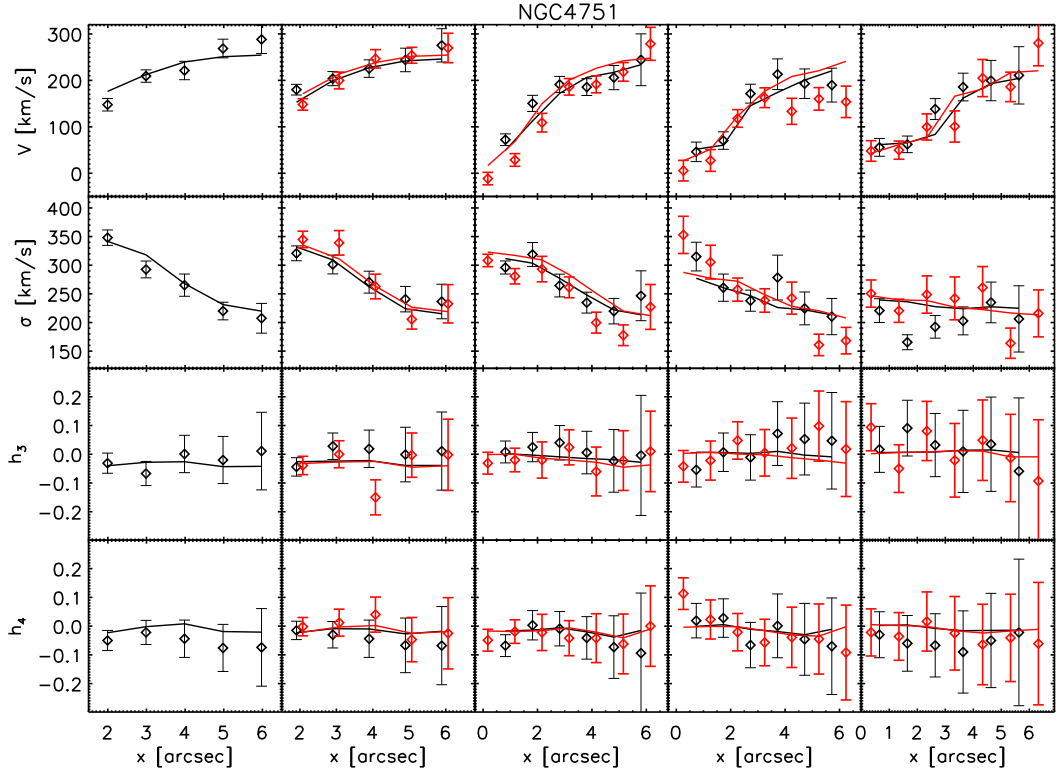}
\caption[]{WiFeS kinematic data (diamonds) and the best-fit model (lines) for NGC\,4751. Note that the Gauss-Hermite moments are not plotted as a function of radius but as a function of x-axis values (or projected radius onto the x-axis) of the datapoints. The leftmost to the rightmost columns show the pseudoslits in an ascending order of y-axis values. The black diamonds and lines correspond to the set of ``pseudoslit'' data having an angle of 5\degree\ and the red color corresponds to the one with an angle of -5\degree\ (see Section \ref{additionalkinematics} and Fig. \ref{ori4751}).\\}
\label{kin4751_slit}
\end{figure*}

\begin{figure*}[h!]
\centering
  \includegraphics[scale=0.85]{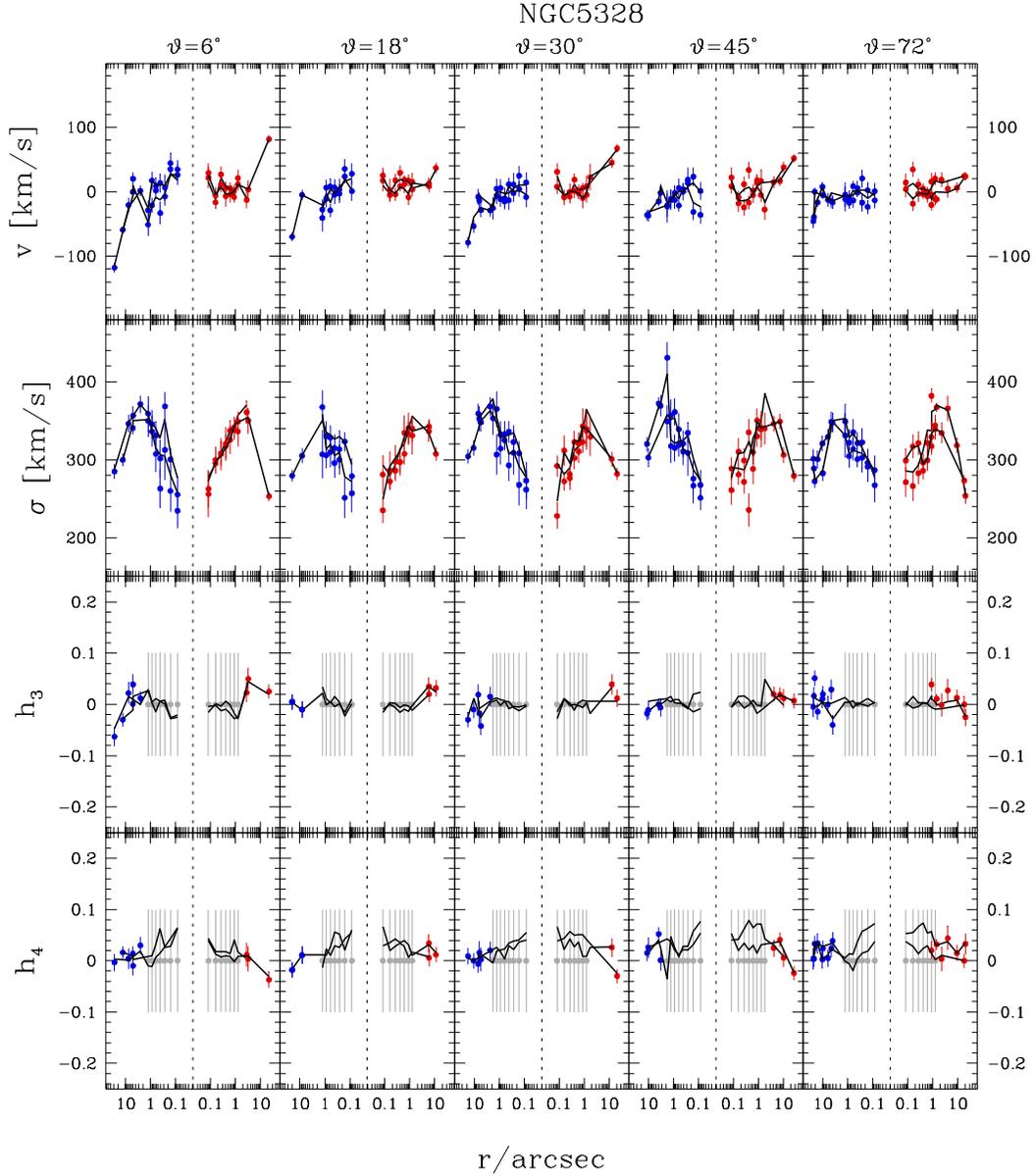}
  \caption[]{Kinematic data (circles) and the best-fit model
      (lines) for NGC\,5328. The five columns represent the angular
      binning ($\vartheta$ is the angle between the bin and the
      major-axis).  The VIRUS-W data points are included according to
      their bin centers. Over the SINFONI FoV we do not model the
      measured $h_3$ and $h_4$ but use $h_3=h_4=0$, $\Delta h_3=\Delta
      h_4=0.1$ (indicated in grey).\\}
\label{kin5328_all}
\end{figure*}

\begin{figure*}[h!]
\centering
  \includegraphics[scale=0.54]{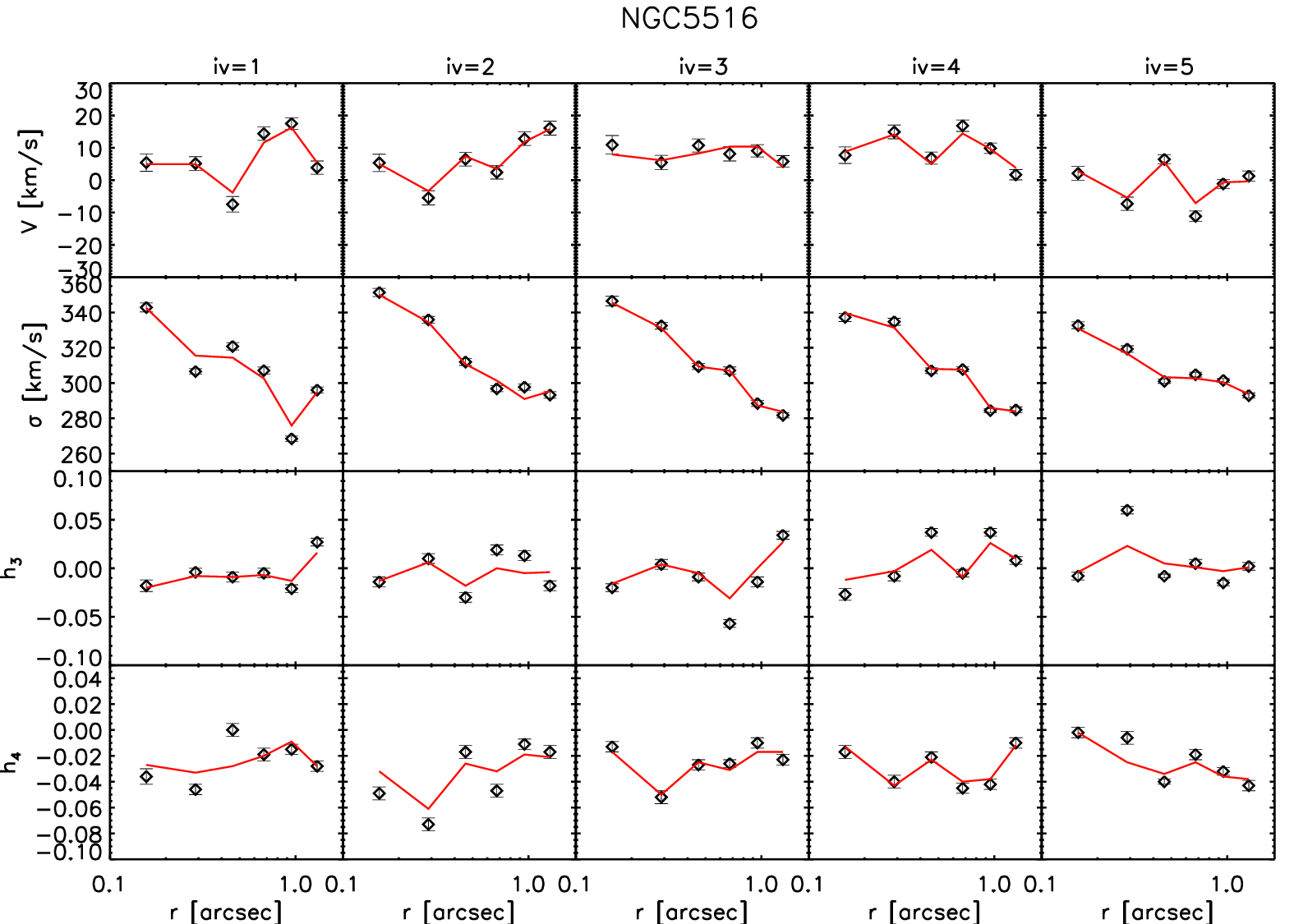}
  \caption[]{The same as Fig. \ref{kin1374} for NGC\,5516. The complementary data are shown in the next figure.\\}
\label{kin5516_sinfo}
\end{figure*}

\begin{figure*}[h!]
\centering
  \includegraphics[scale=0.54]{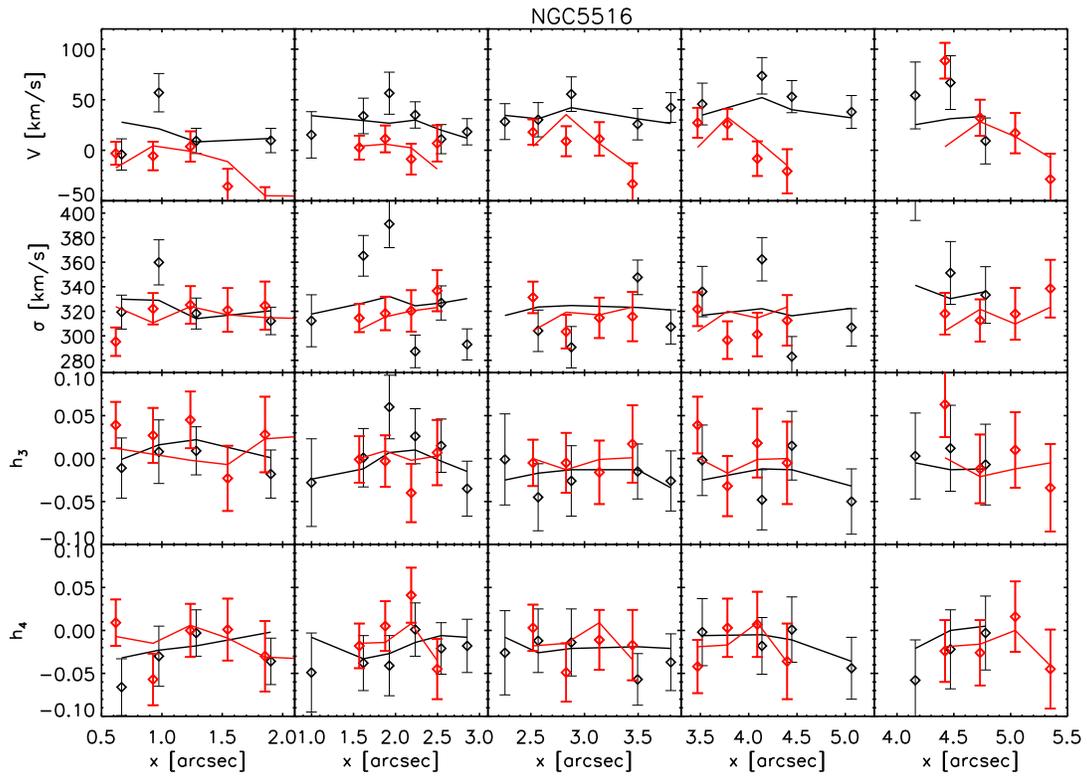}
  \caption[]{The same as Fig. \ref{kin4751_slit} for NGC\,5516. The black diamonds and lines correspond to the set of ``pseudoslit'' data having an angle of 108\degree\ and the red color corresponds to the one with an angle of 72\degree\ (see Section \ref{additionalkinematics} and Fig. \ref{ori5516}.)\\}
\label{kin5516_slit}
\end{figure*}

\begin{figure*}[h!]
\centering
  \includegraphics[scale=0.555]{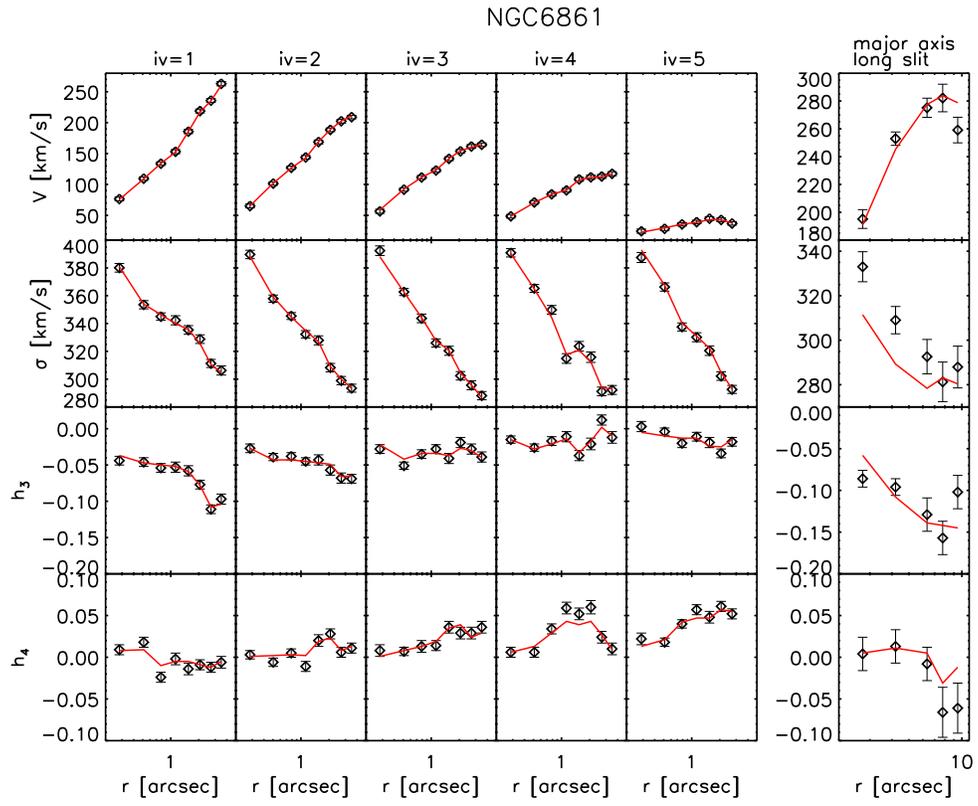}
  \caption[]{The same as Fig. \ref{kin1374} for NGC\,6861. The 6th column shows the slit data along the major axis.\\}
\label{kin6861}
\end{figure*}

\begin{figure*}[h!]
\centering
  \includegraphics[scale=0.555]{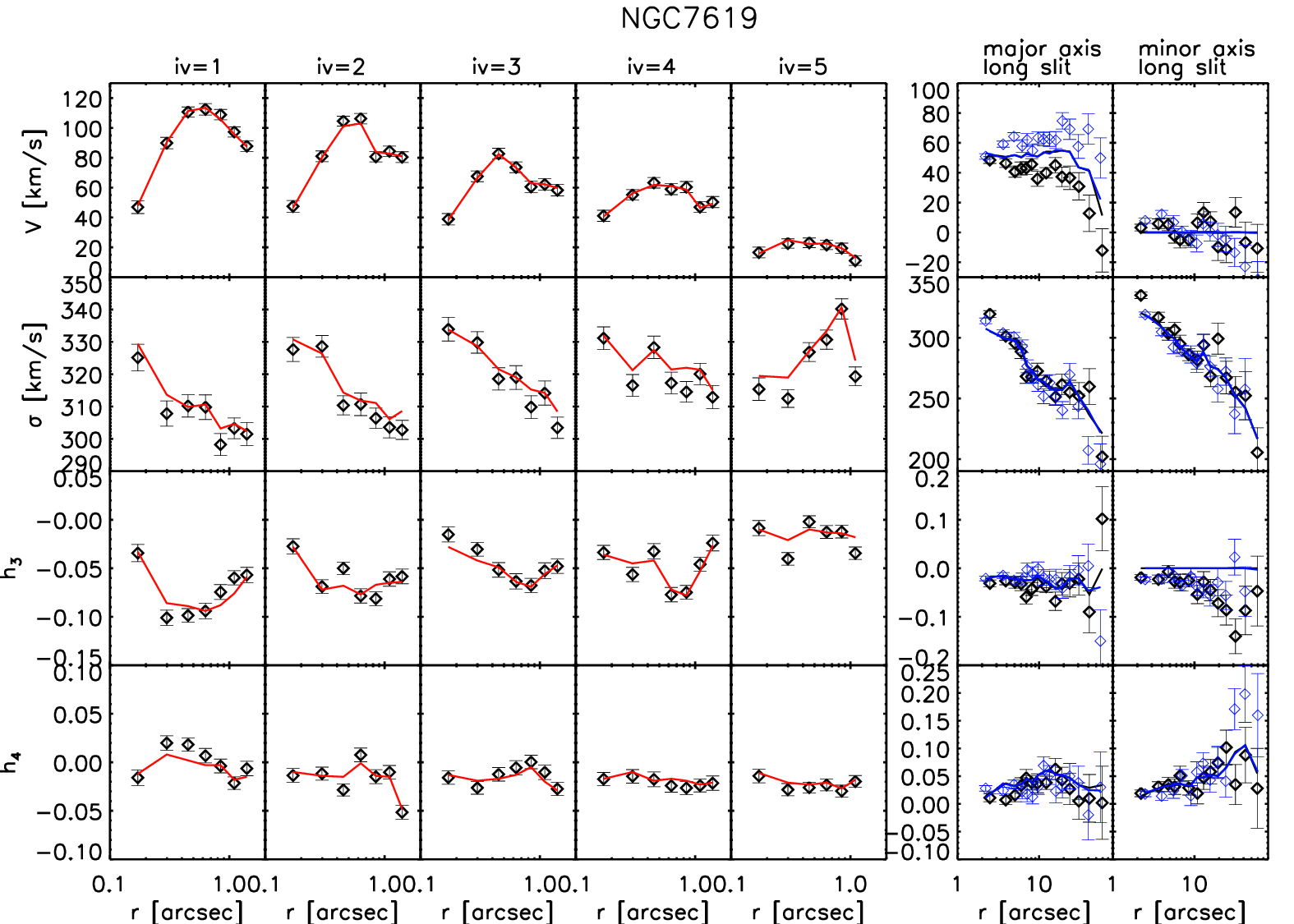}
  \caption[]{Kinematic data (diamonds) and the best-fit model (lines) for NGC\,7619. The first to fifth columns display SINFONI data with increasing angle from major towards minor axis. The 6th column shows both sides of the major axis data from \citet{Pu-10}. The 7th column show both sides of minor axis data from the same paper. The different colors represent data from different sides.\\}
\label{kin7619}
\end{figure*}

We show the kinematic fit of the best-fit model with DM to the data in
Figs. \ref{kin1374}--\ref{kin7619}. The kinematic data are represented
by the Gauss-Hermite moments up to $h_4$. For each galaxy, the SINFONI
averaged kinematics is plotted in five columns, corresponding to the
five angular bins. The leftmost column is for the major axis and the
rightmost column for the minor axis. The complementary (non-SINFONI) kinematic data
that are used in the modeling are also shown with the model fit. The
details of these data can be found in Section
\ref{additionalkinematics}. The diamonds are the datapoints and the
lines are the model (not always plotted in the same color as the
diamonds).

\clearpage

\bibliographystyle{aj}

\end{document}